\documentclass{article}
\usepackage{slashed}
\usepackage{amsmath}
\usepackage{amssymb}
\usepackage{mathrsfs,dsfont}
\usepackage{bbm}
\usepackage{float,color}
\usepackage{youngtab}
\usepackage{accents}
\usepackage{graphicx}
\usepackage{epstopdf}
\makeatletter
\@addtoreset{equation}{section}
\makeatother

\newcommand{\eq}[1]{\begin{equation}#1\end{equation}}
\newcommand{\al}[1]{\begin{align}#1\end{align}}
\newcommand{\spl}[1]{\begin{split}#1\end{split}}
\newcommand{\hoch}[1]{$\, ^{#1}$}
\newcommand{\ubar}[1]{\underaccent{\bar}{#1}}

\newcommand{\au}{{\ubar{\alpha}}}
\newcommand{\bu}{{\ubar{\beta}}}
\newcommand{\gu}{{\ubar{\gamma}}}
\newcommand{\du}{{\ubar{\delta}}}

\def\ba{\begin{array}}
\def\ea{\end{array}}
\def\bea{\begin{eqnarray}}
\def\eea{\end{eqnarray}}
\def\nn{\nonumber}
\newcommand{\ft}[2]{{\textstyle\frac{#1}{#2}}}
\def\fft#1#2{{\frac{#1}{#2}}}

\def\AS{{\tiny\begin{tabular}{|c|}
\hline
\\
\hline
\\
\hline
\end{tabular}}}

\def\SY{
{\tiny\begin{tabular}{|l|l|}
\hline
&\\
\hline
\end{tabular}}}

\def\MAT{
{\tiny\begin{tabular}{|l|l|}
\hline
\!\!$\times$\!\!&\\
\hline
\end{tabular}}}
\newcommand{\be}{\begin{equation}}
\newcommand{\ee}{\end{equation}}
\def\del{\partial}
\begin{document}

\vspace{25pt}

\begin{center}
{\Large {\bf 
ABJ Quadrality
}}

\vspace{10pt}

Masazumi Honda\hoch{1}, Yi Pang\hoch{2} and Yaodong Zhu\hoch{3}

\vspace{10pt}

\hoch{1} {\it
Department of Particle Physics and Astrophysics,
Weizmann Institute of Science, Rehovot 7610001, Israel}

\hoch{2} {\it Max-Planck-Insitut f\"{u}r Gravitationsphysik (Albert-Einstein-Institut)
Am M\"{u}hlenberg 1, DE-14476 Potsdam, Germany}

\hoch{3} {\it George and Cynthia Woods Mitchell Institute for Fundamental Physics and Astronomy,\\
Texas A\&M University, College Station, TX 77843, USA}

\vspace{40pt}

\underline{ABSTRACT}

\end{center}
We study physical consequences of adding orientifolds to the ABJ triality,
which is among 
3d ${\cal N}=6$ superconformal Chern-Simons theory known as ABJ theory,
type IIA string in $AdS_4 \times\mathbb{CP}^3 $
and ${\cal N}=6$ supersymmetric (SUSY) Vasiliev higher spin theory in $AdS_4$.
After adding the orientifolds,
it is known that
the gauge group of the ABJ theory becomes $O(N_1 )\times USp(2N_2 )$
while the background of the string theory is replaced by
$AdS_4 \times\mathbb{CP}^3 /\mathbf{Z}_2$,
and the supersymmetries in the both theories reduce to $\mathcal{N}=5$.
We propose that
adding the orientifolds to the ${\cal N}=6$ Vasiliev theory leads 
to $\mathcal{N}=5$ SUSY Vasiliev theory.
It turns out that
the $\mathcal{N}=5$ case is more involved
because
there are two formulations of the $\mathcal{N}=5$ Vasiliev theory 
with either $O$ or $USp$ internal symmetry.
We show that
the two $\mathcal{N}=5$ Vasiliev theories can be understood 
as certain projections of the $\mathcal{N}=6$ Vasiliev theory,
which we identify with the orientifold projections in the Vasiliev theory.
We conjecture that 
the $O(N_1 )\times USp(2N_2)$ ABJ theory has
the two vector model like limits: $N_2 \gg N_1$ and $N_1 \gg N_2 $ 
which correspond to the semi-classical ${\cal N}=5$ Vasiliev theories 
with $O(N_1 )$ and $USp(2N_2)$ internal symmetries respectively. 
These correspondences together with the standard AdS/CFT correspondence
comprise the ABJ quadrality
among the $\mathcal{N}=5$ ABJ theory, string/M-theory
and two $\mathcal{N}=5$ Vasliev theories.
We provide a precise holographic dictionary for the correspondences
by comparing correlation functions of stress tensor and flavor currents.
Our conjecture is supported by various evidence such as
agreements of the spectra, one-loop free energies and SUSY enhancement on the both sides.
We also predict 
the leading free energy of the ${\cal N}=5$ Vasiliev theory from the CFT side.
As a byproduct, we give a derivation of the relation 
between the parity violating phase in the $\mathcal{N}=6$ Vasiliev theory
and the parameters in the $\mathcal{N}=6$ ABJ theory,
which was conjectured in \cite{Chang:2012kt}.
\thispagestyle{empty}

\vfill
\noindent 
{ WIS/01/17-AUG-DPPA}

\pagebreak
\setcounter{page}{1}

\tableofcontents
\addtocontents{toc}{\protect\setcounter{tocdepth}{2}}

\section{Introduction}
At extremely high energy scale, string theory has been expected to exhibit 
a huge gauge symmetry 
as infinitely many massless higher spin (HS) particles emerge 
in the spectrum \cite{Gross:1988ue}.
Then the usual string scale $1/\sqrt{\alpha'}$ might arise as a dynamical scale via Higgsing the HS gauge symmetry.
While these expectations are still speculative,
there exist a self-consistent description of interacting HS gauge fields
known as Vasiliev theory \cite{Vasiliev:2003ev} independently of string theory.
It is then natural to explore 
the  relation between string theory and Vasiliev theory.
The answer to this question remains largely open 
despite some attempts were made
to directly connect Vasiliev theory to the tensionless limit of string (field) theory \cite{Sagnotti:2013bha}.
One of the indirect but steady steps towards answering this question is
to reinterpret stringy objects or concepts in the framework of the Vasiliev theory.
In this paper
we aim at understanding orientifolds
in the context of higher spin $AdS_4/{\rm CFT}_3$ correspondence 
between Vasiliev theory in $AdS_4$ 
and 3d conformal field theory (CFT) \cite{Klebanov:2002ja},
which generalizes the usual AdS/CFT correspondence \cite{Maldacena:1997re}.

\begin{figure}[t]
\begin{center}
\includegraphics[width=11cm]{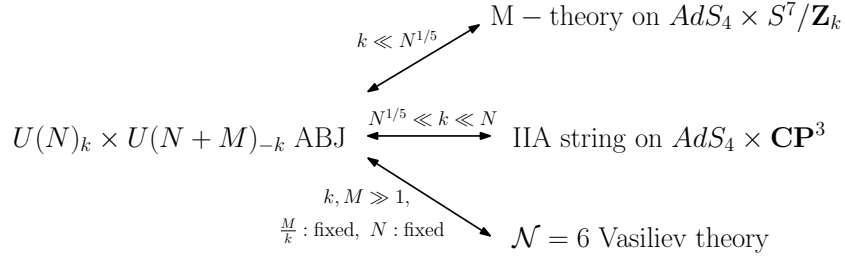}
\end{center}
\caption{
Summary of the $\mathcal{N}=6$ ABJ triality.
}
\label{fig:sumamryN6}
\end{figure}

To be specific,
we study physical consequences of adding orientifolds 
into the setup of ABJ triality \cite{Chang:2012kt, Giombi:2011kc},
which relates three apparently distinct theories 
as summarized in Fig.~\ref{fig:sumamryN6}. 
It involves 
{\it i}) 3d $\mathcal{N}=6$ superconformal Chern-Simons (CS) theory 
called ${\cal N}=6$ ABJ theory \cite{Aharony:2008ug,Aharony:2008gk},
which is the $U(N)_k \times U(N+M)_{-k}$ CS matter theory 
coupled to two bi-fundamental hyper multiplets; 
{\it ii}) Type IIA string theory in $AdS_4 \times \mathbb{CP}^3$; 
{\it iii}) Parity-violating $\mathcal{N}=6$ supersymmetric (SUSY) Vasiliev theory 
with $U(N)$ internal symmetry in $AdS_4$.
The $\mathcal{N}=6$ ABJ theory 
is expected to describe
low energy dynamics of $N$ coincident M2-branes probing\footnote{
The $\mathbb{Z}_k$ orbifolding acts 
on the $\mathbb{C}^4$ coordinate $(z_1,z_2,z_3,z_4)$
as $(z_1,z_2,z_3,z_4)$ $\sim$ $e^{\frac{2\pi i}{k}} (z_1, z_2, z_3, z_4)$.
}  
$\mathbf{C}^4 /\mathbb{Z}_k$, together with $M$ coincident fractional M2-branes localized at the singularity. 
The M-theory background associated with this setup
is $AdS_4 \times S^7 /\mathbf{Z}_k$ 
with the nontrivial 3-form holonomy $\int C_3 \sim M/k $. 
For $k\ll N^{1/5}$,
the M-theory circle shrinks and
the M-theory
is well approximated by type IIA string on $AdS_4 \times \mathbb{CP}^3$. 
It is conjectured in \cite{Chang:2012kt,Giombi:2011kc,Sezgin:2012ag} that
the ${\cal N}=6$ ABJ theory is also dual to
the ${\cal N}=6$ Vasiliev theory with $U(N)$ internal symmetry,
in which the Newton constant $G_{N}\sim 1/M$.
Especially the semi-classical approximation of the Vasiliev theory becomes accurate
in the following limit of the ABJ theory
\be
M,\,|k|\rightarrow\infty\quad
{\rm with}\quad t\equiv \frac{M}{|k|}:{\rm finite}\quad 
{\rm and }\quad N:{\rm finite}~.
\label{eq:HSlimit6}
\ee
In this limit,
the ABJ theory approaches 
a vector-like model which is
the $U(N+M)$ SUSY CS theory coupled to $2N$ fundamental hyper multiplets with a weakly gauged $U(N)$ symmetry.
This correspondence is a generalization of 
the duality between Vasiliev theory and $U(M)$ CS vector model \cite{Giombi:2011kc,Aharony:2011jz,Giombi:2012ms,Aharony:2012ns}
to the case with weakly gauged flavor symmetries\footnote{
There is also a study on this type of correspondence 
for non-SUSY cases \cite{Gurucharan:2014cva}.
}. 
In the ABJ triality,
the fundamental string in the string theory is expected to be realized 
as ``flux tube" solution or ``glueball"-like bound state  
in the Vasiliev theory when the bulk coupling is large. 
The ${\cal N}=6$ ABJ triality was further investigated 
in \cite{Hirano:2015yha,Honda:2015sxa}.

\begin{figure}[t]
\begin{center}
\includegraphics[width=7cm]{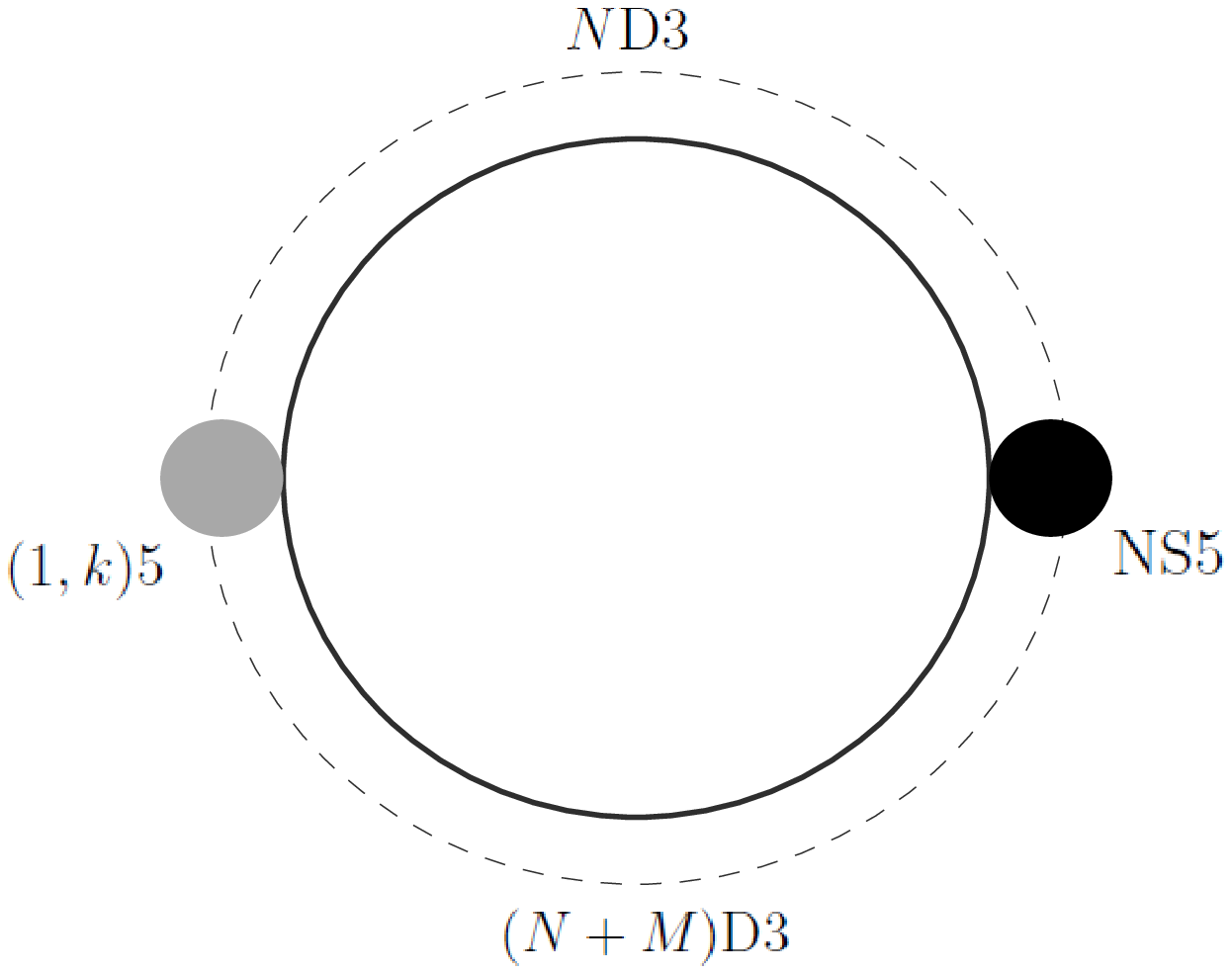}
\includegraphics[width=7cm]{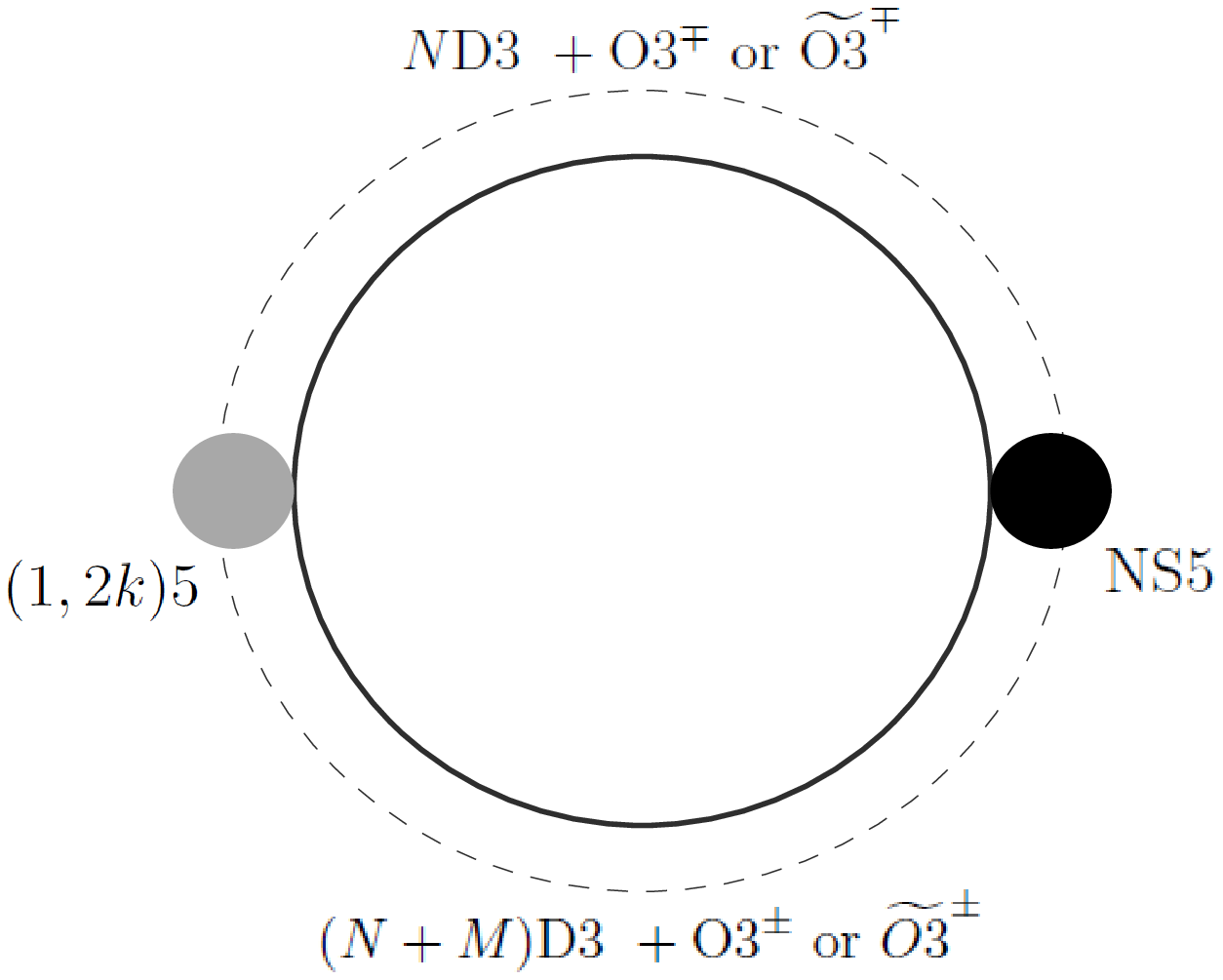}
\end{center}
\caption{
The type IIB brane constructions for the ABJ theories.
All the objects share three common dimensions 
and the D3-branes wind the $S^1$-direction.
[Left] The $\mathcal{N}=6$ case with the gauge group $U(N)_k \times U(N+M)_{-k}$.
[Right] The $\mathcal{N}=5$ case with the gauge group $O(N_1 )_{2k} \times USp(2N_2 )_{-k}$
where $({\rm rank}[O(N_1 )] ,N_2 )$ $=$ $(N,N+M)$ or $(N+M,N)$.
}
\label{fig:brane}
\end{figure}

Now we add orientifolds into this scenario.
For this purpose,
it is convenient to begin with the type IIB brane construction
of the $\mathcal{N}=6$ ABJ theory shown in Fig. \ref{fig:brane} (see \cite{Aharony:2008ug} for detail).
There are four ways to consistently add orientifold 3-planes in this setup.
Recall that there are four orientifold 3-planes\footnote{
$\widetilde{O3}^-$ can be regarded as $O3^-$ plane with a half D3-brane.
$\widetilde{O3}^+$ and $O3^+$ planes are equivalent perturbatively 
but different non-perturbatively \cite{Witten:1998xy}.
}
$O3^-$, $O3^+$, $\widetilde{O3}^-$ and $\widetilde{O3}^+$,
whose combinations with $N$ D3-branes lead to the gauge groups
$O(2N)$, $USp(2N)$, $O(2N+1)$ and $USp(2N)$ respectively. Specifically,
consistently adding $O3^\pm$ into the ${\cal N}=6$ set up
with $k\rightarrow 2k$ leads to the $\mathcal{N}=5$ ABJ theory
with the gauge group $O(N_1 )_{2k} \times USp(2N_2 )_{-k}$,
where $N_1$ is an even integer. The odd $N_1$ case is obtained by adding $\widetilde{O3}^\pm$.
In summary, 
the ${\cal N}=5$ ABJ theory can have
the four types of the gauge group:
\begin{enumerate}
\item $O(2N)_{2k} \times USp(2N+2M)_{-k}$~,
\item $O(2N+2M)_{2k} \times USp(2N)_{-k}$~,
\item $O(2N+1)_{2k} \times USp(2N+2M)_{-k}$~,
\item $O(2N+2M+1)_{2k} \times USp(2N)_{-k}$~.
\label{gaugegroups}
\end{enumerate}
The M-theory background dual to the ${\cal N}=5$ ABJ theory
is given by\footnote{
$\hat{\mathbf{D}}_k$ is the binary dihedral group
which consists of the $\mathbb{Z}_{2k}$ orbifolding and
$(z_1 ,z_2 ,z_3 ,z_4)$  $\sim$ $(iz_2^\ast , -iz_1^\ast ,iz_4^\ast ,-iz_3^\ast )$.
} $AdS_4 \times S^7 /\hat{\mathbf{D}}_k$.
Similar to the $\mathcal{N}=6$ case,
the M-theory circle shrinks for $k\ll N^{1/5}$
and the M-theory 
is well approximated by the type IIA string in $AdS_4 \times\mathbb{CP}^3 /\mathbf{Z}_2$
with the NS-NS 2-form holonomy $\int B_2 \propto M/k $.
While this is well known, 
inspired by the $\mathcal{N}=6$ ABJ triality
it is natural to ask
whether the $\mathcal{N}=5$ ABJ theory also admits 
some dual higher spin description.
To the best of our knowledge, 
this aspect has not been studied in literature.
The focus of this paper is to establish 
the AdS/CFT correspondence 
among the $\mathcal{N}=5$ ABJ theory, 
type IIA string in $AdS_4 \times\mathbb{CP}^3 /\mathbf{Z}_2$
and $\mathcal{N}=5$ Vasiliev theory in $AdS_4$ with internal symmetry.


\begin{figure}[t]
\begin{center}
\includegraphics[width=11.5cm]{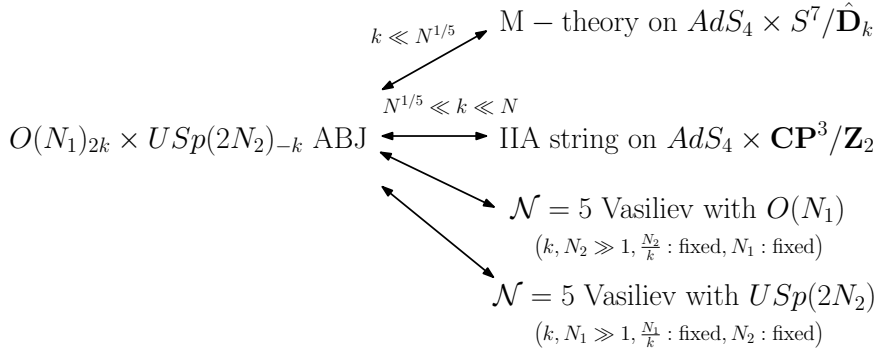}
\end{center}
\caption{
Summary of our proposal on the ABJ quadrality
among the $\mathcal{N}=5$ ABJ theory, string/M-theory and 
two $\mathcal{N}=5$ Vasiliev theories. 
The parameter $N$ is the rank of ``smaller" gauge group,
namely $N={\rm min}({\rm rank}(O(N_1)), N_2)$.
The main difference from the $\mathcal{N}=6$ case is that
we have two higher spin limits corresponding to
the $\mathcal{N}=5$ Vasiliev theories with different internal symmetries. 
}
\label{fig:summaryN5}
\end{figure}

We carry out this by first constructing the ${\cal N}=5$ Vasiliev theory.
As shown in sec.~\ref{basicHS}, 
there are two types of allowed internal symmetry for the ${\cal N}=5$ HS theory, 
which is either $O$ or $USp$ group. 
These two possibilities should correspond to two vector limits of the ${\cal N}=5$ ABJ theory.
Recalling that the gauge group of the ${\cal N}=5$ ABJ theory is $O(N_1) \times USp(N_2)$, 
we first propose that
the $\mathcal{N}=5$ ABJ theory 
is dual to the semi-classical $\mathcal{N}=5$ Vasiliev theory 
with $O(N_1 )$ internal symmetry
in the following limit
\be
N_2=|O(N_1)|+M\,,\quad M,\,|k|\rightarrow\infty\quad{\rm with}\quad t\equiv \frac{M}{|k|}\quad {\rm and }\quad N_1 :{\rm finite}~,
\label{vlimit1}
\ee
where $|O(N_1 )|$ is the rank of $O(N_1)$.
We also propose that
the second limit corresponding 
to the semi-classical Vasiliev theory with $USp(2N_2 )$ internal symmetry is
\be
|O(N_1)|=N_2+M\,,\quad M,\,|k|\rightarrow\infty\quad{\rm with}\quad t\equiv \frac{M}{|k|}\quad {\rm and }\quad N_2 : {\rm finite}~.
\label{vlimit2}
\ee
The correspondence between the HS and CFT parameters  is as follows.
As the $\mathcal{N}=5$ ABJ theory has the three parameters $(k,M,N)$,
the $\mathcal{N}=5$ Vasiliev theory also has 
the three parameters $(G_N ,\theta , N)$,
where $G_N$ is the Newton constant,
$\theta$ is the parity-violating phase 
and $N$ is the rank of the internal symmetry group.
We derive the precise holographic dictionary
by matching correlation functions of stress tensor and flavor symmetry currents,
which we compute on the CFT side
by SUSY localization \cite{Pestun:2007rz}.
As we will discuss in sec.~\ref{sec:comparison},
the analysis of the stress tensor correlation function 
suggests that the Newton constant $G_N$ is related to $M$ by
\be
\frac{G_N}{L_{\rm AdS}^2}=\frac{t}{M\sin\pi t} ,
\label{eq:GN}
\ee
while the comparison of the flavor current correlation function indicates that 
the parity-violating phase $\theta$ is related to $t$ by
\begin{equation}
\theta =\frac{\pi t}{2} .
\label{eq:theta}
\end{equation}
We also show that 
the relation \eqref{eq:theta} is true also for the $\mathcal{N}=6$ ABJ triality,
where \eqref{eq:theta} was conjectured but not proven in \cite{Chang:2012kt}.
In the limit \eqref{vlimit1},
the $\mathcal{N}=5$ ABJ theory approaches 
the $USp(2N_2 )$ SUSY CS theory 
coupled to $N_1$ fundamental hyper multiplets with a weakly gauged $O(N_1 )$ symmetry
while the limit \eqref{vlimit2} provides
the $O(N_1 )$ SUSY CS theory 
coupled to $N_2$ fundamental hyper multiplets with a weakly gauged $USp(2N_2 )$ symmetry.
Our correspondence is 
a generalization of the duality between Vasiliev theory and 
$O(M)$ or $USp(2M)$ CS vector model \cite{Aharony:2011jz,Giombi:2013fka}
to the case with weakly gauged flavor symmetries\footnote{
There are also proposals on dS/CFT correspondence
between Vasiliev theory in $dS_4$ and $USp(2M)$ CS vector model coupled to matters with wrong statistics \cite{Chang:2013afa}
(see also \cite{Anninos:2011ui}).
}.
As in the $\mathcal{N}=6$ case,
we expect that 
the fundamental string in the dual string theory 
is realized as a ``flux tube" in the $\mathcal{N}=5$ Vasiliev theory.
Combined with the standard AdS/CFT correspondence,
we conjecture the duality-like relations among the four apparently different theories,
namely the $\mathcal{N}=5$ ABJ theory, string/M-theory and
two $\mathcal{N}=5$ Vasiliev theories with $O$ and $USp$ internal symmetries.
Thus we shall call it ABJ quadrality as summarized in Fig.~\ref{fig:summaryN5}.
Since the $\mathcal{N}=5$ Vasiliev theories with $O$ and $USp$ internal symmetries
have the bulk 't Hooft couplings $\sim N_1 /N_2$ and $\sim N_2 /N_1$ respectively,
the relation between the two Vasiliev theories
looks like a strong-weak duality of the bulk 't Hooft coupling as a result.

We have various evidence for the proposed correspondence
between the $\mathcal{N}=5$ ABJ theory and Vasiliev theory.
First we will see in sec.~\ref{sec:spectrum} that the
spectrum of higher spin particles in the ${\cal N}=5$ Vasiliev theory 
agrees with that of the higher spin currents in the ${\cal N}=5$ ABJ theory.

Second, there is a non-trivial consistency
among the spectra, $\mathcal{N}=6$ ABJ triality and ``orientifold projection". 
It is known \cite{Aharony:2008gk} that
the ${\cal N}=5$ ABJ theory can be understood  
as a certain projection of the ${\cal N}=6$ ABJ theory.
We show in sec.~\ref{6to5bside} that
one can also derive the ${\cal N}=5$ Vasiliev theory
by applying a projection on the ${\cal N}=6$ Vasiliev theory,
which we identify with the counterpart of the orientifold projection
in the Vasiliev theory.
Roughly speaking,
the projection acts 
on both the $R$-symmetry part and the internal symmetry part of master fields\footnote{
This projection for the $O(N)$ internal symmetry case
is SUSY generalization of a known projection
between non-SUSY Vasiliev theories 
with $U(N)$ and $O(N)$ internal symmetries,
which are dual to $U(M)$ and $O(M)$ CS theories coupled 
to $N$ fundamental scalars or fermions at fixed points.
One of differences is that our projection acts also on the $R$-symmetry part.
}
and preserves the $USp(4)\subset SU(4)$ $R$-symmetry.
More precisely, this is achieved by projection conditions \eqref{eq:orientifoldV}
induced by two automorphisms of the ${\cal N}=6$ HS algebra. 
Then we prove in sec.~\ref{sec:projections} that
the action of the projection on the higher spin currents in the ABJ theory
is the same as the one on the Vasiliev theory.
For example,
the ${\cal N}=5$ Vasiliev theory contains two short multiplets: 
a usual supergravity (SUGRA) multiplet and gravitino multiplet. 
The gravitino multiplet carries 
adjoint representation of $O$ or $USp$ internal symmetry. 
These two short ${\cal N}=5$ supermultiplets appear 
once imposing the projection conditions 
on the $U(N)$ adjoint ${\cal N}=6$ SUGRA multiplet 
in the $\mathcal{N}=6$ Vasiliev theory.

Third, SUSY enhancement occurs on the both sides under the same circumstance
as discussed in sec.~\ref{sec:enhancement}.
It is known \cite{Hosomichi:2008jb,Schnabl:2008wj} that
the SUSY of the $O(N_1 )_{2k}\times USp(2N_2 )_{-k}$ ABJ theory 
is enhanced from $\mathcal{N}=5$ to $\mathcal{N}=6$ when $N_1 =2$.
Interestingly
the dual $\mathcal{N}=5$ Vasiliev theory with the $O(N_1 )$ internal symmetry
has also enhanced $\mathcal{N}=6$ SUSY in the $O(2)$ case 
as explained in sec.~\ref{sec:construction}.

\begin{figure}[t]
\begin{center}
\includegraphics[width=7.4cm]{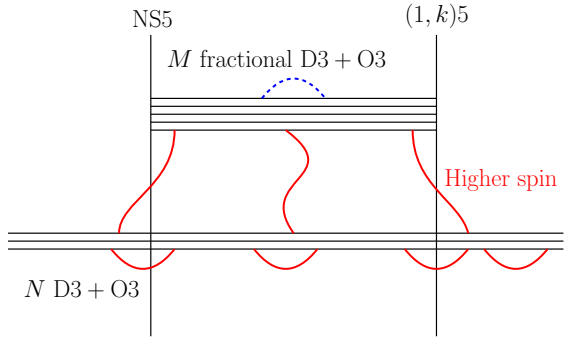}
\end{center}
\caption{
Identification of degrees of freedom
corresponding to those of the Vasiliev theory
from the viewpoint of the brane construction.
Strings denoted by the red solid lines are the HS degrees of freedom
while the blue dashed line is what we are subtracting.
}
\label{fig:dof}
\end{figure}

Finally we find agreement of the sphere free energies on the both sides at $\mathcal{O}(\log{G_N})$
up to a subtlety in the comparison.
The subtlety is that
the free energy of the ABJ theory behaves as $\mathcal{O}(M^2 )$
while the one of the Vasiliev theory should behave
as $\mathcal{O}(G_N^{-1})=\mathcal{O}(M)$. 
Therefore the ABJ theory has apparently more degrees of freedom 
than the Vasiliev theory
and we have to subtract some degrees of freedom appropriately.
This problem appears 
also in CS matter theory coupled to fundamental matters \cite{Giombi:2011kc,Aharony:2012ns}.
and the $\mathcal{N}=6$ ABJ theory \cite{Hirano:2015yha}.
We propose that
the free energy which should be compared to the one in Vasiliev theory is
\be
F^{\rm vec}_{N,M}\equiv-\log\frac{|Z_{{\rm G}_{N,M}}|}{|Z_{{\rm G}_{0,M}}|}~,
\label{fvabj}
\ee
where ${\rm G}_{N,M}$ denotes the gauge group of each case in \eqref{sdua}
and $Z_{{\rm G}_{N,M}}$ is
the sphere partition function of the ABJ theory 
with the gauge group ${\rm G}_{N,M}$. 
This quantity satisfies the following reasonable properties:
i) $1/M$-expansion starts at $\mathcal{O}(M)$;
ii) Invariance under Seiberg-like duality;
iii) The $\mathcal{O}(\log{M})$ term agrees with that in
the one-loop free energy of the $\mathcal{N}=5$ Vasiliev theory. 
Our proposal implies that 
the open string degrees of freedom corresponding to the Vasiliev theory
are given by Fig.~\ref{fig:dof} from the viewpoint of the brane construction.
Utilizing localization method and matrix model technique, 
we compute $F^{\rm vec}_{N,M}$ up to the ${\cal O}(1)$ term in $1/M$ expansion 
but exact in $t$. 
Using this result and our holographic dictionary,
we propose that 
the free energy of the ${\cal N}=5$ Vasiliev theory with $O(N_1 )$ or $USp (2N_2 )$ internal symmetry 
takes the form in the small $G_N$ expansion
\be
F_{\rm HS}
=\frac{8L^2_{\rm AdS}I(\theta)}{G_N\pi\sin2\theta}
-\frac{{\rm min}\{{\rm dim}O(N_1 ),{\rm dim}USp(2N_2 )\}}{2}\log{G_N} +\mathcal{O}(1)~,
\label{hsfe}
\ee
where\footnote{
$I(x)$ also has the integral representation $I(x)=-\int_0^x dy\ \log{\tan{y}}$
and satisfies $I(\pi /2 -x)=I(x)$.
}
\be
I(x) = {\rm Im}\Bigl[ {\rm Li}_2 ({\rm i}\tan{x}) \Bigr] -x \log{\tan{x}}~.
\ee
The first term in \eqref{hsfe} should correspond to the tree level action of Vasiliev theory evaluated on $AdS_4$
which we cannot currently compare with any results in literature, 
since the full action of the Vasiliev theory has not been constructed.
Hence we regard our result as prediction to 
the on-shell action of the $\mathcal{N}=5$ Vasiliev theory in $AdS_4$.
As mentioned above,
the second term agrees  
with the one-loop free energy of the Vasiliev theory on $AdS_4$, 
which is free of logarithmic divergences \cite{Pang:2016ofv} 
and equal to
$(-1/2)$ times the number of bulk spin-1 gauge fields obeying the mixed boundary condition \cite{Giombi:2013yva}.

In Section 6, we summarize and discuss possible extensions of this work.

\section{$\mathcal{N}=5$ supersymmetric Vasiliev theory}
\label{basicHS}
In this section,
we explain some details on the $\mathcal{N}=5$ SUSY Vasiliev theory.
First
we construct the $\mathcal{N}=5$ Vasiliev theory
for the two cases with $O(N)$ and $USp(2N)$ internal symmetries.
Next
we linearize the $\mathcal{N}=5$ Vasiliev theory
around the $AdS_4$ vacuum preserving $\mathcal{N}=5$ SUSY.
We explicitly write down the equations of motion, gauge transformations and SUSY transformations around the $AdS_4$ vacuum.

\subsection{Construction}
\label{sec:construction}
Here we construct the $\mathcal{N}=5$ Vasiliev theory.
The ${\cal N}=5$ Vasiliev theory
is based on $husp(4;4|4)$ SUSY higher spin algebra \cite{Konstein:1989ij},
which contains the maximal compact subalgebra  $usp(4)\oplus usp(4)$. 
As we will explain,
this theory admits either $O(N)$ or $USp(2N)$ as an internal symmetry.
We begin with aspects
which are common between the two cases and
then specify the internal symmetries.
Four dimensional Vasiliev theory
is realized by introducing the spinorial oscillators\footnote{
See appendix \ref{conv} for some details.
} $(Y,Z)=(y,\bar{y},z,\bar{z})$
with the associative but non-commutative $\star$-product defined as
\eq{
\star\equiv\exp[{\rm i}C^{\alpha\beta}(\overset{\leftarrow}{\partial}_{y^\alpha}
+\overset{\leftarrow}{\partial}_{z^\alpha})(\overset{\rightarrow}{\partial}_{y^\beta}
-\overset{\rightarrow}{\partial}_{z^\beta})+{\rm i}C^{\dot{\alpha}\dot{\beta}}(\overset{\leftarrow}{\partial}_{\bar{y}^{\dot{\alpha}}}
+\overset{\leftarrow}{\partial}_{\bar{z}^{\dot{\alpha}}})(\overset{\rightarrow}{\partial}_{\bar{y}^{\dot{\beta}}}
-\overset{\rightarrow}{\partial}_{\bar{z}^{\dot{\beta}}})]~,
}
where $C^{\alpha\beta}=-i\epsilon^{\alpha\beta}$
and $C^{\dot{\alpha}\dot{\beta}}=-i\epsilon^{\dot{\alpha}\dot{\beta}}$.
The indices $\alpha, \beta =1,2$ and $\dot{\alpha},\dot{\beta} =1,2$
serve as indices of two-component spinors.
According to this definition,
we have the following identities
\eq{
\spl{
y^\alpha\star y^\beta=y^\alpha y^\beta+{\rm i}C^{\alpha\beta}~&,~~z^\alpha\star z^\beta=z^\alpha z^\beta-{\rm i}C^{\alpha\beta}~,\\
[y^\alpha,f]_\star=2{\rm i}C^{\alpha\beta}\partial_{y^\beta}f~&,~~
[z^\alpha,f]_\star=-2{\rm i}C^{\alpha\beta}\partial_{z^\beta}f~,\\
\{y^\alpha,f\}_\star=2y^\alpha f-2{\rm i}C^{\alpha\beta}\partial_{z^\beta}f~&,~~
\{z^\alpha,f\}_\star=2z^\alpha f+2{\rm i}C^{\alpha\beta}\partial_{y^\beta}f~ ,
}
\label{eq:star}
}
where $f$ is arbitrary function of $(x,Y,Z)$.

In the $\mathcal{N}=5$ Vasiliev theory,
we take fields to be $8N\times 8N$ matrices,
which are tensor products of $8\times 8$ and $N\times N$ parts.
Roughly speaking,
the $8\times 8$ part is needed to introduce fermions
and the size of this part depends on the type of SUSY
while the $N\times N$ part $\mathcal{M}$ describes internal symmetry
and properties of $\mathcal{M}$ depend on the internal symmetry under consideration.
We describe the $8\times 8$ part
in terms of the six Grassmannian variables $(\xi^1,...,\xi^5 ,\eta )$
which commute with $(Y,Z)$
and satisfy the Clifford algebra\footnote{
Strictly speaking,
the products here are $\star$ product
but we drop the $\star$ product symbol regarding $(\xi^i ,\eta )$ for simplicity.
}
\eq{
\{\xi^i,\xi^j\} =\delta^{ij}~,~~ (\eta )^2 =1~,~~
\{\eta,\xi^i\} =0~.
}
Viewing $(\xi^i ,\eta )$ as the $SO(6)$ gamma matrices,
we can realize the $8\times 8$ part as a sum of products of $(\xi^i ,\eta )$.

The Vasiliev system is described by
so-called master fields,
which consist of the connection 1-form $A$ in $(x,Z)$ space and the 0-form $\Phi$ given by
\begin{equation}
A=A(x,y,\bar{y},z,\bar{z},\xi^i,\eta)
=\Omega_\mu dx^\mu
+S_{\alpha}dz^{\alpha}+S_{\dot{\alpha}}d\bar{z}^{\dot{\alpha}}~,\quad
\Phi=\Phi(x,y,\bar{y},z,\bar{z},\xi^i,\eta)~.
\end{equation}
They obey the spin-statistics condition
\eq{
\pi\bar{\pi}\pi_{\xi}\pi_{\eta}(A,\Phi)=(A,\Phi)~,
}
where $\pi$'s
are the homomorphisms of the $\star$-product defined by
\begin{eqnarray}
\pi(y,\bar{y},z,\bar{z}) =(-y,\bar{y},-z,\bar{z}),\quad
\bar{\pi}(y,\bar{y},z,\bar{z})=(y,-\bar{y},z,-\bar{z}),\quad
\pi_{\xi}(\xi^i)=-\xi^i,~\pi_{\eta}(\eta)=-\eta .
\end{eqnarray}
The master fields contain both dynamical and auxiliary degrees of freedom.
The physical degrees of freedom
are contained in the $Z$ independent part of $\Omega_\mu$ and $\Phi$
while $S_\alpha$ and $S_{\dot{\alpha}}$ have only auxiliary degrees of freedom.
When the fields carry non-trivial representations of the internal symmetry,
the $Z$ independent parts of $\Omega_\mu$ and $\Phi$ have the general expansions
\eq{
\spl{
&\Omega_\mu|_{Z=0}=\sum_{\substack{p,q\geq0\\ k=0,..., 5}}\frac{1}{k!}\Big(\Omega_{\mu,i_1\cdots i_k}(p,q)\xi^{i_1\cdots i_k}+\Omega'_{\mu,i_1\cdots i_k}(p,q)\xi^{i_1\cdots i_k}\eta\Big)\otimes\mathcal{M}~,\\
&\Phi|_{Z=0}=\sum_{\substack{p,q\geq0\\ k=0,..., 5}}\frac{1}{k!}\Big(\Phi_{i_1\cdots i_k}(p,q)\xi^{i_1\cdots i_k}+\Phi'_{i_1\cdots i_k}(p,q)\xi^{i_1\cdots i_k}\eta\Big)\otimes\mathcal{M}~,
}
}
where  $\xi^{i_1\cdots i_k} = \xi^{i_1} \cdots \xi^{i_k}$ and
\eq{
{\cal P}(p,q)=\frac{1}{p!q!}{\cal P}_{\alpha_1\cdots\alpha_p\dot{\alpha}_1\cdots\dot{\alpha}_q}y^{\alpha_1}\cdots y^{\alpha_p}\bar{y}^{\dot{\alpha}_1}\cdots\bar{y}^{\dot{\alpha}_q}~.
}
The spin $s$ gauge fields are described by the $p+q=2s-2$ components of $\Omega_\mu|_{Z=0}$,
in which the $p=q$ and $|p-q|=1$ components 
give rise to the (generalized) vierbein and gravitini respectively, 
while the $|p-q|>1$ components correspond to the spin connections.
The matter fields with spin $s\leq\frac{1}{2}$ 
arise as components of $\Phi|_{Z=0}$ with $p+q\leq1$.
The remaining components in $\Phi|_{Z=0}$ are auxiliary and  related to the Weyl tensors of the physical fields and their derivatives via equations of motion.

\subsubsection{$O(N)$ internal symmetry}
Let us specify our internal symmetry to $O(N)$. 
First we take $\mathcal{M}$ to be 
the $N\times N$ real matrix associated with the internal symmetry $O(N)$,
which can be decomposed into symmetric and antisymmetric parts.
Next we define the $\tau$ map as
\eq{
\label{tauO}
\spl{
\tau(y,\bar{y},z,\bar{z})=({\rm i}y,{\rm i}\bar{y},-{\rm i}z,-{\rm i}\bar{z})~,\quad
\tau(\xi^i)={\rm i}\xi^i~,\quad
\tau(\eta)=-{\rm i}\eta~,~~
}
}
and
\begin{equation}
\tau(\mathcal{M})=\mathcal{M}^T~ .
\label{eq:M_O}
\end{equation}
The conditions \eqref{tauO} will be imposed also for the case with $USp(2N)$ internal symmetry
while the condition for $\mathcal{M}$ will differ from \eqref{eq:M_O}.
Then we require the master fields to satisfy the reality condition 
\begin{equation}
A^\dag=-A\,,\quad\Phi^\dag=\pi(\Phi)\Gamma~,
\end{equation}
and the $\tau$-condition
\begin{equation}
\tau(A)=-A\,,\quad\tau(\Phi)=\bar{\pi}(\Phi)~,
\label{taurec}
\end{equation}
where $\Gamma={\rm i}\xi^1\cdots\xi^5\eta$ and $\Gamma^2=1$.
The $\dag$ acts on ($Y,Z,\xi^i,\eta$) and $\mathcal{M}$ according to\footnote{
We follow the notation of \cite{Sezgin:2012ag},
which is different from the one in \cite{Chang:2012kt}: 
$ y^{\alpha\dag} |_{\rm there}=\bar{y}_{\dot{\alpha}} |_{\rm there}$, 
$\bar{y}^{\dot{\alpha}\dag} |_{\rm there}= -y_{\alpha} |_{\rm there}$.
}
\al{
 y^{\alpha\dag}=\bar{y}^{\dot{\alpha}}~,~~
z^{\alpha\dag}=-\bar{z}^{\dot{\alpha}}~,~~
\xi^{i\dag}=\xi^i~,~~
\eta^\dag=\eta~,~~
(\xi^i\xi^j)^\dag=\xi^j\xi^i~,~~
(\xi^i\eta)^\dag=\eta\xi^i ~,~~
\mathcal{M}^\dag = (\mathcal{M}^T )^\ast .
}

The $\tau$- and reality conditions 
affect the spectrum of physical degrees of freedom.
We now analyze their consequences on $\Omega_\mu|_{Z=0}$.
First let us consider symmetric part of $\mathcal{M}$,
which corresponds to two index symmetric representation of $O(N)$.
Noting that acting $\tau$ on $\Omega_{\mu ,i_1 \cdots ,i_k}\xi^{i_1 \cdots i_k} $ gives the extra factor $i^{p+q+k}$,
as a consequence, the $\tau$-condition requires\footnote{
From now on
we do not explicitly write the matrix $\mathcal{M}$ for succinctness.
}
\eq{
\label{symo}
\spl{
&\Omega_\mu|_{Z=0}
=\sum^\infty_{n=0}\Big\{\sum_{p+q=4n}\Big(\ft{1}{2!}\Omega_{\mu ,ij}(p,q)\xi^{ij}+\ft{1}{3!}\Omega'_{\mu ,ijk}(p,q)\xi^{ijk}\eta\Big)\\
&\quad+\sum_{p+q=4n+1}\Big( \Omega_{\mu ,i}(p,q)\xi^{i}+\ft{1}{5!}\Omega_{\mu ,i_1\cdots i_5}(p,q)\xi^{i_1\cdots i_5}+\ft{1}{2!}\Omega'_{\mu ,ij}(p,q)\xi^{ij}\eta\Big)\\
&\quad+\sum_{p+q=4n+2}\Big( \Omega_\mu(p,q)+\ft{1}{4!}\Omega_{\mu ,i_1\cdots i_4}(p,q)\xi^{i_1\cdots i_4}+\Omega'_{\mu ,i}(p,q)\xi^{i}\eta+\ft{1}{5!}\Omega'_{\mu ,i_1\cdots i_5}(p,q)\xi^{i_1\cdots i_5}\eta\Big)\\
&\quad+\sum_{p+q=4n+3}\Big( \ft{1}{3!}\Omega_{\mu ,ijk}(p,q)\xi^{ijk}+\Omega'_\mu(p,q)\eta+\ft{1}{4!}\Omega'_{\mu ,i_1\cdots i_4}(p,q)\xi^{i_1\cdots i_4}\eta\Big)\Big\}\,.
}
}
The reality condition further requires 
\begin{equation}
{\Omega}^{\dagger}_{\mu ,i_1\cdots i_k}(p,q)=(-1)^{\frac{k(k-1)}{2}+1}\Omega_{\mu ,i_1\cdots i_k}(q,p)~,\quad
{\Omega'}^{\dagger}_{\mu ,i_1\cdots i_k}(p,q)=(-1)^{\frac{k(k+1)}{2}+1}\Omega'_{\mu ,i_1\cdots i_k}(q,p)~.
\end{equation}
The analysis for $\Phi$ is similar and
the result for $p\geq q$ is
\eq{
\label{symp}
\Phi|_{Z=0}=\left\{
\begin{matrix}
&~\Phi(p,q)+\ft{1}{4!}\Phi_{i_1\cdots
i_{4}}(p,q)\xi^{i_1\cdots i_{4}}+\Phi'_i(p,q)\xi^i\eta+\ft{1}{5!}\Phi'_{i_1\cdots i_5}(p,q)\xi^{i_1\cdots i_5}\eta
&{ p-q=0\ \mbox{mod}\ 4}\\
&~\ft{1}{3!}\Phi_{ijk}(p,q)\xi^{ijk} +\Phi'(p,q)\eta+\ft{1}{4!}\Phi'_{i_1\cdots i_4}(p,q)\xi^{i_1\cdots i_4}\eta
&{ p-q=1\ \mbox{mod}\ 4}\\
&~\ft{1}{2!}\Phi_{ij}(p,q)\xi^{ij} +\ft{1}{3!}\Phi'_{ijk}(p,q)\xi^{ijk}\eta
&{ p-q=2\ \mbox{mod}\ 4}\\
&~\Phi_{i}(p,q)\xi^{i}+\ft{1}{5!}\Phi_{i_{1}\cdots i_{5}}(p,q)\xi^{i_{1}\cdots
i_{5}}+\ft{1}{2!}\Phi'_{ij}(p,q)\xi^{ij}\eta
&{ p-q=3\ \mbox{mod}\ 4}
\end{matrix}
\right.
}
where due to the reality condition\footnote{
As an example, for spin-$1/2$ fields,
${\Phi}^{\dagger}_{\dot{\alpha},ijk} 
=-\ft{{\rm i}}{2!}\varepsilon_{ijklm} \Phi'^{\,lm}_{\dot{\alpha}}$,
${\Phi}'^{\dagger}_{\dot{\alpha}}
=\ft{{\rm i}}{5!}\varepsilon_{ijklm} \Phi^{\,ijklm}_{\dot{\alpha}}$,
${\Phi}'^{\dagger}_{\dot{\alpha},ijkl}={\rm i}\varepsilon_{ijklm} \Phi^{\,m}_{\dot{\alpha}}$.
},
\begin{equation}
{\Phi}^{\dagger}(p,p)=\ft{{\rm i}}{5!} (-1)^{p+1}\varepsilon^{ijklm}\Phi'_{ijklm}(p,p)~,\quad
{\Phi}^{\dagger}_{ijkl}(p,p)={\rm i}(-1)^{p+1}\, \varepsilon_{ijklm}\Phi'^m(p,p)~.
\end{equation}
The $p<q$ components of $\Phi|_{Z=0}$ are related to the $p>q$ ones via the reality condition (\ref{taurec}).
The $SO(5)$ indices are raised and lowered by $\delta^{ij}$.
We summarize the final result in table \ref{t1}.
Note that 
in SUSY Vasiliev theory with internal symmetry, 
fields in the usual SUGRA multiplet are extended to matrices,
and
only the singlet components under the internal symmetry,
namely the trace part, 
are related to the operators inside the dual CFT stress tensor multiplet 
via holography.
The Konishi multiplet and other higher spin multiplets exhibit 
the standard long multiplet pattern with the spin range being $\ft52$.

\begin{table}[t]
\begin{center}
{\footnotesize \tabcolsep=1mm
\begin{tabular}{|c|cccccccccccccc|}\hline
& & & & & & & & & & & & &
& \\
{\large${}_{\ell}\backslash s$} & $0$ & $\frac{1}{2}$ & $1$ &
$\frac{3}{2}$ & $2$ & $\frac{5}{2}$ & $3$ & $\frac{7}{2}$ & $4$ &
$\frac{9}{2}$ & $5$ & $\frac{11}{2}$ &
$6$ & $\cdots$ \\
& & & & & & & & & & & & & & \\ \hline & & & & & & & & & & & & &
& \\
$0$ & $\underline{5}+\underline{5}$ & $\underline{1}+\underline{10}$ & $\underline{10}+10$ & $\underline{5}+10$ & $\underline{1}+5$ & $1$
& & & & & & & & \\
$1$ & $1+1$ & $5$ & & $1$ & $5+1$ & $10+5$ & $10+10$ & $5+10$ & $1+5$ & $1$ & & & &  \\
$2$ & & & & & & & & $1$ & $5+1$
& $10+5$ & $10+10$ & $5+10$ & $1+5$ & $\cdots$ \\
$3$ & & & & & & & & & & & & $1$ & $5+1$
& $\cdots$ \\
$\vdots$ & & & & & & & & & & & & & &  \\ \hline
\end{tabular}}
\end{center}
\caption{
The spectrum of physical fields carrying symmetric $\cal{M}$
in the $\mathcal{N}=5$ Vasiliev theory with the $O(N)$ internal symmetry
in the language of $SO(5)$ representations. For $s\ge1$ fields, the level $\ell$ is related to $s$ by
$s=2\ell+2-k/2+r/2$ where $k$ is the number of $\xi$s and $r$ is the number of $\eta$. The values of
$\ell$ are assigned to spin-0, 1/2 fields such that fields belonging to the same supermultiplet are labeled by the same $\ell$. 
The underlines denote the fields in the $\mathcal{N}=5$ SUGRA multiplet.
This table also provides
the spectrum of physical fields associated with antisymmetric ${\cal J}\cal{M}$
in the case with $USp(2N)$ internal symmetry.
}
\label{t1}
\end{table}

Next we consider anti-symmetric part of $\mathcal{M}$
corresponding to two index anti-symmetric representation of $O(N)$.
Then imposing $\tau$-condition leads to
\eq{
\label{ato}
\spl{
&\Omega_\mu|_{Z=0}=\sum^\infty_{n=0}\Big\{\sum_{p+q=4n+1}\Big(\ft{1}{3!} \Omega_{\mu ,ijk}(p,q)\xi^{ijk}+\Omega'_\mu(p,q)\eta+\ft{1}{4!}\Omega'_{\mu ,i_1\cdots i_4}(p,q)\xi^{i_1\cdots i_4}\eta\Big)\\
&\quad+\sum_{p+q=4n+2}\Big(\ft{1}{2!}\Omega_{\mu ,ij}(p,q)\xi^{ij}+\ft{1}{3!}\Omega'_{\mu ,ijk}(m,n)\xi^{ijk}\eta\Big)\\
&\quad+\sum_{p+q=4n+3}\Big(\Omega_{\mu ,i}(p,q)\xi^i+\ft{1}{5!}\Omega_{\mu ,i_1\cdots i_5}(p,q)\xi^{i_1\cdots i_5}+\ft{1}{2!}\Omega'_{\mu ,ij}(p,q)\xi^{ij}\eta\Big)\\
&\quad+\sum_{p+q=4n}\Big(\Omega_\mu(p,q)+\ft{1}{4!}\Omega_{\mu ,i_1\cdots i_4}(p,q)\xi^{i_1\cdots i_4}+\Omega'_{\mu ,i}(p,q)\xi^i\eta+\ft{1}{5!}\Omega'_{\mu ,i_1\cdots i_5}(p,q)\xi^{i_1\cdots i_5}\eta\Big) \Big\}\,,
}
}
and the reality conditions requires
\begin{equation}
{\Omega}^{\dagger}_{\mu ,i_1\cdots i_k}(p,q)=(-1)^{\frac{k(k-1)}{2}}\Omega_{\mu ,i_1\cdots i_k}(q,p)~,\quad
{\Omega}'^{\dagger}_{\mu ,i_1\cdots i_k}(p,q)=(-1)^{\frac{k(k+1)}{2}}\Omega'_{\mu ,i_1\cdots i_k}(q,p)~.
\end{equation}
Similarly, $\Phi|_{Z=0}$ for $p\geq q$ possesses the expansion
\begin{equation}
\label{atp}
\Phi|_{Z=0}=\left\{
\begin{matrix}
&~\ft{1}{2!}\Phi_{ij}(p,q)\xi^{ij}+\ft{1}{3!}\Phi'_{ijk}(p,q)\xi^{ijk}\eta
&{ p- q=0\ \mbox{mod}\ 4}\\
&~\Phi_i(p,q)\xi^i+\ft{1}{5!}\Phi_{i_1\cdots i_5}(p,q)\xi^{i_1\cdots i_5}+\ft{1}{2!}\Phi'_{ij}(p,q)\xi^{ij}\eta
&{ p-q=1\ \mbox{mod}\ 4}\\
&~\Phi(p,q)+\ft{1}{4!}\Phi_{i_1\cdots i_4}(p,q)\xi^{i_1\cdot i_4}+\Phi'_i(p,q)\xi^i\eta+\ft{1}{5!}\Phi'_{i_1\cdots i_5}(p,q)\xi^{i_1\cdots i_5}\eta
&{ p- q=2\ \mbox{mod}\ 4}\\
&~\ft{1}{3!}\Phi_{ijk}(p,q)\xi^{ijk}+\Phi'(p,q)\eta+\ft{1}{4!}\Phi'_{i_1\cdots i_4}(p,q)\xi^{i_1\cdots i_4}\eta
&{ p-q=3\ \mbox{mod}\ 4}
\end{matrix}\right.
\end{equation}
where the reality condition constrains\footnote{
For example, for spin-$1/2$ fields,
${\Phi}^{\dagger}_{\dot{\alpha},i}=-\ft{{\rm i}}{4!}\varepsilon_{ijklm}
\Phi'^{\,jklm}_{\dot{\alpha}}$,
${\Phi}^{\dagger}_{\dot{\alpha},ijklm}
={\rm i}\varepsilon_{ijklm}\Phi'_{\dot{\alpha}}$,
${\Phi}'^{\dagger}_{\dot{\alpha},ij}
=-\ft{{\rm i}}{3!}\varepsilon_{ijklm}\Phi^{\,klm}_{\dot{\alpha}}$.
}
\eq{
{\Phi}^{\dagger}_{ij}(p,p)=\ft{{\rm i}}{3!}(-1)^p\,\varepsilon_{ijklm} \Phi'^{klm}(p,p)~.
}
The $p<q$ components of $\Phi|_{Z=0}$ are related to the $p>q$ ones through the reality condition (\ref{taurec}).
The final result is summarized in Table \ref{t2}.
Especially we have the gravitino multiplet,
which is underlined in Table \ref{t2}.
The gravitino multiplet for $N=2$ is special 
because the two-index anti-symmetric representation of $O(N)$ is singlet.
Together with the $O(2)$-singlet $\mathcal{N}=5$ SUGRA multiplet, 
it comprises the $\mathcal{N}=6$ SUGRA multiplet singlet under the internal symmetry.
This indicates that
the supersymmetry of the $O(2)$ case
is enhanced from $\mathcal{N}=5$ to $\mathcal{N}=6$.
For $N\neq 2$, 
the existence of the gravitino multiplet does not imply 
the SUSY enhancement
since SUSY generators should be singlet under the internal symmetry
and the gravitino multiplet does not contain any singlet parts.
We will come back to this point in sec.~\ref{sec:enhancement}.

\begin{table}[t]
\begin{center}
{\footnotesize \tabcolsep=1mm
\begin{tabular}{|c|cccccccccccccc|}\hline
& & & & & & & & & & & & &
& \\
{\large${}_{\ell}\backslash s$} & $0$ & $\frac{1}{2}$ & $1$ &
$\frac{3}{2}$ & $2$ & $\frac{5}{2}$ & $3$ & $\frac{7}{2}$ & $4$ &
$\frac{9}{2}$ & $5$ & $\frac{11}{2}$ &
$6$ & $\cdots$ \\
& & & & & & & & & & & & & & \\ \hline & & & & & & & & & & & & &
& \\
$0$& $\underline{10}+\underline{10}$ & $\underline{5}+\underline{10}$ & $\underline{1}+\underline{5}$ & $\underline{1}$ & & & & & & & & & & \\
$1$ & & $1$ & $5+1$ & $10+5$ & $10+10$ & $5+10$
& $1+5$ & $1$ & & & & & & \\
$2$ & & & & & & $1$ & $5+1$ & $10+5$ & $10+10$ & $5+10$ & $1+5$ & $1$ & &\\
$3$ & & & & & & & & & & $1$ & $5+1$
& $10+5$ & $10+10$ & $\cdots$ \\
$\vdots$ & & & & & & & & & & & & & &  \\ \hline
\end{tabular}}
\end{center}
\caption{  
The spectrum of physical fields carrying antisymmetric $\cal{M}$
in the case with $O(N)$ internal symmetry.  For $s\ge1$ fields, the level $\ell$ is related to $s$ by
$s=2\ell+1-k/2+r/2$ . The values of
$\ell$ are assigned to spin-0, 1/2 fields such that fields belonging to the same supermultiplet are labeled by the same $\ell$. 
We have underlined the fields belonging to the gravitino multiplet.
This table also provides
the spectrum of physical fields with symmetric ${\cal J}{\cal M}$
in the case with $USp(2N)$ internal symmetry.
}
\label{t2}
\end{table}

In summary, for bosonic fields carrying symmetric $\mathcal{M}$, 
the even spins are always in the $1+1+5+5$ representations of $SO(5)$, 
and the odd spins are in the $10+10$ representations. 
For bosonic fields carrying antisymmetric $\mathcal{M}$, 
the situation is reversed. 
The even spins are always in the $10+10$ representations of $SO(5)$, 
while the odd spins are in the $1+1+5+5$ representations. 
The fermions are always in the $1+5+10$ representations of $SO(5)$, 
regardless of their representations under $O(N)$.

\subsubsection{$USp(2N)$ internal symmetry}
\label{sec:HSUSp}
Next we consider the $\mathcal{N}=5$ HS theory with $USp(2N)$ internal symmetry.
Construction for this case is similar to the $O(N)$ case except two points.
First we take the internal symmetry part $\mathcal{M}$ of the master fields to be $2N\times 2N$ hermitian matrices.
Second we take $\tau$-condition for $\mathcal{M}$ as
\eq{
\label{tauUsp}
\tau(\mathcal{M})=({\cal J}\mathcal{M}{\cal J}^T)^T~,
}
where ${\cal J}$ is the $USp(2N)$ invariant tensor explicitly given by
\eq{
{\cal J}=\begin{pmatrix}
0&\mathbf{1}_{N\times N}\\
-\mathbf{1}_{N\times N}&0
\end{pmatrix}~.
}

Now let us figure out the spectrum of physical fields constrained by the $\tau$-condition.
For this purpose,
it is convenient to decompose $\mathcal{M}$ according to the symmetry property of ${\cal J}\mathcal{M}$ as in the $O(N)$ case.
\begin{itemize}
    \item If $({\cal J}\mathcal{M})^T=({\cal J}\mathcal{M})$, then $\tau(\mathcal{M})=-\mathcal{M}$. 
    The full $\tau$-condition hence implies that
    $\Omega_\mu|_{Z=0}$ and $\Phi|_{Z=0}$ takes the same forms as (\ref{ato}) and (\ref{atp}).
    Hence the spectrum for this case is the same as 
    those for the $O(N)$ case with $\mathcal{M}^T =-\mathcal{M}$ 
    given in Table~\ref{t2}.

    \item If $({\cal J}\mathcal{M})^T=-({\cal J}\mathcal{M})$, then $\tau(\mathcal{M})=\mathcal{M}$. 
    The $\tau$-condition makes $\Omega_\mu|_{Z=0}$ and $\Phi|_{Z=0}$ the same forms (\ref{symo}) and (\ref{symp}) respectively.
    Therefore the spectrum for this case is 
    those for the $O(N)$ case with $\mathcal{M}^T =\mathcal{M}$
    summarized in Table~\ref{t1}.
\end{itemize}
In summary,
for bosonic fields carrying symmetric ${\cal J}\cal{M}$, 
the even spins are always in the $10+10$ representations of $SO(5)$, 
and the odd spins are in the $1+1+5+5$ representations. 
For bosonic fields carrying antisymmetric $J\mathcal{M}$, 
the even spins are in the $1+1+5+5$ representations of $SO(5)$, 
while the odd spins are in the $10+10$ representations. 
The fermions are always in the $1+5+10$ representations of $SO(5)$, 
regardless of their representations under $USp(2N)$. 
The consequence of the reality condition here is slightly different 
from the $O(N)$ case. 
The reality condition imposed on the master fields 
acts on the internal symmetry matrix as hermitian conjugation. 
Thus the the reality conditions induced on the component fields are 
solely determined by the number of $(Y,Z,\xi,\eta)$  
and can be easily obtained from those in the $O(N)$ case 
by adding an extra sign to the ones associated with antisymmetric $\cal{M}$.

\subsection{Analysis of equations of motion and supersymmetry transformations}
\label{sec:EOM}
In this subsection, we first linearize Vasiliev equations 
around the $AdS_4$ vacuum preserving ${\cal N}=5$ SUSY. 
We show that fields comprising the ${\cal N}=5$ SUGRA multiplet 
indeed satisfy the linearized equations of motion of 
the ${\cal N}=5$ $SO(5)$ gauged SUGRA around $AdS_4$. 
We then study the linearized HS gauge transformations
and show that 
the HS gauge transformations generated 
by the Killing spinors of $AdS_4$ relate the fields 
in the ${\cal N}=5$ SUGRA multiplet 
in the same way 
as the linearized SUSY transformation  
of the ${\cal N}=5$ $SO(5)$ gauged SUGRA around $AdS_4$.

\subsubsection{$AdS_4$ vacuum}
The Vasiliev's equations of motion for the master fields are\footnote{
At linearized level, the internal symmetry and $R$-symmetry play no essential roles and 
therefore we suppress their indices when analyzing the linearized Vasiliev's equations.
}
\begin{equation}
dA+A\star A=\frac{{\rm i}}{4}(\hat{V}dz^2+\hat{\bar{V}}d\bar{z}^2)~,\quad
d\Phi+A\star\Phi-\Phi\star\pi (A) =0~,
\label{eq:VasilievEq}
\end{equation}
where $d=\partial_\mu dx^{\mu}+\partial_{z^{\alpha}}dz^{\alpha}+\partial_{\bar{z}^{\dot{\alpha}}}d\bar{z}^{\dot{\alpha}}$,
$x^{\mu}=(x^i,\,r)$. $\hat{V}$ and $\hat{\bar{V}}$ are functions of the master 0-form $\Phi$.
By field redefinition one can reduce $\hat{V}$ and $\hat{\bar{V}}$ to the following form
\eq{
\hat{V}=e^{{\rm i}\theta}\Phi\star\kappa\Gamma~,\quad
\hat{\bar{V}}=e^{-{\rm i}\theta}\Phi\star\bar{\kappa}~,
\label{eq:PVphase}
}
where $\kappa$ and $\bar{\kappa}$ are the Kleinians operators defined as
\eq{
\kappa=e^{{\rm i}y^\alpha z_\alpha}~,\quad
\bar{\kappa}=e^{{\rm i}\bar{y}^{\dot{\alpha}}\bar{z}_{\dot{\alpha}}}~,\quad
\kappa^\dag=\bar{\kappa}~.
}
The parameter $\theta$ in \eqref{eq:PVphase} is called parity violating phase,
which breaks the parity of the Vasiliev theory except for $\theta =0,\pi /2$.
Two models with $\theta$ and $\theta+\pi/2$ are related to each other
by the field redefinition $A\rightarrow \xi^i A\xi^i$, $\Phi\rightarrow {\rm i }\xi^i \Phi\xi^i$ for any $i$ \cite{Chang:2012kt}.
Each component of the first equation in \eqref{eq:VasilievEq} is
\begin{equation}
d_x\Omega+\Omega\star\Omega=0~,\quad
d_zS+S\star S=\frac{{\rm i}}{4}(e^{{\rm i}\theta}\Phi\star\kappa\Gamma dz^2+e^{-{\rm i}\theta}\Phi\star\bar{\kappa}d\bar{z}^2)~,\quad
d_z\Omega+d_xS+\Omega\star S+S\star\Omega=0~,
\end{equation}
where $d_z=\partial_{z^{\alpha}}dz^{\alpha}+\partial_{\bar{z}^{\dot{\alpha}}}d\bar{z}^{\dot{\alpha}}$, $dz^2=dz^\alpha dz_\alpha$ and $d\bar{z}^2=d\bar{z}^{\dot{\alpha}}\bar{z}_{\dot{\alpha}}$. 
The equation of motion of the 0-form read
\eq{
\spl{
d_x\Phi+\Omega\star\Phi-\Phi\star\pi(\Omega)=0~,\quad
d_z\Phi+S\star\Phi-\Phi\star\pi(S)=0~.
}
}
In the Poincar\'e coordinates
\eq{
ds^2=\frac{\eta_{ij}dx^idx^j+dr^2}{r^2}~,
}
the $AdS_4$ background has the following vierbein and spin connection\footnote{The flat and curved indices on $\sigma$ are related by $e^{\alpha\dot{\beta}}=\ft{1}{4{\rm i}}e^a_\mu\sigma^{\alpha\dot{\beta}}_adx^\mu=\ft{1}{4{\rm i}r}\sigma^{\alpha\dot{\beta}}_a\delta^a_\mu dx^\mu=\ft{1}{4{\rm i}r}\sigma^{\alpha\dot{\beta}}_\mu dx^\mu$.}
\al{
&e=\frac{1}{4{\rm i}r}\sigma^{\alpha\dot{\beta}}_\mu y_\alpha\bar{y}_{\dot{\beta}}dx^\mu~,\quad
\omega=\frac{1}{8{\rm i}r}({\sigma_{ir}^{\alpha\beta}}y_\alpha y_\beta+{\bar{\sigma}_{ir}^{\dot{\alpha}\dot{\beta}}}\bar{y}_{\dot{\alpha}} \bar{y}_{\dot{\beta}})dx^i~,
}
which correspond to the exact solution to Vasiliev equations
\eq{
A^{(0)}=e+\omega~,~~\Phi^{(0)}=0~,
\label{eq:AdS4}
}
where $e$ and $\omega$ carry 
the unit matrix of the internal symmetry.

\subsubsection{Linearization}
Let us linearize the equation of motion around the $AdS_4$ vacuum \eqref{eq:AdS4}.
The linearized equations around the $AdS_4$ background 
then take the forms
\eq{
\label{1f}
\spl{
&d_x\Omega^{(1)}+\{\omega +e,\Omega^{(1)}\}_\star =0~,\\
&d_zS^{(1)}=\frac{{\rm i}}{4}(e^{{\rm i}\theta}\Phi^{(1)}\star\kappa\Gamma dz^2+e^{-{\rm i}\theta}\Phi^{(1)}\star\bar{\kappa}d\bar{z}^2)~,\\
&d_z\Omega^{(1)}+d_xS^{(1)}+\{\omega +e,S^{(1)}\}_\star =0~,
}
}
and
\eq{
\label{0f}
d_x\Phi^{(1)}+[\omega,\Phi^{(1)}]_\star+\{e,\Phi^{(1)}\}_\star=0~,~~d_z\Phi^{(1)}=0~.
}
For simplicity, from now on we omit the superscript 
and simply use $\Omega$, $S$ and $\Phi$ to denote the first order master fields. 
The second equation in (\ref{0f}) indicates that $\Phi$ is independent of $z$. Next, from the second equation in (\ref{1f}) $S$ can be solved in terms of $\Phi$:
\eq{
\label{Sphi}
S=\ft{{\rm i}}{2}z^\alpha dz_\alpha e^{{\rm i}\theta}\int^1_0tdt[\Phi\star\, e^{{\rm i}y^\alpha z_\alpha}]\Big|_{z\rightarrow tz}\Gamma+\ft{{\rm i}}{2}\bar{z}^{\dot{\alpha}}d\bar{z}_{\dot{\alpha}} e^{-{\rm i}\theta}\int^1_0tdt[\Phi\star\, e^{{\rm i}\bar{y}^{\dot{\alpha}}z_{\dot{\alpha}}}]\Big|_{z\rightarrow tz}~.
}
where we have chosen the gauge $S|_{Z=0}=0$ and applied the identity (\ref{kl}).
It is useful to split $\Omega$ into the $z$-dependent and independent parts
\eq{
\Omega=W(x,Y,\xi^i,\eta)+W'(x,Y,Z,\xi^i,\eta)~,
}
with $W'|_{Z=0}=0$. $W'$ can be determined from the third equation in (\ref{1f})
\eq{
\label{W'S}
W'=z^\alpha\int^1_0dt(D_0S_\alpha)\Big|_{z\rightarrow tz}+\bar{z}^{\dot{\alpha}}\int^1_0dt(D_0S_{\dot{\alpha}})\Big|_{\bar{z}\rightarrow t\bar{z}}~,
}
where $D_0 S_\alpha=d_xS_\alpha+[\omega+e,S_\alpha]_\star$.
Plugging \eqref{Sphi} into \eqref{W'S}, some explicit calculations give
\eq{
\label{W'}
\spl{
W'=&{\rm i}e^{{\rm i}\theta}z^\alpha\int^1_0dt(1-t)(2{\rm i}\omega_{\alpha\beta}tz^\beta+e_{\alpha\dot{\beta}}C^{\dot{\beta}\dot{\gamma}}\partial_{\bar{y}^{\dot{\gamma}}})\Big(\Phi(x,-tz,\bar{y}) e^{{\rm i}ty^\alpha z_\alpha}\Big)\Gamma\\
&+{\rm i}e^{-{\rm i}\theta}\bar{z}^{\dot{\alpha}}\int^1_0dt(1-t)(2{\rm i}\omega_{\dot{\alpha}\dot{\beta}}t\bar{z}^{\dot{\beta}}+e_{\beta\dot{\alpha}}C^{\beta\gamma}\partial_{y^\gamma})\Big(\Phi(x,y,-t\bar{z}) e^{{\rm i}t\bar{y}^{\dot{\alpha}}\bar{z}_{\dot{\alpha}}}\Big)~.
}
}
Finally, using the results above, 
the first equation in (\ref{1f}) and (\ref{0f}) can be recast as
\al{
 D_0 W
= {\rm i}e^{{\rm i}\theta}e^{\alpha\dot{\beta}}\wedge{e_\alpha}^{\dot{\gamma}}
\partial_{\bar{y}^{\dot{\beta}}}\partial_{\bar{y}^{\dot{\gamma}}}
\Phi(x,0,\bar{y})\Gamma
+{\rm i}e^{-{\rm i}\theta}e^{\beta\dot{\alpha}}\wedge{e^\gamma}_{\dot{\alpha}}\partial_{y^\beta}\partial_{y^\gamma}\Phi(x,y,0)~,\quad
\label{peom}
 \widetilde{D}_0\Phi = 0~,
}
where we have defined
\eq{
\spl{
& D_0
:=\nabla-2{\rm i}e^{\alpha\dot{\alpha}}[y_\alpha\partial_{\bar{y}^{\dot{\alpha}}}
+\bar{y}_{\dot{\alpha}}\partial_{y^\alpha}]~,~~
\widetilde{D}_0
:=\nabla+2e^{\alpha\dot{\alpha}}[y_\alpha\bar{y}_{\dot{\alpha}}
-\partial_{y^\alpha}\partial_{\bar{y}^{\dot{\alpha}}}]~,\\
&\nabla
=d_x-2{\rm i}\omega^{\alpha\beta}(y_\alpha\partial_{y^\beta}
+y_\beta\partial_{y^\alpha})
-2{\rm i}\omega^{\dot{\alpha}\dot{\beta}}
(\bar{y}_{\dot{\alpha}}\partial_{\bar{y}^{\dot{\beta}}}
+\bar{y}_{\dot{\beta}}\partial_{\bar{y}^{\dot{\alpha}}})~.
}
}
%
\subsubsection{Relation to $\mathcal{N}=5$ $SO(5)$ gauged supergravity}
In the following, 
we shall show that the fields comprising the ${\cal N}=5$ SUGRA multiplet indeed satisfy the standard equations of motion
when linearized around $AdS_4$, 
and therefore carry the correct degrees of freedom.
In SUSY Vasiliev theory with internal symmetry,
fields inside SUGRA multiplet are matrix valued
and only the single components under the internal symmetry 
are closely related to operators
inside the dual CFT stress tensor multiplet.
From \eqref{peom} 
we derive the linearized equations of motion for fields in the $\mathcal{N}=5$ SUGRA multiplet,
which are summarized as follows
\begin{itemize}
    \item spin-0\\
    The complex scalars $\Phi^{ijkl}$ are the $Y$-independent components of $\Phi$ and satisfy
    \eq{
    \nabla_\mu\Phi^{ijkl}-2e^{\alpha\dot{\alpha}}_\mu\Phi^{ijkl}_{\alpha\dot{\alpha}}=0~,~~
     \nabla_\mu\Phi^{ijkl}_{\alpha\dot{\alpha}}+2e_\mu{}_{\alpha\dot{\alpha}}\Phi^{ijkl}-2e^{\beta\dot{\beta}}_\mu\Phi^{ijkl}_{\alpha\beta,\dot{\alpha}\dot{\beta}}=0~.
    }
    Taking another covariant derivative of the first equation
    and solving for $\nabla\Phi_{\alpha\dot{\alpha}}$ from the second equation, we arrive at the Klein-Gordon equation
    \eq{
    \nabla^2\Phi^{ijkl}+2\Phi^{ijkl}=0~,
    }
    where we have used $e^{\alpha\dot{\alpha}}_\mu e^\mu{}^{\beta\dot{\beta}}=\epsilon^{\alpha\beta}\epsilon^{\dot{\alpha}\dot{\beta}}$ and $\epsilon^{\alpha\beta}\epsilon^{\dot{\alpha}\dot{\beta}}\Phi_{\alpha\beta,\dot{\alpha}\dot{\beta}}=0$.

    \item spin-$\ft12$\\
    There are two Weyl fermions $\Phi'_\alpha$ and $\Phi^{ijk}_\alpha$. From \eqref{peom} their equations are
    \eq{
    \nabla_\mu\Phi'_\alpha-2e^{\beta\dot{\beta}}_\mu\Phi'_{\alpha\beta,\dot{\beta}}=0~,~~\nabla_\mu\Phi^{ijk}_\alpha-2e^{\beta\dot{\beta}}_\mu\Phi^{ijk}_{\alpha\beta,\dot{\beta}}=0~.
    }
    Multiplying them by $e^\mu{}_{\dot{\gamma}}{}^\alpha$, 
    the second terms of both equations above vanish 
    and we obtain the free Dirac equations
    \eq{
    \sigma^\mu{}_{\dot{\gamma}}{}^{\alpha}\nabla_\mu\Phi'_\alpha=0~,~~\sigma^\mu{}_{\dot{\gamma}}{}^{\alpha}\nabla_\mu\Phi^{ijk}_\alpha=0~.
    }

    \item spin-1\\
    The spin-1 gauge fields denoted as $B^{ij}_\mu$ are the $Y$-independent components of $W$ and obey
    \eq{
    \Phi^{ij}_{\alpha\beta}=2{\rm i}e^{{\rm i}\theta}(dB^{ij})_{\alpha\beta}~,~~
    {\sigma^\mu}_{\lambda\dot{\delta}}\nabla_{\mu}\Phi^{ij}_{\alpha\beta}-{\rm i}\Phi^{ij}_{\alpha\beta\lambda,\dot{\delta}}=0~,
    }
    Multiplying the second equation by $\epsilon^{\lambda\beta}$
     and utilizing the first equation, 
     we obtain the Maxwell equations and the linearized Bianchi identity
    \eq{
    \nabla^\mu(dB^{ij})_{\mu\nu}=0~,~~\sigma^{\mu\nu\rho}\nabla_\rho(dB^{ij})_{\mu\nu}=0~.
    }

    \item spin-$\ft32$\\
    The gravitini $W^i_\alpha$ are in the $\mathbf{5}$ representation of $SO(5)$ and 
    according to \eqref{peom} they obey
    \eq{
    \nabla W^i_\alpha+2{\rm i}{e_\alpha}^{\dot{\beta}}W^i_{\dot{\beta}}=-{\rm i}e^{-{\rm i}\theta}e^{\beta\dot{\alpha}}\wedge {e^\gamma}_{\dot{\alpha}}\Phi^i_{\alpha\beta\gamma}~.
    }
   Multiplying both sides of the equation above by ${{\sigma^{\mu\nu\rho}}_{\dot{\delta}}}^\alpha$, the RHS vanishes and we obtain the linearized
   Rarita-Schwinger equation around $AdS_4$
    \eq{
    {{\sigma^{\mu\nu\rho}}_{\dot{\delta}}}^\alpha(\nabla_\nu W^i_{\rho\,\alpha}+2{\rm i}{e_{\nu\alpha}}^{\dot{\beta}}W^i_{\rho\,\dot{\beta}})=0~.
    }

   \item spin-2\\
   The graviton is described by the vierbein $W_{\alpha\dot{\alpha}}$
    and spin connections $W_{\alpha\beta}$, $W_{\dot{\alpha}\dot{\beta}}$ via a set of first order equations contained 
   in \eqref{peom}
    \bea
    &&\nabla W_{\alpha\dot{\alpha}}+2{\rm i}{e_\alpha}^{\dot{\beta}}W_{\dot{\alpha}\dot{\beta}}+2{\rm i}{e^{\beta}}_{\dot{\alpha}}W_{\alpha\beta}=0~,\label{tosionfree}\\
    &&\nabla W_{\alpha\beta}+4{\rm i}{e_\alpha}^{\dot{\alpha}}W_{\beta\dot{\alpha}}={\rm i}e^{-{\rm i}\theta}e^{\gamma\dot{\gamma}}\wedge {e^\lambda}_{\dot{\gamma}}\Phi_{\alpha\beta\gamma\lambda}~.
   \eea
  Multiplying the second equation by ${{\sigma^{\mu\nu}}_\alpha}^\beta$ leads to
  \eq{
    {{\sigma^{\mu\nu}}_\alpha}^\beta\nabla_\mu W_{\nu\,\beta\gamma}+4{\rm i}{{\sigma^{\mu\nu}}_\alpha}^\beta {e_{\mu\,\beta}}^{\dot{\alpha}}W_{\nu\,\gamma\dot{\alpha}}=0~.
  }
This equation together with \eqref{tosionfree} amounts to 
the usual linearized Einstein equation with a negative cosmological constant.
\end{itemize}

The fields above form 
a supermultiplet of $OSp(5|4)$ which is a subalgebra of the $husp(4;4|4)$ HS algebra
when the background is fixed to $AdS_4$.
Therefore the linearized SUSY transformation relating different spins in the ${\cal N}=5$ SUGRA multiplet
can be read off from the HS gauge transformation around $AdS_4$.
The HS gauge transformation of the master 1-form is given as
\eq{
\delta A=d\epsilon+[A,\epsilon]_\star~,
}
where $d=\partial_\mu dx^{\mu}+\partial_{z^{\alpha}}dz^{\alpha}+\partial_{\bar{z}^{\dot{\alpha}}}d\bar{z}^{\dot{\alpha}}$ 
and the gauge parameter $\epsilon$ is in general a function of ($x$, $y$, $\bar{y}$) and $\xi^i$.
The parameters generating SUSY transformations are 
the components of $\epsilon$ linear in $y$ and $\bar{y}$
which we denote as $\Lambda_\alpha y^\alpha+\bar{\Lambda}_{\dot{\alpha}}\bar{y}^{\dot{\alpha}}$.
$\Lambda$ and $\bar{\Lambda}$ are chosen such that the $AdS_4$ solution is invariant under the gauge transformation
\eq{
d_x(\Lambda_\alpha y^\alpha+\bar{\Lambda}_{\dot{\alpha}}\bar{y}^{\dot{\alpha}})+[e+\omega,\Lambda_\alpha y^\alpha+\bar{\Lambda}_{\dot{\alpha}}\bar{y}^{\dot{\alpha}}]_\star=0~.
}
In fact $\Lambda_\alpha$ and $\Lambda_{\dot{\alpha}}$ 
correspond to the Killing spinors of $AdS_4$.
For the $\mathcal{N}=5$ case, they are linear in $\xi^i$,
\eq{
\Lambda_\alpha=\Lambda^i_\alpha\xi^i~,~~\bar{\Lambda}_{\dot{\alpha}}=\bar{\Lambda}^i_{\dot{\alpha}}\xi^i~,\\
}
where $\Lambda^i_\alpha$ and $\bar{\Lambda}^i_{\dot{\alpha}}$ are fermionic. 
Around the $AdS_4$ vacuum, the master 1-form transforms according to
\begin{equation}
\delta(\Omega^{(0)}+\Omega)
=\delta \Omega
=[\Omega,\Lambda_\alpha y^\alpha+\bar{\Lambda}_{\dot{\alpha}}\bar{y}^{\dot{\alpha}}]_\star~,\quad
\delta S
=[S,\Lambda_\alpha y^\alpha+\bar{\Lambda}_{\dot{\alpha}}\bar{y}^{\dot{\alpha}}]_\star~.
\end{equation}
We focus on the first transformation, which is physical
while the second one is auxiliary. 
Since $\Omega=W(x,y,\bar{y})+W'(x,y,\bar{y},z,\bar{z})$, we have
\eq{
\label{tW}
\delta W=[W,\Lambda_\alpha y^\alpha+\bar{\Lambda}_{\dot{\alpha}}\bar{y}^{\dot{\alpha}}]_\star+[W',\Lambda_\alpha y^\alpha+\bar{\Lambda}_{\dot{\alpha}}\bar{y}^{\dot{\alpha}}]_\star\Big|_{Z=0}~.
}
The solution of $W'$ is given in (\ref{W'}). It is of the form $W'=z^\alpha H_\alpha+\bar{z}^{\dot{\alpha}}\bar{H}_{\dot{\alpha}}$, where $H$ and $\bar{H}$ are functions of $(Y,Z)$. 
Because of the properties \eqref{eq:star} of the $\ast$-product,
the second term on the RHS of (\ref{tW}) may contribute when $W'$, $\Lambda$ and $\bar{\Lambda}$ depend on internal anticommuting parameters.
For the master 0-form we have the twisted HS gauge transformation
\eq{
\label{tP}
\spl{
\delta\Phi
=&[\Phi,\Lambda_\alpha y^\alpha]_\star-\{\Phi,\bar{\Lambda}_{\dot{\alpha}}\bar{y}^{\dot{\alpha}}\}_\star~.
}
}
Substituting the component expansion of $W$ and $\Phi$ to \eqref{tW} and \eqref{tP}, 
we then read off the linearized SUSY transformations of the fields inside the ${\cal N}=5$ SUGRA multiplet as follows
\al{
&\text{spin-2}:~~~~~&&\delta W_{\mu\,\alpha\dot{\alpha}}=2W^i_{\mu\,\alpha}\bar{\Lambda}^i_{\dot{\alpha}}-2\bar{W}^i_{\mu\,\dot{\alpha}}\Lambda^i_\alpha~,\nonumber\\[0.55em]
&\text{spin-$\ft32$}:&&\delta W^i_{\mu\,\alpha}=2{\rm i}\bar{\Lambda}^{i,\dot{\alpha}}W_{\mu\,\alpha\dot{\alpha}}+2{\rm i}W_{\mu\,\alpha\beta}\Lambda^{i,\beta}+B^{ij}_\mu\Lambda^j_\alpha -2{\rm i}e^{\beta\dot{\alpha}}_\mu F^{ij}_{\beta\alpha}\bar{\Lambda}^j_{\dot{\alpha}}~,\nonumber\\[0.55em]
&\text{spin-1}:&&\delta B^{ij}_\mu=-4{\rm i}W^{[i,\alpha}_\mu\Lambda^{j]}_\alpha-e^{-{\rm i}\theta}e^{\alpha\dot{\alpha}}_\mu\Phi^{ijk}_\alpha\bar{\Lambda}^k_{\dot{\alpha}}+{\rm h.c.}~,\nonumber\\[0.55em]
&\text{spin-$\ft12$}:
&&\delta\Phi^{ijk}_\alpha=2\Phi^{ijkl}\Lambda^l_\alpha-2i\Phi^{ijkl}_{\alpha\dot{\alpha}}\bar{\Lambda}^{l,\dot{\alpha}}-12e^{{\rm i}\theta}F^{[ij}_{\alpha\beta}\Lambda^{k],\beta}~,\quad 
\delta\Phi'_\alpha=-2\Phi'^{m}\Lambda^m_\alpha+2{\rm i}\Phi'^m_{\alpha\dot{\alpha}}\bar{\Lambda}^{m,\dot{\alpha}}~,\nonumber\\[0.55em]
&\text{spin-0}:&&\delta
\Phi^{ijkl}=8{\rm i}\Phi^{[ijk}_\alpha\Lambda^{l],\alpha}-2\epsilon^{ijklm}\Phi'_{\alpha}\Lambda^{m,\alpha}~,
}
where $F^{ij}_{\alpha\beta}$ is 
the anti-self-dual part of the field strength of $B^{ij}_\mu$ 
and $\Phi'^m_{\alpha\dot{\alpha}}$ is the gradient of the scalar $\Phi'^m$. 
After proper rescaling and expressing them in terms of the vector basis, 
the transformation above can be recast into the familiar form
\bea
\delta e_\mu{}^a &=& {\bar\epsilon}^i \gamma^a\psi_{\mu i} + \mbox{h.c.}
\nn\\[1mm]
\delta\psi_\mu^i  &=& \ft12\omega^{(L)ab}_{\mu}\gamma_{ab}\epsilon^i-2gB_{\mu}^i{}_j\epsilon^j
+\frac{1}{2} F^-_{\rho\sigma}{}^{ij} \gamma^{\rho\sigma} \gamma_\mu\epsilon_j
+ g \delta^{ij}\gamma_\mu \epsilon_j\ ,
\nn\\[1mm]
\delta B_\mu{}^{ij} &=& -\left(e^{-{\rm i}\theta} {\bar\epsilon}_k\gamma_\mu \chi^{ijk}
+ 2 {\bar\epsilon}^i\psi_\mu{}^j \right)+ \mbox{h.c.}\ ,
\nn\\[1mm]
\delta\chi^{ijk} &=& -\partial_{\mu}\phi^{ijkl}\gamma^\mu\epsilon_l + \tfrac32 \gamma^{\mu\nu} e^{{\rm i}\theta}F^-_{\mu\nu}{}^{[ij}\epsilon^{k]}
+g \phi^{ijkl} \epsilon^l    \ ,
\nn\\[1mm]
\delta\chi &=& -\partial_{\mu}\phi^l\gamma^\mu\epsilon_l
+g \phi^l \epsilon^l\,, \qquad \phi^i= -\fft1{24}\,\epsilon^{ ijkl m}\phi_{jkl m}   \,
\nn\\[1mm]
\delta \phi^{ijkl} &=&  -8 \left( \bar\epsilon^{[i}\chi^{jkl]}+\tfrac1{24}\varepsilon^{ijkl m}\,
\bar\epsilon_m \chi\right)\ .
\eea
When $\theta=0$, this transformation reproduces those of the linearized $\mathcal{N}=5$ SUGRA around $AdS_4$.

\section{$\mathcal{N}=5$ Vasiliev theory from $\mathcal{N}=6$ Vasiliev theory}
\label{6to5bside}
In this section
we discuss that
the $\mathcal{N}=5$ HS theory constructed in the last section
can be understood as certain projections of the $\mathcal{N}=6$ HS theory.
Then using this result,
we obtain supersymmetric boundary conditions 
for the $\mathcal{N}=5$ HS theory.

\subsection{Projections of the $\mathcal{N}=6$ Vasiliev theory}
Before we discuss the projection, 
we quickly review the formulation of the $\mathcal{N}=6$ Vasiliev theory.
The $\mathcal{N}=6$ Vasiliev theory is based 
on the $hu(4;4|4)$ HS algebra \cite{Konstein:1989ij},
which contains $u(4)\oplus u(4)$ as the maximal compact subalgebra.
The master fields in the $\mathcal{N}=6$ HS theory
are also tensor products of 
$8\times 8$ matrices described by the Clifford algebra
and the $N\times N$ matrices $\mathcal{M}$ 
associated with the internal symmetry.
In contrast to the $\mathcal{N}=5$ case,
we take 
the internal symmetry part $\mathcal{M}$ to be $N\times N$ hermitian matrices
and do not impose the $\tau$-condition,
while we take formally the same reality and spin-statistics conditions:
\eq{
A^\dag=-A~,\quad  \Phi^\dag=\pi(\Phi)\Gamma~, \quad \pi\bar{\pi}\pi_{\xi}(A,\Phi)=(A,\Phi) ,
}
which determine the allowed internal symmetry to be $U(N)$ \cite{Konstein:1989ij,Sezgin:2012ag}. 
The above conditions determine
the spectrum of the ${\cal N}=6$ Vasiliev theory with $U(N)$ internal symmetry 
summarized in Table \ref{n6rep}. 
In particular, all the fields
carry the adjoint representation of the internal symmetry $U(N)$.

\begin{table}[t]
\begin{center}
{\footnotesize \tabcolsep=1mm
\begin{tabular}{|c|c|c|c|}\hline
&&&\\
\ Supersymmetry\  & \ Internal\  & $s=0,1,2,...$ & $s=\frac12,\frac32,...$ \\
&&&\\
\hline
&&&\\
 ${\cal N}=6$  & $U(N)$ & 
  \MAT\ $\oplus$\MAT' $\oplus$\MAT$^{[IJ]}\oplus$\MAT $^{\prime [IJ]}$ &
  \MAT $^I \oplus$\MAT $^{\prime I}\oplus$\MAT $^{[IJK]}$ \\
  &&&\\
 \hline
\end{tabular}}
\end{center}
\caption{
The spectra of the $\mathcal{N}=6$ Vasiliev theories 
with the $U(N)$ internal symmetry.
The indices $I,J,K$
label the fundamental representation of $SO(6)_R$.
The Young tableaux with the cross denote the adjoint representation of $U(N)$.
}
\label{n6rep}
\end{table}

Now we consistently truncate 
the $\mathcal{N}=6$ Vasiliev theory to the $\mathcal{N}=5$ theory
following the approach\footnote{
Conventions in this subsection closely follow those in \cite{Konstein:1989ij}.
} of \cite{Konstein:1989ij}.
Generally, in order to truncate SUSY Vasiliev theory consistently, 
one needs an automorphism $\rho$ defined on the original theory as
\eq{
\label{atm}
\rho(P)=-{\rm i}^{\pi(P)}\sigma(P)~,
}
where $P$ is any component of the master fields, $\pi(P)$ is 0 (1)
if $P$ is bosonic (fermionic) and $\sigma$ is an anti-automorphism defined on $P$ as
\eq{
\label{aato}
{\sigma\big(P(y,\bar{y})\big)_\au}^\bu
=\mathcal{S}^{\bu\gu}P({\rm i}y,{\rm i}\bar{y})_\gu{}^\du(\mathcal{S}^{-1})_{\du\au}~,
}
where $\au,\,\bu,\,\gu$ denote the combined indices for the $R$-symmetry and internal symmetry.
The matrix $\mathcal{S}^{\au\bu}$ projects the original $R$-symmetry and internal symmetry to their subgroups preserving $\mathcal{S}^{\au\bu}$. 
For the $\mathcal{N}=6$ HS theory 
$P$ has the following structure
\begin{equation}
{P_\au}^\bu
=\begin{pmatrix}
{\mathbb{B}_\alpha}^\beta &{\mathbb{F}_\alpha}^{\beta'}\cr
{\mathbb{F}_{\alpha'}}^\beta&{\mathbb{B}_{\alpha'}}^{\beta'}
\end{pmatrix}
\otimes\mathcal{M}_{N\times N}~,
\end{equation}
where 
the diagonal blocks $\mathbb{B}$ are bosons 
while the off-diagonal blocks $\mathbb{F}$ are fermions. 
The $SU(4)$ indices $\alpha,\beta,\alpha',\beta'$ run from $1$ to $4$ and
$\mathcal{M}_{N\times N}$ denotes 
the $N\times N$ matrix transforming under the adjoint representation of the internal symmetry $U(N)$. 
Using the $SO(6)$ gamma matrix, 
the $SU(4)$ basis can be converted to the $SO(6)$ basis spanned by $\xi^i$ 
(see App.~\ref{so52usp4} for details).

To obtain the $\mathcal{N}=5$ Vasiliev theory, 
we impose the following condition on the $\mathcal{N}=6$ HS fields
\eq{
\rho(A|_{Z=0})=A|_{Z=0}~,~~\rho(\Phi|_{Z=0})=-\bar{\pi}(\Phi|_{Z=0})~,
\label{eq:orientifoldV}
}
where $\bar{\pi}(y,\bar{y})=(y,-\bar{y})$ and
\eq{
\mathcal{S}=\begin{pmatrix}
J_{4\times 4}&0\\
0&J_{4\times 4}
\end{pmatrix}\otimes g_{N\times N}~.
}
Here $J_{4\times 4}$ is the invariant matrix of $USp(4)$ group, 
and will reduce the $R$-symmetry group from $SU(4)$ to $USp(4)\simeq SO(5)$.
$g_{N\times N}$ is the metric defined 
on the representation space of $U(N)$ internal symmetry group.
According to \cite{Konstein:1989ij}, the only non-trivial $g_{N\times N}$ is
either the symmetric $\delta_{N\times N}$ or the anti-symmetric $\mathcal{J}_{N\times N}$ (when $N$ is even)
and this choice determines
whether the internal symmetry is $O(N)$ or $USp(N)$ 
as we will see soon.
The $Z$-dependent components related to the $Z$-independent components via equations of motion 
are subject to similar projections.

\subsubsection{$O(N)$ internal symmetry}
Let us first choose $g_{N\times N}$ to be $\delta_{N\times N}$.
This projects the internal symmetry to $O(N)$.
We focus on the consequence of the projection on the master 1-form.
For bosonic fields, the projection condition implies\footnote{
We have suppressed the spinor indices of the master field
since the projection trivially acts on the indices.
For example,
if we denote the spinor indices by $\alpha_1,\alpha_2,...$ and $\dot{\alpha}_1,\dot{\alpha}_2,...$,
then the first condition in \eqref{eq:Obos_P} is 
\[
-{\rm i}^{m+n}{J^{\beta\gamma}\delta^{bc}
{\mathbb{B}_{\gamma,\, c}}^{\delta,\, d}}_{\alpha_1,...,\alpha_m,\dot{\alpha}_1,...,\dot{\alpha}_n} J_{\alpha\delta}\delta_{da}
={{\mathbb{B}_{\alpha,\, a}}^{\beta,\, b}}_{\alpha_1,...,\alpha_m,\dot{\alpha}_1,...,\dot{\alpha}_n} .
\]
The spinor indices in other equations of this section can be recovered similarly.
}
\eq{
\label{eq:Obos_P}
\spl{
& -{\rm i}^{m+n}{J^{\beta\gamma}\delta^{bc}
{\mathbb{B}_{\gamma,\, c}}^{\delta,\, d}}(m,n)
J_{\alpha\delta}\delta_{da}
={{\mathbb{B}_{\alpha,\, a}}^{\beta,\, b}}(m,n) ,\\
& -{\rm i}^{m+n}{J^{\beta'\gamma'}\delta^{bc}
{\mathbb{B}_{\gamma',\, c}}^{\delta',\, d}(m,n)}
J_{\alpha'\delta'}\delta_{da}
={{\mathbb{B}_{\alpha',\, a}}^{\beta',\, b}}(m,n) ,
}
}
where 
$\alpha,\beta,...,\alpha',\beta',...$ denote the vector indices of $USp(4)$
and
$a,b,...$ stand for the vector indices of $O(N)$.
The projection condition on bosons requires
\begin{equation}
\label{pjc}
{\rm i}^{m+n}{\mathbb{B}^{\alpha\beta,\, ab}} (m,n)
={\mathbb{B}^{\beta\alpha,\, ba}}(m,n) ,\quad
{\rm i}^{m+n}{\mathbb{B}^{\alpha'\beta',\, ab}}(m,n)
={\mathbb{B}^{\beta'\alpha',\, ba}}(m,n) ,
\end{equation}
where we have used $J^{\alpha\beta}$ and $\delta^{ab}$ to raise and lower the vector indices of $USp(4)$ and $O(N)$ respectively.
When $m+n=0,4,8,\cdots$ corresponding to odd spins, we have two cases with
\begin{itemize}
    \item both $(\alpha,\beta)$ and $(a,b)$ being symmetric. 
    This corresponds to the adjoint representation of $USp(4)$ and the symmetric representation of $O(N)$ group. 
    The number of fields is then $$\big(\mathbf{10}+\mathbf{10}'\big)_{USp(4)}\times [\ft12N(N+1)]_{O(N)}~;$$

    \item both $(\alpha,\beta)$ and $(a,b)$ being also antisymmetric. 
    This corresponds to the antisymmetric representation of $USp(4)$
     and the adjoint representation of $O(M)$. 
     Then the number of fields is $$\big(\mathbf{1}+\mathbf{1}'+\mathbf{5}+\mathbf{5}'\big)_{USp(4)}\times [\ft12N(N-1)]_{O(N)}~.$$
\end{itemize}
When $m+n=2,6,10,\cdots$ or even spins, we have two cases with
\begin{itemize}
    \item $(\alpha,\beta)$ being symmetric and $(a,b)$ being antisymmetric. 
    This corresponds to the adjoint representations of both $USp(4)$ and $O(N)$, which leads us to the number of fields $$\big(\mathbf{10}+\mathbf{10}'\big)_{USp(4)}\times [\ft12N(N-1)]_{O(N)}~;$$
    
    \item $(\alpha,\beta)$ being antisymmetric and $(a,b)$ being symmetric. 
    This gives the antisymmetric representation of $USp(4)$
     and the symmetric representation of $O(N)$. 
     Then the number of fields is  $$\big(\mathbf{1}+\mathbf{1}'+\mathbf{5}+\mathbf{5}'\big)_{USp(4)}\times [\ft12N(N+1)]_{O(N)}~.$$
\end{itemize}
The projection conditions for fermions are
\eq{
\label{ccf}
\spl{
& -{\rm i}^{m+n+1}{J^{\beta'\gamma'}\delta^{bc}
{\mathbb{F}_{\gamma',\, c}}^{\delta,\, d}}(m,n)
J_{\alpha\delta}\delta_{da}
={{\mathbb{F}_{\alpha,\, a}}^{\beta',\, b}}(m,n) ,\\
& -{\rm i}^{m+n+1}{J^{\beta\gamma}\delta^{bc}
{\mathbb{F}_{\gamma,\, c}}^{\delta',\, d}(m,n) }
J_{\alpha'\delta'}\delta_{da}
={{\mathbb{F}_{\alpha',\, a}}^{\beta,\, b}}(m,n) ,
}
}
which relate the two sets of complex fermions. Therefore, for each half-integer spin, the number of fields is given by $16\times N^2$. The $4\times 4$ $SU(4)$ matrix decomposes under $USp(4)$ to $\mathbf{1}+\mathbf{5}+\mathbf{10}$ representations.
Putting bosons and fermions together, 
we see that the spectrum matches 
with that of the $\mathcal{N}=5$ HS theory with $O(N)$ internal symmetry 
given by tables~\ref{t1} and \ref{t2} in sec.~\ref{basicHS}.
We can also similar analysis for the master 0-form
and 
the results match with the spectrum given in sec.~\ref{basicHS}.

\subsubsection{$USp(2N)$ internal symmetry}
If we choose $g_{2N\times 2N}=\mathcal{J}_{2N\times 2N}$, 
then the internal symmetry is reduced to $USp(2N)$.
Similar to the previous case, the conditions on bosons now read
\eq{
\label{uspbk}
\spl{
& -{\rm i}^{m+n}J^{\beta\gamma}\mathcal{J}^{bc}
{\mathbb{B}_{\gamma,\, c}}^{\delta,\, d}{}(m,n)
J_{\alpha\delta}\mathcal{J}_{ad}
={\mathbb{B}_{\alpha,\, a}}^{\beta,\, b}{}(m,n) ,\\
& -{\rm i}^{m+n}J^{\beta'\gamma'}\mathcal{J}^{bc}
{\mathbb{B}_{\gamma',\, c}}^{\delta',\, d}{}(m,n)
J_{\alpha'\delta'}\mathcal{J}_{ad}
={\mathbb{B}_{\alpha',\, a}}^{\beta',\, b}{}(m,n) .
}
}
After raising and lowering the indices by $J^{\alpha\beta}$ and $\mathcal{J}^{ab}$, we find
\begin{equation}
\label{pjcc}
-{\rm i}^{m+n}\mathbb{B}^{\alpha\beta,\, ab}{}(m,n)
=\mathbb{B}^{\beta\alpha,\, ba}{}(m,n) ,\quad
-{\rm i}^{m+n}\mathbb{B}^{\alpha'\beta',\, ab}{}(m,n)
=\mathbb{B}^{\beta'\alpha',\, ba}{}(m,n) .
\end{equation}
When $m+n=0,4,8,\cdots$, or odd spins, we have the two cases with
\begin{itemize}
    \item $(\alpha,\beta)$ being symmetric and $(a,b)$ being antisymmetric.
This corresponds to the adjoint representation of $USp(4)$ 
and the (reducible) antisymmetric representation of $USp(2N)$.
The number of fields is then 
$$\big(\mathbf{10}+\mathbf{10}'\big)_{USp(4)}\times [N(2N-1)]_{USp(2N)}~;$$

    \item$(\alpha,\beta)$ being antisymmetric while $(a,b)$ being symmetric. 
    This corresponds to the antisymmetric representation of $USp(4)$
    and adjoint representation of $USp(2N)$. 
    Hence we have the number of fields  $$\big(\mathbf{1}+\mathbf{1}'+\mathbf{5}+\mathbf{5}'\big)_{USp(4)}\times [N(2N+1)]_{USp(2N)}~.$$
\end{itemize}
For $m+n=2,6,10,\cdots$ or even spins, we have the two cases with
\begin{itemize}
    \item $(\alpha,\beta)$ and $(a,b)$ being symmetric. 
    This corresponds to the adjoint representations both in $USp(4)$
    and $USp(2N)$, which give the number of fields as
     $$\big(\mathbf{10}+\mathbf{10}'\big)_{USp(4)}\times[N(2N+1)]_{USp(2N)}~;$$

    \item $(\alpha,\beta)$ and $(a,b)$ being antisymmetric. 
    This gives the antisymmetric representation of $USp(4)$ 
    and the (reducible) antisymmetric representation of $USp(2N)$. 
    The number of fields is then $$\big(\mathbf{1}+\mathbf{1}'+\mathbf{5}+\mathbf{5}'\big)_{USp(4)}\times[N(2N-1)]_{USp(2N)}~.$$
\end{itemize}
For fermions, the projection conditions read
\eq{
\label{uspfbk}
\spl{
& -{\rm i}^{m+n+1}J^{\beta'\gamma'}\mathcal{J}^{bc}
{\mathbb{F}_{\gamma',\, c}}^{\delta,\, d}{}(m,n)
J_{\alpha\delta}\mathcal{J}_{ad}
={\mathbb{F}_{\alpha,\, a}}^{\beta',\, b}{}(m,n) ,\\
& -{\rm i}^{m+n+1}J^{\beta\gamma}\mathcal{J}^{bc}
{\mathbb{F}_{\gamma,\, c}}^{\delta',\, d}{}(m,n)
J_{\alpha'\delta'}\mathcal{J}_{ad}
={\mathbb{F}_{\alpha',\, a}}^{\beta,\, b}{}(m,n) .
}
}
Again this condition simply relates the two sets of complex fermions. The number of fermions for each half-integer spin is then $16\times (2N)^2$. The $4\times 4$ $SU(4)$ matrix decomposes under $USp(4)$ to $\mathbf{1}+\mathbf{5}+\mathbf{10}$ representations.
Putting bosons and fermions together, 
we see that the spectrum matches with that of the $\mathcal{N}=5$ HS theory 
with $USp(2N)$ internal symmetry 
summarized in tables~\ref{t2} and \ref{t1} in sec.~\ref{basicHS}. 
Similar analysis can be done for the master 0-form $\Phi$ and the results match with the spectrum given in sec.~\ref{basicHS}.

\subsection{Supersymmetric boundary conditions}
In the previous subsection, 
we have shown that the ${\cal N}=5$ Vasiliev theory 
can be obtained from the consistent truncations of
the ${\cal N}=6$ theory. 
Therefore 
the SUSY boundary conditions of the ${\cal N}=5$ models 
inherit those of the ${\cal N}=6$ models.
The pure $AdS_4$ vacuum in the ${\cal N}=6$ Vasiliev theory 
preserves the full ${\cal N}=6$ SUSY. 
The linear boundary conditions imposed on the fluctuations of fields around this vacuum have been analyzed in \cite{Chang:2012kt}, 
in which 
the $R$-symmetry neutral spin-1 gauge field 
inside the SUGRA obeys the mixed boundary condition 
with the mixing angle related to the $\theta$-parameter\footnote{
Similar phenomenon was discovered in the $\omega$-deformed ${\cal N}=6$ supergravity \cite{Borghese:2014gfa}. 
There due to nonlinear effects, the mixing angle takes discreet values.
}. 
This can be easily seen from the linearized SUSY transformations for the ${\cal N}=6$ SUGRA multiplet given below
\bea
\delta e_\mu{}^a &=& {\bar\epsilon}^I \gamma^a\psi_{\mu I} + \mbox{h.c.}
\nn\\[1mm]
\delta\psi_\mu^I  &=& \ft12\omega^{(L)ab}_{\mu}\gamma_{ab}\epsilon^r-2gA_{\mu}^{IJ}\epsilon^J+\ft{1}{2\sqrt 2} F^-_{\rho\sigma}{}^{IJ} \gamma^{\rho\sigma} \gamma_{\mu}\epsilon_J
+ g \delta^{IJ}e_\mu{}^a\gamma_a \epsilon_J\ ,
\nn\\[1mm]
\delta A_\mu{}^{IJ} &=&-\left(e^{-{\rm i}\theta}{\bar\epsilon}_K\gamma_\mu \chi^{IJK}
+ 2{\sqrt 2} {\bar\epsilon}^I\psi_\mu{}^J \right)+ \mbox{h.c.}\ ,
\nn\\[1mm]
\delta A_\mu &=& -2e^{-{\rm i}\theta}{\bar\epsilon}_I\gamma_\mu \chi^{I}
+  \mbox{h.c.}\ ,
\nn\\[1mm]
\delta\chi^{IJK} &=& -\partial_{\mu}\phi^{IJKL}\gamma^\mu\epsilon_L+ \ft32  e^{{\rm i}\theta}\gamma^{\mu\nu} F^-_{\mu\nu}{}^{[IJ}\epsilon^{K]}
+g\phi^{IJKL} \epsilon^L \ ,
\nn\\[1mm]
\delta\chi^{I} &=& -\partial_{\mu}\phi^{IJ}\gamma^\mu\epsilon_J+ \ft12e^{{\rm i}\theta} \gamma^{\mu\nu} F^-_{\mu\nu}\epsilon^I
+g\phi^{IJ} \epsilon^J    \ ,
\nn\\[1mm]
 \delta \phi^{IJKL}  &=&  2\sqrt{2} \left( \bar\epsilon^{[I}\chi^{JKL]}+\ft1{4}\varepsilon^{IJKLMN}\,
\bar\epsilon_M \chi_{N}\right)\ ,
\eea
where $I,J\ldots=1,\ldots,6$ are the SO(6) indices, fermions carrying upper and lower SO(6) indices have the opposite chiralities with respect to $\gamma_5$ and
\begin{equation}
F^-_{\mu\nu}=\ft12(F_{\mu\nu}+{\rm i}*F_{\mu\nu})~,\quad\phi^{IJ}=\ft1{24}\varepsilon^{IJKLMN}\phi_{KLMN}~,\quad \phi_{KLMN}=(\phi^{KLMN})^*.
\end{equation}
In terms of the new variables
\begin{equation}
\widetilde{\phi}^{IJKL}\equiv\,e^{-{\rm i}\theta}\phi^{IJKL}\,,\quad \widetilde{F}^-_{\mu\nu}\equiv\,e^{2{\rm i}\theta}F^-_{\mu\nu}\,,\quad
\widetilde{\chi}^{IJK}\equiv\,e^{-{\rm i}\theta}\chi^{IJK}\,,\quad \widetilde{\chi}^I\equiv\,e^{{\rm i}\theta}\chi^I\,,
\end{equation}
the SUSY transformations above 
can be recast to the standard form 
independent of the $\theta$-parameter\footnote{
Using the linearized equation of motion for $\chi^I$, 
one can show that the super-covariant field strength ${\cal F}_{\mu\nu}=F_{\mu\nu}+\cdots$ satisfies $\delta\widetilde{{\cal F}}^-_{\mu\nu}=4{\bar\epsilon}_I\gamma_\mu \partial_{\nu}\widetilde{\chi}^{I}$.
}.
Therefore, the Fefferman-Graham
expansion leads to the mixed boundary conditions for the original fields. 
In particular, 
the bulk spin-1 gauge field 
satisfies
\begin{equation}
{\rm Re}[e^{2{\rm i}\theta}F^-_{ij}]\Big|_{r=0}=0\,,
\end{equation}
which is equivalent to
\begin{equation}
\sin2\theta\,F_{ri}\Big|_{r=0}=\ft12\cos2\theta\,\varepsilon_{ijk}F^{jk}\Big|_{r=0}\,.
\label{mbc}
\end{equation}
Fields of spin $s>1$ must satisfy the Dirichlet boundary conditions in order to avoid the propagating HS gauge fields in the dual boundary theory. 
There is 
another $R$-symmetry neutral spin-1 gauge field 
belonging to a spin-4 supermultiplet. 
It appears in the transformation of gravitini and 
therefore does not admit any mixed boundary condition. 
Decomposing the ${\cal N}=6$ SUGRA multiplet under $OSp(5|4)$ 
leads to an ${\cal N}=5$ SUGRA multiplet 
and an ${\cal N}=5$ gravitino multiplet, consisting of the fields
\bea
(e_\mu{}^a,\,\psi^I_{\mu},\,A^{IJ}_{\mu},\,\chi^{IJK},\,\chi^6,\,\phi^{I6})\oplus(\psi^6_{\mu},\,A^{I6}_{\mu},\,A_{\mu},\,\chi^{IJ6},\,\chi^I,\,\phi^{IJ})\,,\quad I=1,\ldots, 5~.
\eea
Therefore, in the ${\cal N}=5$ Vasiliev theory, 
the spin-1 gauge fields 
satisfying mixed boundary conditions belongs to the gravitino multiplet. 
According to Table \ref{t2}, 
there are $\frac{N}{2}(N-1)$ 
such spin-1 gauge fields when the internal symmetry is $O(N)$, 
while there are $N(2N+1)$ of them for the $USp(2N)$ internal symmetry.

\section{ABJ quadrality}
\label{6ABJto5ABJ}
In this section
we propose
the AdS/CFT correspondence between
the ${\cal N}=5$ Vasiliev theoryon $AdS_4$ 
and the ${\cal N}=5$ ABJ theory.
Combining this with the standard AdS/CFT correspondence, 
we arrive at ABJ quadratlity.
We provide a precise holographic dictionary
and various evidence for this correspondence.
We finally give a prediction of the leading free enrgy 
from the ABJ theory to the bulk side.

\subsection{ABJ theory and its string/M-theory dual}
Here we review 
some properties of the ABJ theory
and the standard AdS/CFT correspondence between the ABJ theory and string/M-theory.

\subsubsection{$\mathcal{N}=6$ case}
The $\mathcal{N}=6$ ABJ theory \cite{Aharony:2008ug,Aharony:2008gk} 
is the 3d $\mathcal{N}=6$ superconformal CS matter theory 
with the gauge group $U(N_1 )_k \times U(N_2 )_{-k}$
coupled to two bi-fundamental hyper multiplets.
If we decompose the bi-fundamental hypers
into pairs of 3d $\mathcal{N}=2$
bi-fundamental chiral multiplets $A_{1,2}$ and  
anti-bi-fundamental chirals $B_{1,2}$,
the superpotential of this theory is given by 
\begin{equation}
W \propto {\rm Tr}\left( 
\mathcal{A}_1 B_1 \mathcal{A}_2 B_2
-\mathcal{A}_1 B_2 \mathcal{A}_2 B_1   \right)  .
\end{equation}
The $\mathcal{N}=6$ ABJ theory is expected to describe 
the low energy dynamics of $N$ coincident M2-branes 
probing $\mathbf{C}^4 /\mathbb{Z}_k$, 
together with $M$ coincident fractional M2-branes localized at the singularity. 
The M-theory background associated with the M2-brane configuration is  
\begin{equation}
 ds_{11}^2 
=\frac{R^2}{4}ds^2_{AdS_4} +R^2ds^2_{S^7 /\mathbb{Z}_{k}},\quad
\frac{1}{2\pi} \int_{S^3 /\mathbb{Z}_k \subset S^7 /\mathbb{Z}_k} C_3 = \frac{M}{k} -\frac{1}{2} ,
\end{equation}
where\footnote{
The factor "$1/2$" has been corrected in \cite{Aharony:2009fc}.} 
in the unit of the Planck length $\ell_p$ the radius $R$ is 
given by $R/\ell_p  =(32\pi^2 k N)^{\frac{1}{6}} $.
If we identify the M-theory circle
with the orbifolding direction by $\mathbb{Z}_k$,
then the M-theory circle radius $R_{11}$ is given by 
\begin{align}
\frac{R_{11}}{\ell_p}
=\frac{R}{k\ell_p} =\left(\frac{32\pi^2 N}{k^5}\right)^{\frac{1}{6}} .
\end{align}
As the M-theory circle shrinks for $k \gg N^{1/5}$ 
the M theory 
is well approximated by the type IIA string on $AdS_4 \times \mathbb{CP}^3$
with the B-field holonomy
\begin{equation}
\frac{1}{2\pi} \int_{\mathbb{CP}^1 \subset\mathbb{CP}^3} B_2 
= \frac{M}{k}-\frac{1}{2} . 
\end{equation}
The radius of $\mathbb{CP}^3$ in the unit of string length $\ell_s$
and the string coupling constant $g_s$ are given by
\begin{align}
\frac{R_{\mathbb{CP}^3}}{\ell_s}
=\left(\frac{32\pi^2N}{k}\right)^{\frac{1}{4}},\quad
g_{s}^2=\left(\frac{32\pi^2 N}{k^5}\right)^{\frac{1}{2}}.
\end{align}
Therefore the approximation by the type IIA SUGRA is accurate
for $N^{1/5}\ll k \ll N$.
There are several tests of this correspondence
at classical level (see e.g.~\cite{Marino:2011nm})
and some tests at one-loop level\footnote{
Localization of the supergravity \cite{Dabholkar:2014wpa}
reproduced 
full $1/N$ corrections of $S^3$ partition function 
for $M=0$ \cite{Fuji:2011km}
up to renormalization of Newton constant 
and non-perturbative corrections of the $1/N$ expansion
(the results of \cite{Liu:2016dau} seem to suggest that bulk one-loop free energy contributed by the supergravity KK modes alone are not sufficient to reproduce
the ${\cal O}(N^0)$ term in the CFT free energy).
} \cite{Fuji:2011km,Bhattacharyya:2012ye,Matsumoto:2013nya}.

The ``braneology" associated with Fig.~\ref{fig:brane} [Left]
suggests some interesting properties of the ABJ theory \cite{Aharony:2008gk}.
First,
the brane configuration implies that
SUSY is broken for $M>|k|$ \cite{Kitao:1998mf}
as it follows from so-called ``s-rule" \cite{Hanany:1996ie},
which forbids multiple D3-branes 
from ending on a NS5/D5-brane pair (now we have $|k|$ such pairs).
This statement is also supported 
by some field theory computations
on Witten index \cite{Kitao:1998mf} and sphere partition function \cite{Awata:2012jb}.
It was also argued in \cite{Aharony:2008gk} that
the theory with $M>|k|$ should not be unitary
by carefully taking into account the
CS level shift \cite{Witten:1988hf} at low-energy.
Second, the brane configuration also indicates the Seiberg-like duality
between two ABJ theories with the gauge groups
\begin{equation}
U(N+M)_{k} \times U(N)_{-k}
\quad \longleftrightarrow \quad
U(N+k -M)_{-k} \times U(N)_{k}~ ,
\end{equation}
following from the brane-creation effect \cite{Hanany:1996ie},
which means a D3-branes is created when
an NS5-brane and a D5-brane cross from each other.
This duality has already been checked 
for the sphere partition function \cite{Kapustin:2010mh,Willett:2011gp,Awata:2012jb,Matsumoto:2013nya}.

\subsubsection{$\mathcal{N}=5$ case}
The $\mathcal{N}=5$ ABJ theory
is the 3d $\mathcal{N}=5$ superconformal CS theory 
with the gauge group $O(N_1 )_{2k} \times USp (2N_2 )_{-k}$
coupled to one bi-fundamental hyper multiplet.
The $\mathcal{N}=5$ ABJ theory can be obtained 
by the following projection of the $\mathcal{N}=6$ ABJ theory 
with the gauge group $U(N_1 )_{2k}\times U(2N_2 )_{-2k}$:
\eq{
\label{CFT65}
\mathcal{B}_1=\mathcal{J}\mathcal{A}^T_1~,~~\mathcal{B}_2=\mathcal{J}\mathcal{A}^T_2~,
}
where $\mathcal{J}$ is the invariant tensor of $USp(2N_2 )$.
Then superpotential of the $\mathcal{N}=5$ theory in 3d $\mathcal{N}=2$ language 
is given by
\begin{equation}
W \propto {\rm Tr}\left( 
\mathcal{A}_1 \mathcal{J} \mathcal{A}_1^T \mathcal{A}_2 \mathcal{J} \mathcal{A}_2^T 
-\mathcal{A}_1 \mathcal{J} \mathcal{A}_2^T 
\mathcal{A}_2 \mathcal{J} \mathcal{A}_1^T   \right) .
\end{equation}

The $\mathcal{N}=5$ ABJ theory is expected to be 
low-energy effective theory of $N$ M2-branes 
probing $\mathbb{C}^4 /\hat{\mathbf D}_k$
with $M$ fractional D3-branes.
The M-theory background associated with this setup
is $AdS_4 \times S^7 /\hat{\mathbf{D}}_k$
with the 3-form background $\int C_3 \sim \frac{M}{k}$. 
As in the $\mathcal{N}=6$ case,
for $k\gg N^{1/5}$,
the M-theory circle shrinks
and
the M-theory is well approximated 
by type IIA string on $AdS_4 \times\mathbb{CP}^3 /\mathbf{Z}_2$
with the B-field holonomy $\int B_2 \sim M/k $.
There are some checks of this correspondence
at classical level \cite{Gulotta:2012yd,Cheon:2012be} 
and one-loop level \cite{Bhattacharyya:2012ye,Moriyama:2015asx,Okuyama:2016xke,Moriyama:2016xin,Moriyama:2016kqi}.

As in the $\mathcal{N}=6$ case,
the brane physics associated with Fig.~\ref{fig:brane} [Right] implies
some nontrivial properties of the $\mathcal{N}=5$ ABJ theory.
Firstly, the ``s-rule" suggests that
the SUSY is broken if
\begin{eqnarray}
 M>|k|+1 &{\rm for}& O(2N+2M)_{2k} \times USp(2N)_{-k}, \nonumber\\
 M>|k|-1 &{\rm for}& USp(2N+2M)_{k} \times O(2N)_{-2k}, \nonumber\\
 M>|k|   &{\rm for}& O(2N+2M+1)_{2k} \times USp(2N)_{-k}, \nonumber\\ 
 M>|k|   &{\rm for}& USp(2N+2M)_{k} \times O(2N+1)_{-2k} .
\end{eqnarray}
This statement is also supported 
by computations of the sphere partition function 
on the field theory side \cite{Okuyama:2016xke,Moriyama:2016xin,Moriyama:2016kqi},
which showed vanishing of the partition function in the parameter regime above.
The argument based on CS level shift also implies that
the theory is non-unitary in the parameter regime above.
Secondly, 
compared to the $\mathcal{N}=6$ case,
the brane creation effect suggests that
the $\mathcal{N}=5$ ABJ theory possesses richer Seiberg-like dualities:
\bea
O(2N+2M)_{2k} \times USp(2N)_{-k} &\longleftrightarrow &
O(2N+2(k-M+1))_{-2k} \times USp(2N)_{k}~, \nn\\
USp(2N+2M)_{k} \times O(2N)_{-2k} &\longleftrightarrow &
USp(2N+2(k-M-1))_{-k} \times O(2N)_{2k}~, \nn\\
O(2N+2M+1)_{2k} \times USp(2N)_{-k} &\longleftrightarrow &
O(2N+2(k-M)+1)_{-2k} \times USp(2N)_{k}~, \nn\\
USp(2N+2M)_{k} \times O(2N+1)_{-2k} &\longleftrightarrow &
USp(2N+2(k-M))_{-k} \times O(2N+1)_{2k}~.
\label{sdua}
\eea
Some checks on these dualities 
for sphere partition function\footnote{
Strictly speaking,
the $O(2N+2M)_{2k} \times USp(2N)_{-k}$ case with $M=0$ and $M=|k|+1$
has not been checked due to a technical reason \cite{Moriyama:2016xin}.
In appendix~\ref{app:duality},
we give another argument to support these dualities.
} can be found in \cite{Moriyama:2016xin,Moriyama:2016kqi}.

\subsection{Proposal for the AdS/CFT correspondence between ABJ theory and SUSY Vasiliev theory}
\label{cftresults}
\subsubsection{$\mathcal{N}=6$ case}
First we review the $\mathcal{N}=6$ ABJ triality \cite{Chang:2012kt}.
It is conjectured in \cite{Giombi:2011kc, Chang:2012kt} that
the $U(N)_k \times U(N+M)_{-k}$ ABJ theory is
dual to parity violating $\mathcal{N}=6$ Vasiliev theory in $AdS_4$.
Especially, in this conjecture, 
semi-classical approximation of the Vasiliev theory becomes good
in the following limit of the ABJ theory
\[
M,\,|k|\rightarrow\infty\quad{\rm with}\quad 
t\equiv \frac{M}{|k|}:{\rm finite}\quad {\rm and }\quad N:{\rm finite}~ .
\]
Indeed it has been shown that
the spectrum of the bulk fields matches 
with that of the single trace primary operators in the vector limit of the ABJ theory.

Correspondence between parameters in the two theories is as follows.
As the $\mathcal{N}=6$ ABJ theory has the three parameters $(k,M,N)$,
the $\mathcal{N}=6$ Vasiliev theory also has 
the three parameters $(G_N ,\theta , N)$,
where $G_N$ is the Newton constant,
$\theta$ is the parity-violating phase 
and $N$ is the rank of the $U(N)$ internal symmetry.
First the Newton constant $G_N$ 
is roughly related to $M$ by $G_N \sim 1/M$
and analysis of stress tensor correlator on the CFT side 
suggests the more precise relation \cite{Honda:2015sxa}:
\be
\frac{G_N}{L_{\rm AdS}^2}=\frac{2t}{M\sin\pi t} .
\ee
It was conjectured in \cite{Chang:2012kt} that
the parity-violating phase $\theta$ is related to $t$ by
\begin{equation}
\theta =\frac{\pi t}{2} ,
\end{equation}
which we will justify in sec.~\ref{sec:comparison}.
The higher spin symmetry in this setup
is broken by $1/M$ effects
since divergences of higher spin currents 
are given by double trace operators \cite{Maldacena:2011jn,Chang:2012kt}.

In this scenario, 
the fundamental string in the dual string theory 
is expected to be realized as a ``flux tube" string
or a ``glueball"-like bound state in the Vasiliev theory.
While a single string state in the string theory corresponds
to the CFT operator $\sim{\rm tr}(ABAB\cdots ABAB)$ schematically,
the field in the Vasiliev theory corresponds to the CFT operator of the form $\sim AB$.
Thus as the 't Hooft coupling in Vasiliev theory increases,
we expect the bound states to form the string excitations.

There is a subtlety 
in the comparison of the bulk and boundary free energies.
This is because the free energy of the ABJ theory 
in the limit \eqref{eq:HSlimit6}
behaves as $\mathcal{O}(M^2 )$ due to the $U(N+M)$ vector multiplet
while Vasiliev theory is dual to vector model in general,
whose leading free energy should behave linearly in $M$.
Therefore the ABJ theory has apparently more degrees of freedom than the Vasiliev theory
and we have to subtract some degrees of freedom appropriately for the comparison.
This issue was addressed in \cite{Hirano:2015yha},
which proposed the definition of the free energy for ABJ theory in the vector limit as
\begin{equation}
F_{\rm vec}^{\mathcal{N}=6}
= -\log{ \frac{\left| Z_{U(N)_k \times U(N+M)_{-k}}\right|}
{\left| Z_{U(M)_{-k}}\right|}} ,
\label{eq:F6}
\end{equation}
where $Z_{U(M)_{-k}}$ is the partition function for the $N=0$ case and is the same as that of
the $\mathcal{N}=3$ SUSY pure CS theory with the gauge group $U(M)_{-k}$.
The quantity $F_{\rm vec}$ satisfies the following three properties:
\begin{enumerate}
\item $1/M$-expansion starts at $\mathcal{O}(M)$;

\item Invariance under Seiberg-like duality: $M\rightarrow |k|-M$, $k\rightarrow -k$
because this acts on the denominator $\left| Z_{U(M)_{-k}}\right|$ as the level-rank duality of the pure CS theory;

\item The $\mathcal{O}(\log{M})$ term matches the $\mathcal{O}(\log{G_N})$ term in the one-loop free energy of the Vasiliev theory.
\end{enumerate}
Especially the second point excludes
a possibility to divide by $Z_{U(N+M)_{-k}}$ rather than $Z_{U(M)_{-k}}$,
which is a naive expectation from the story of CS theory coupled to fundamental matters \cite{Giombi:2011kc,Aharony:2012ns}.
Indeed it is known that the
``mirror" representation of $Z_{U(N)_k \times U(N+M)_{-k}}$ 
factorizes into $Z_{U(M)_{-k}}$ and a $N$-dimensional integral \cite{Awata:2012jb} which also supports the division by $Z_{U(M)_{-k}}$.

\subsubsection{$\mathcal{N}=5$ case}
Now we propose the ABJ quadrality.
We have seen in sec.~\ref{basicHS} that 
the $\mathcal{N}=5$ SUSY Vasiliev theory admits 
the two choices of internal symmetries, $O(N)$ and $USp(2N)$.
This implies that 
there are two limits of the $\mathcal{N}=5$ ABJ theory
which are dual to semi-classical approximations of the two $\mathcal{N}=5$ Vasiliev theories.

We first propose that
the $O(N_1)_{2k}\times USp(2N_2)_{-k}$ ABJ theory is dual to
the semi-classical $\mathcal{N}=5$ Vasiliev theory with $O(N_1 )$ internal symmetry
in the following limit\footnote{
For general $(k,M,N)$,
more appropriate definitions of $t$ are
$\frac{M-1/2}{k},\frac{M+1/2}{k},\frac{M}{k}$ and $\frac{M}{k}$, respectively.
These differences may be neglected in the higher spin limits according to the purpose of the study.
}
\[
N_2=|O(N_1)|+M\,,\quad M,\,|k|\rightarrow\infty\quad{\rm with}\quad t\equiv \frac{M}{|k|}\quad {\rm and }\quad N_1 :{\rm finite}~,
\]
where $|O(N_1 )|$ is the rank of $O(N_1)$, specifically, $|O(2N )|=|O(2N+1)|=N$.
The second limit corresponding to the Vasiliev theory with $USp(2N_2 )$ internal symmetry is
\[
|O(N_1)|=N_2+M\,,\quad M,\,|k|\rightarrow\infty\quad{\rm with}\quad t\equiv \frac{M}{|k|}\quad {\rm and }\quad N_2 : {\rm finite}~.
\]

As we will discuss in sec.~\ref{sec:comparison}, for both cases,
the Newton constant $G_N$ is related to $M$ by 
\[
\frac{G_N}{L_{\rm AdS}^2}=\frac{t}{M\sin\pi t} ,
\]
and the parity-violating phase $\theta$ is related to $t$ by
\[
\theta =\frac{\pi t}{2} .
\]
As in the $\mathcal{N}=6$ case,
we expect that 
the fundamental string in the dual string theory 
is realized as a ``flux tube" string or ``glueball"-like bound state in the Vasiliev theory,
and strong coupling dynamics of the Vasiliev theory 
exhibits the bound states to form the string excitations.
Thus, as summarized in Fig.~\ref{fig:summaryN5},
our ABJ quadrality relates the four apparently different theories:
the $\mathcal{N}=5$ ABJ theory, string/M-theory and $\mathcal{N}=5$ Vasiliev theories with $O$ and $USp$ internal symmetries.

Comparison of free energies encounters a similar issue to the $\mathcal{N}=6$ case.
Namely the free energy in the ABJ theory 
behaves as $\mathcal{O}(M^2 )$ rather than $\mathcal{O}(M )$,
due to the vector multiplet associated with the ``larger gauge group".
Therefore 
we have to subtract something appropriately as in the $\mathcal{N}=6$ case \cite{Hirano:2015yha}.
In the end, we propose
\[
F^{\rm vec}_{N,M}\equiv-\log\frac{|Z_{{\rm G}_{N,M}}|}{|Z_{{\rm G}_{0,M}}|}~,
\]
where we used the shorthand notation ${\rm G}_{N,M}$ 
to represent the gauge group\footnote{
Note that this definition includes also the $\mathcal{N}=6 case$\eqref{eq:F6}
if we parameterize $G_{N,M}=U(N)_k \times U(N+M)_{-k}$.
} of each case in \eqref{sdua}. 
Indeed we will see that 
this quantity behaves as $\mathcal{O}(M)$ in the higher spin limits
and contains a $\mathcal{O}(\log{M})$ term, 
which agrees with the $\mathcal{O}(\log{G_N})$ term in the one-loop free energy of the $\mathcal{N}=5$ Vasiliev theory.
Here invariance under the Seiberg-like duality \eqref{sdua} is more complicated than in the $\mathcal{N}=6$ case
since we now have four types of $\mathcal{N}=5$ ABJ theory.
For the two types with $O(2N+2M)_{2k}\times USp(2N)_{-k}$ and $O(2N+2M+1)_{2k}\times USp(2N)_{-k}$ gauge groups,
$Z_{{\rm G}_{0,M}}$ is nothing but the partition function of 
$\mathcal{N}=2$ $O(2M)_{2k}$ and $O(2M+1)_{2k}$ pure CS theories, respectively,
whose level-rank dualities are\footnote{
These dualities are essentially level-rank dualities of pure bosonic CS theory.
The main difference is that 
the pure bosonic CS theory has CS level shift: $k_{\rm eff}=k+h_G {\rm sign}(k)$,
where $h_G$ is dual coxeter number of gauge group $G$ and
we have $h_{O(N)}=N-2$ and $h_{USp(2N)}=N+1$. 
If we take $k\rightarrow k_{\rm eff}$ in \eqref{eq:dual_pureO} and \eqref{eq:dual_pureUSp},
then the duality is nothing but exchange of bare CS level and rank.
}
\bea
SO(N)_{2k}&\longleftrightarrow & SO(2|k|+2-N)_{-2k}~.
\label{eq:dual_pureO}
\eea
The Seiberg-like dualities act on $Z_{{\rm G}_{0,M}}$ exactly like this
and hence the ratio is duality invariant.
Similarly for the $O(2N)_{2k}\times USp(2N+2M)_{-k}$ type,
$Z_{{\rm G}_{0,M}}$ is the one of $\mathcal{N}=2$ $USp(2M)_{-k}$ pure CS theory
satisfying the level-rank duality
\bea
USp(2N)_{k}&\longleftrightarrow &USp(2|k|-2N-2)_{-k}~ ,
\label{eq:dual_pureUSp}
\eea
which is the same action as the Seiberg-like duality.
The most subtle case is the $O(2N+1)_{2k}\times USp(2N+2M)_{-k}$ case,
where $Z_{{\rm G}_{0,M}}$ is the partition function of the $O(1)_{2k}\times USp(2M)_{-k}$ theory.
Although the $O(1)$ sector does not have gauge degrees of freedom,
it gives an additional fundamental hyper multiplet of $USp(2M)_{-k}$ 
because of the ``zero-root" in $O(1)$.
However using localization results, one can show that 
the partition function $Z_{{\rm G}_{0,M}}$ is the same\footnote{
This seems accidental for the round $S^3$ partition function.
For instance, this statement is not true for squashed $S^3$ partition function.
} as that of
the $\mathcal{N}=2$ $O(2M+1)_{2k}$ pure CS theory
and hence the Seiberg-like duality acts on $Z_{{\rm G}_{0,M}}$ 
as the level-rank duality \eqref{eq:dual_pureO} for the $O(2M+1)_{2k}$ case.
Thus the ratio \eqref{fvabj} is invariant under the Seiberg-like duality for all the cases. 
This implies that the open
string degrees of freedom underlying the vector limits of ${\cal N}=5$ ABJ theory
are given by Fig.~\ref{fig:dof} from the viewpoint of the brane construction.

\subsection{Matching of spectrum}
\label{sec:spectrum}
In this section
we find agreement between the
spectrum of the HS currents in the ${\cal N}=5$ ABJ theory in the vector limits
and that of the HS fields in the ${\cal N}=5$ Vasiliev theory.

\subsubsection{$O(N)$ internal symmetry}
We have proposed that
the $O(N_1 )_{2k}\times USp(2N_2 )_{-k}$ ABJ theoery is dual to
the semiclassical $\mathcal{N}=5$ Vasiliev theory 
with the $O(N_1 )$ internal symmetry 
in the limit \eqref{vlimit1}.
Then the dynamical higher spin gauge fields in the bulk
should be dual to  
gauge invariant single trace operators in the sense of $USp(2N_2 )$,
which can be expressed in terms of the scalars and fermions 
in the ABJ theory: $\phi_{\alpha, ra}$ and $\psi_{\beta, sb}$,
where $r,s=1,\cdots ,2N_2$ label $USp(2N_2 )$, 
$\alpha,\beta =1,\cdots , 4$ label the $R$-symmetry $USp(4)\simeq SO(5)$ indices, 
and $a,b =1,\cdots ,N_1$ label $O(N_1 )$. 
The scalars and fermions are subject to the symplectic real condition
\eq{
\label{sympc}
\phi^{*\alpha, ra}
=J^{\alpha\beta} \mathcal{J}^{rs}\phi_{\beta, sb}\delta^{ab},\quad
 (\psi^c)^{\alpha, ra}
 =J^{\alpha\beta}\mathcal{J}^{rs}\delta^{ab}\psi_{\beta, sb}~,
}
where $\mathcal{J}^{rs}$ and $J^{\alpha\beta}$ are $USp(2N_2 )$ and $USp(4)$ invariant tensors respectively, 
and $\psi^c$ is the charge conjugation of $\psi$.
In the limit \eqref{vlimit1}, 
the $O(N_1 )$ is weakly gauged and
the operators dual to the bulk fields 
are bilinear in $\phi_{\alpha, ra}$ and $\psi_{\beta, rb}$, 
since it must be invariant under the gauge group $USp(2N_2 )$. 
For example, the operators dual to the bulk scalars are
\eq{
\phi_{[\alpha, (a} \cdot \phi_{\beta], b)},\quad  \bar{\psi}_{[\alpha, (a} \cdot \psi_{\beta], b)}, \quad
\phi_{(\alpha, [a}  \cdot \phi_{\beta), b]},\quad  \bar{\psi}_{(\alpha, [a} \cdot \psi_{\beta), b]}~ ,
}
where we use the following notation for contraction of the $USp(2N_2)$ indices in this subsubsection:
\begin{equation}
\phi_{\alpha ,a} \cdot \phi_{\beta ,b} 
\equiv \phi_{\alpha ,ra} \mathcal{J}^{rs} \phi_{\beta ,sb}  .
\label{eq:contractionO}
\end{equation}
The symmetry properties of the indices are chosen 
such that the operators do not vanish identically\footnote{
Note that we have $\bar{\psi}\chi$=$\bar{\chi}\psi$, 
and $\bar{\psi}\gamma_\mu\chi=-\bar{\psi}\gamma_\mu\chi$ in 3d.
}. 
It is straightforward to see that the first two operators belong to the representations
\eq{
\big(\mathbf{1}+\mathbf{1}+\mathbf{5}+\mathbf{5}\big)_{SO(5)}\otimes  [\ft{1}{2}N_1 (N_1 +1)]_{\, O(N_1 )}~,
}
and the last two are in the representations
\eq{
\big(\mathbf{10}+\mathbf{10}\big)_{SO(5)}\otimes 
[\ft{1}{2}N_1 (N_1 -1)]_{\, O(N_1 )}~.
}
Likewise, other even spin single trace operators can be constructed. 
In odd spin cases, for example, the operators for $s=1$ take the form
\eq{
\phi_{[\alpha, [a} \cdot \partial_\mu\phi_{\beta], b]},\quad 
\bar{\psi}_{[\alpha , [a} \cdot \gamma_\mu\psi_{\beta ], b]},\quad
\phi_{(\alpha, (a} \cdot \partial_\mu\phi_{\beta), b)},\quad 
\bar{\psi}_{(\alpha, (a} \cdot \gamma_\mu\psi_{\beta ), b)}~.
}
Other choices of the symmetry give rise to operators 
which are written as total derivatives of other operators, 
meaning that they are descendants. 
One can see that the first two operators lie in the representation
\eq{
\big(\mathbf{1}+\mathbf{1}+\mathbf{5}+\mathbf{5}\big)_{SO(5)}\otimes
 [\ft{1}{2}N_1 (N_1 -1)]_{\, O(N_1 )}~,
}
while the last two are of the representations
\eq{
\big(\mathbf{10}+\mathbf{10}\big)_{SO(5)}\otimes 
[\ft{1}{2}N_1 (N_1 +1)]_{\, O(N_1 )}~.
}
For other odd spin operators, the construction is the similar. 
The fermionic operators are constructed 
from one $\phi_{\alpha, ra}$ and one $\psi_{\beta, sb}$. 
For instance, the spin-$1/2$ operators are $\psi_{\alpha, a}\cdot \phi_{\beta, b}$.
The product of two $USp(4)$ fundamental representations yields 
$\mathbf{16}=\mathbf{1}+\mathbf{5}+\mathbf{10}$ representations of $USp(4)\simeq SO(5)$. 
The product of two $O(N)$ indices gives rise to
\eq{
[\ft{1}{2}N_1 (N_1 +1)]_{\, O(N_1 ))}
\oplus[\ft{1}{2}N_1 (N_1 -1)]_{\, O(N_1 )}~,
}
where the symmetric representation includes the trace part. 
A simple way to obtain to obtain all the half-integer spin operators 
is to replace the scalar field $\phi_{\alpha, ra}$ 
by a chiral superfield in the integer spin operators.
The gauge invariant HS operators 
in the $\mathcal{N}=5$ ABJ theory in the limit \eqref{vlimit1}
are summarized in Table \ref{t6}.
The spectrum coincides with that of 
the $\mathcal{N}=5$ Vasiliev theory the $O(N_1 )$ internal symmetry.

\begin{table}[t]
\begin{center}
{\footnotesize \tabcolsep=1mm
\begin{tabular}{|c|c|c|c|c|}\hline
&&&&\\
\ Supersymmetry\  & \ Internal\  & $s=0,2,4,...$ & $s=1,3,5...$ &$s=\frac12,\frac32,\frac52...$ \\
&&&&\\
\hline
&&&&\\
 ${\cal N}=5$  & $O(N)$ & 
 \SY\ $\oplus$\SY $^{\,i}\oplus\AS^{\,[ij]}$ & 
 \AS\ $\oplus\ \AS^{\,i}\oplus$\SY $^{\,[ij]}$ & 
 \SY\ $\oplus$\SY$^{\,i}\oplus$\SY $^{\,[ij]}$ \\[0.70em]
 &&\SY'$\oplus$\SY $^{\prime i}\oplus\AS'^{[ij]}$ & 
 $\AS'\oplus\AS'^{i}\oplus$\SY $^{\prime [ij]}$ &
  \AS\ $\oplus\ \AS^{\,i}\oplus\AS^{\,[ij]}$ \\
  &&&&\\
 \hline
\end{tabular}}
\end{center}
\caption{
The spectrum of the HS primary operators 
in the $\mathcal{N}=5$ ABJ theory in the limit \eqref{vlimit1}. 
The indices $i,j$ label the fundamental representation of $SO(5)_R$. 
The Young tableaux denotes the representation of $O(N)$.
}
\label{t6}
\end{table}

\subsubsection{$USp(2N)$ internal symmetry}
Let us take the limit \eqref{vlimit2}
corresponding to the semi-classical $\mathcal{N}=5$ Vasiliev theory 
with the $USp(2N_2)$ internal symmetry.
Since the $USp(2N_2 )$ symmetry is weakly gauged in this limit,
construction of HS primary operators dual to the bulk HS gauge fields 
is analogous to the previous case 
and they should be invariant under $O(N_1 )$ gauge symmetry in the present case.
Hence we use the following notation 
for contraction of the $O(N_1 )$ indices in this subsubsection:
\begin{equation}
\phi_{\alpha ,r} \cdot \phi_{\beta ,s} 
\equiv \phi_{\alpha ,ra} \delta^{ab} \phi_{\beta ,sb}  .
\end{equation}
When the 't Hooft coupling is small, the primary operators dual to the bulk scalars are
\eq{
\phi_{[\alpha, [r} \cdot \phi_{\beta], s]},\quad \bar{\psi}_{[\alpha, [r}\cdot \psi_{\beta], s]} ~,\quad
\phi_{(\alpha, (r}\cdot \phi_{\beta), s)}, \quad \bar{\psi}_{(\alpha, (r} \cdot \psi_{\beta), s)} ~.
}
Other even spin operators possess the same symmetry properties.
It is straightforward to see that the first two operators carry the representations
\eq{
\big(\mathbf{1}+\mathbf{1}+\mathbf{5}+\mathbf{5}\big)_{SO(5)}\otimes  
[N_2 (2N_2 -1)]_{\, USp(2N_2 )}~,
}
and the last two are in the representations
\eq{
\big(\mathbf{10}+\mathbf{10}\big)_{SO(5)}\otimes 
[N_2 (2N_2 +1)]_{\, USp(2N_2 )}~.
}
For odd spin case, for example we have the spin-1 primary operators
\eq{
\phi_{[\alpha, (r} \cdot \partial_\mu\phi_{\beta], s)},\quad  \bar{\psi}_{[\alpha, (r}\cdot \gamma_\mu\psi_{\beta], s)}~,\quad
\phi_{(\alpha, [r}\cdot \partial_\mu\phi_{\beta), s]},\quad  \bar{\psi}_{(\alpha, [r}\cdot \gamma_\mu\psi_{\beta), s]} ~.
}
It is straightforward to see that the first two operators carry the representations
\eq{
\big(\mathbf{1}+\mathbf{1}+\mathbf{5}+\mathbf{5}\big)_{SO(5)}\otimes  
[N_2 (2N_2 +1)]_{\, USp(2N_2 )}~,
}
and the last two are in the representations
\eq{
\big(\mathbf{10}+\mathbf{10}\big)_{SO(5)}\otimes
 [N_2 (2N_2 -1)]_{\, USp(2N_2 )}~.
}
Other spin odd operators have the same index structure.
The fermionic operators are constructed 
from one $\phi_{\alpha, ra}$ and one $\psi_{\beta, sb}$. 
For instance, the $s=\frac12$ operators are
$\psi_{\alpha, r}\cdot \phi_{\beta, s}$.
The product of two $USp(4)$ fundamental representations yields 
$\mathbf{16}=\mathbf{1}+\mathbf{5}+\mathbf{10}$ representations of $USp(4)\simeq SO(5)$. 
The product two $USp(2N_2 )$ indices gives rise to
\eq{
[N_2 (2N_2 +1)]_{\, USp(2N_2 )}\oplus[N_2 (2N_2 -1)]_{\, USp(2N_2 )}~,
}
where the antisymmetric representation includes the trace part. 

\begin{table}[t]
\begin{center}
{\footnotesize \tabcolsep=1mm
\begin{tabular}{|c|c|c|c|c|}\hline
&&&&\\
\ Supersymmetry\  & \ Internal\  & $s=0,2,4,...$ & $s=1,3,5...$ &$s=\frac12,\frac32,\frac52...$ \\
&&&&\\
\hline
&&&&\\
 ${\cal N}=5$  & $USp(2N)$ & 
 \AS\ $\oplus\ \AS^{\,i}\oplus$\SY{}$^{\,[ij]}$ & 
 \SY\ $\oplus$\SY $^{\,i}\oplus\AS^{\,[ij]}$  & 
 \SY\ $\oplus$\SY $^{\,i}\oplus$\SY $^{\,[ij]}$ \\[0.70em]
 &&$\AS'\oplus\AS'^{i}\oplus$\SY $'^{[ij]}$ & 
 \SY' $\oplus$\SY $^{\prime i}\oplus\AS'^{[ij]}$ &
 \AS\ $\oplus\ \AS^{\,i}\oplus \AS^{\,[ij]}$ \\
 &&&&\\
 \hline
\end{tabular}}
\end{center}
\caption{
The spectrum of HS primary operators 
in the $\mathcal{N}=5$ ABJ theory in the limit \eqref{vlimit2}.
The Young tableaux denotes the representation of $USp(2N)$.
}
\label{t7}
\end{table}
%

\subsection{Relating HS and CFT projections}
\label{sec:projections}
The $\mathcal{N}=6$ ABJ theory with gauge group $U(N_1)_{2k}\times U(2N_2)_{-2k}$ 
have the two pairs of fundamental chirals $\mathcal{A}_{1,2}$
and anti-bi-fundamental chirals $\mathcal{B}_{1,2}$, 
which carry the $(N_1, \overline{2N}_2)$ and $(\overline{N}_1, 2N_2)$ representations of the gauge groups respectively. 
Regarding $\mathcal{A}_{1,2}$ as $N_1 \times 2N_2$ matrix 
and $\mathcal{B}_{1,2}$ as $2N_2\times N_1$ matrix, 
the reduction from ${\cal N}=6$ to ${\cal N}=5$ is achieved by imposing the projection condition 
$\mathcal{B}_1=\mathcal{J}\mathcal{A}^T_1 $,
$\mathcal{B}_2=\mathcal{J}\mathcal{A}^T_2 $ \cite{Hosomichi:2008jb}.
These conditions restricts the gauge groups to be $ O(N_1)\times USp(2N_2)$. $\mathcal{A}_{1,2}$ and $\mathcal{B}^\dagger_{1,2}$ can be assembled 
into a vector transforming as the fundamental representation of the $SU(4)$ $R$-symmetry
\eq{
{\cal C}\equiv(\mathcal{A}_1,\mathcal{A}_2,\mathcal{B}^\dag_1,\mathcal{B}^\dag_2)=(\mathcal{A}_1,\mathcal{A}_2,-\mathcal{A}^*_1\mathcal{J},-\mathcal{A}^*_2\mathcal{J})~,
}
${\cal C}$ obeys the symplectic reality condition
${\cal C}_{\alpha}=J_{\alpha\beta}{\cal C}^{*\beta}\mathcal{J}$
which amounts to \eqref{sympc} and reduces the $R$-symmetry from $SU(4)$ to $USp(4)$. Recall that $J$ is the $USp(4)$ invariant matrix
\eq{
J_{\alpha\beta}=\begin{pmatrix}
0&\mathbf{1}_{2\times2}\\
-\mathbf{1}_{2\times2}&0
\end{pmatrix}~.
}

In terms of components, 
the complex matter fields can be represented as
$\phi_{\alpha, ra}$ and $\psi_{\beta, sb}$,
where $r,s$ run from 1 to $2N_2$ of $U(2N_2)$, 
$\alpha,\beta$ run from 1 to 4 of $SU(4)$, and $a,b$ run from 1 to $N_1$ of $U(N_1)$. 
The simplest chiral primary operators in the ${\cal N}=6$ ABJ theory are the mass operators
\eq{
\label{massop}
\bar{\phi}^{\alpha, ra}\phi_{\beta, ra}~,\quad
\bar{\psi^c}^{\alpha, ra}\psi_{\beta, ra}~.
}
In sec.~\ref{6to5bside}, 
we have shown that
the $\mathcal{N}=5$ Vasiliev theories can be obtained 
from the $\mathcal{N}=6$ theory 
by imposing the automorphisms (\ref{atm}). 
Here we relate the projections on the CFT side to that on the bulk side.

\subsubsection{$O(N)$ internal symmetry}
In the limit \eqref{vlimit1},
the $U(N_1)$ symmetry is weakly gauged 
and therefore one can liberates the $U(N_1)$ indices in the
mass operators \eqref{massop}
\eq{
\mathcal{O}^{\alpha\, a}{}_{\beta\,b} := \bar{\phi}^{\alpha, a}\cdot \phi_{\beta, b}~,~~
\mathcal{O}'^{\alpha\, a}{}_{\beta\,b} :=\bar{\psi^c}^{\alpha, a}\cdot \psi_{\beta, b}~,
}
where we are using the notation \eqref{eq:contractionO} for the contraction in this subsubsection.
When the projection condition \eqref{CFT65} is imposed, we have
\eq{
\label{msym}
\mathcal{O}^{\alpha\, a}{}_{\beta\,b}
=J^{\alpha\delta} \delta^{ac} \phi_{\delta, c} \cdot J_{\beta\gamma} \delta_{bd}\phi^{*\gamma, d}
=J_{\beta\delta}\delta_{bd}J^{\alpha\gamma}\delta^{ac}\mathcal{O}^{\delta\, d}{}_{\gamma\,c}~,
}
and same for the fermion mass operator $\mathcal{O}'$. This can be rewritten in a more compact form
\eq{
\mathcal{O}=(\mathcal{S}\mathcal{O}\,\mathcal{S}^{-1})^T~,\quad\mathcal{O}'=(\mathcal{S}\mathcal{O}'\,\mathcal{S}^{-1})^T~,\quad\mathcal{S}_{\alpha\, a,\,\beta\, b}:=J_{\alpha\beta}\delta_{ab}~,
}
or equivalently
\eq{
\label{es}
\mathcal{O}^{(\prime)}{}^{\alpha\,\beta,a\,b}=-\mathcal{O}^{(\prime)}{}^{\beta\,\alpha,b\,a}~,
}
where the $USp(2N_2)$ and $O(N_1)$ indices are raised and lowered by $J^{rs}$ and $\delta^{ab}$ respectively. 
Other even spin operators are constructed 
by inserting even number of derivatives between $\phi\phi$ and $\psi\psi$ 
with the similar $USp(4)$ and $O(N_1)$ index structure. 
For operators of odd spins, 
their analogs of \eqref{msym} have an additional minus sign. 
For example, the spin-$1$ operators are
\eq{
\mathcal{O}^{\alpha\, a}_\mu{}_{\beta\,b} :=\phi^{*\alpha, a}\cdot \partial_\mu\phi_{\beta, b}~,~~
\mathcal{O}'^{\alpha\, a}_\mu{}_{\beta\,b} :=\bar{\psi^c}^{\alpha, a}\cdot \gamma_\mu\psi_{\beta, b}~,
}
which satisfy
\eq{
\mathcal{O}^{(\prime)}{}^{\alpha\, a}_\mu{}_{\beta\,b}=-J_{\beta\delta}\delta_{bd}J^{\alpha\gamma}\delta^{ac}\mathcal{O}^{(\prime)}{}^{\delta\, d}_\mu{}_{\gamma\,c}~,
}
where for the bosonic spin-1 operator $\mathcal{O}^{\alpha\,\beta,a\,b}_\mu$, 
we have identified two operators differing by a total derivative. Equivalently, this can be written as
\eq{
\label{ess}
\mathcal{O}^{(\prime)}_\mu=-(\mathcal{S}\mathcal{O}^{(\prime)}_\mu\,\mathcal{S}^{-1})^T~,\quad\text{or}\quad\mathcal{O}^{(\prime)}{}^{\alpha\,\beta,a\,b}_\mu=\mathcal{O}^{(\prime)}{}^{\beta\,\alpha,b\,a}_\mu~.
}
This symmetry property holds also for other operators with odd spins. 
As for complex spin-$\ft12$ operators which are bilinear in bosons and fermions, we have
\be
\mathcal{O}_{F}{}^{\alpha\, a}{}_{\beta\,b} :=\psi^c{}^{\alpha, a} \cdot \phi_{\beta, b}~,
\ee
which are related to its complex conjugate by
\be
\mathcal{O}_{F}{}^{\alpha\, a}{}_{\beta\,b}
=J^{\alpha\gamma} \psi_{\gamma, a}\cdot J_{\beta\delta} \bar{\phi}^{\delta, b}
=J^{\alpha\gamma}J_{\beta\delta}C\bar{{\cal O}}_{F\,\gamma\,a}{}^{\delta\,b}~.
\ee
Therefore 
the number of spin-$\ft12$ operators is reduced from $32M^2$ to $16M^2$.

All these conditions on the bilinear operators are equivalent to those 
on the Vasiliev theory side given in \eqref{ccf} and \eqref{pjc}. 
On the HS side the spin is characterized by the number of $Y$-oscillators. 
The condition $(y,\bar{y})\rightarrow({\rm i}y,{\rm i}\bar{y})$ imposed by the anti-automorphisms then distinguishes 
the symmetry properties of even and odd spin operators 
in the same way as in \eqref{es} and \eqref{ess}. 
The number of fermionic operators are also constrained in the same way 
as the fermionic HS fields.

\subsubsection{$USp(2N)$ internal symmetry}
In the other limit \eqref{vlimit2},
the $U(2N_2)$ symmetry is weakly gauged and
one can liberates the $U(2N_2)$ indices in the mass operators \eqref{massop}. 
The projection condition \eqref{CFT65} then implies
\eq{
\mathcal{O}^{(2s)}=(\mathcal{S}\mathcal{O}^{(2s)}\,\mathcal{S}^{-1})^T~,\quad\mathcal{O}^{(2s+1)}=-(\mathcal{S}\mathcal{O}^{(2s+1)}\,\mathcal{S}^{-1})^T~,\quad\mathcal{S}_{\alpha\, r,\,\beta\, s}:=J_{\alpha\beta}\mathcal{J}_{rs}~,
}
which are equivalent to \eqref{uspbk}. 
The projection condition \eqref{CFT65} also constrains the number of fermionic operators in a way similar to \eqref{uspfbk}.

\subsection{SUSY enhancement}
\label{sec:enhancement}
In sec.~\ref{sec:construction}
we have seen that
the $\mathcal{N}=5$ Vasiliev theory with the $O(N_1 )$ internal symmetry
has enhanced $\mathcal{N}=6$ SUSY in the $O(2)$ case
since the two index anti-symmetric representation of $O(N_1 )$ becomes trivial
and the gravitino multiplet combined with the $\mathcal{N}=5$ SUGRA multiplet 
comprises the $\mathcal{N}=6$ SUGRA multiplet.
Interestingly similar phenomenon occurs also in the CFT side \cite{Hosomichi:2008jb,Schnabl:2008wj}.
It was shown that
Gaiotto-Witten type theory \cite{Gaiotto:2008sd} with $\mathcal{N}=5$ SUSY 
has enhanced $\mathcal{N}=6$ SUSY
if representations of matters can be decomposed into a complex representation and its conjugate.
Now the $\mathcal{N}=5$ ABJ theory with the gauge group $O(2)_{2k}\times USp(2N_2 )_{-k}$
belongs to this class and 
therefore SUSY of the $\mathcal{N}=5$ ABJ theory is enhanced to $\mathcal{N}=6$
when the gauge subgroup $O(N_1 )$ is $O(2)$. 
Thus the analysis in sec.~\ref{sec:construction} shows that
the $\mathcal{N}=5$ Vasiliev theory with the $O(N)$ internal symmetry
knows about the SUSY enhancement in the $\mathcal{N}=5$ ABJ theory.
This is a strong evidence for our proposal.

\subsection{Correlation functions and free energy of ABJ theory in higher spin limit}
\label{sec:correlation}
Here we compute two-point functions of a $U(1)$ flavor symmetry current
and stress tensor,
and sphere free energy in the $\mathcal{N}=5$ ABJ theory.
In 3d CFT on flat space,
the two-point function of $U(1)$ flavor symmetry current $j_\mu$ 
is constrained as
\be
\langle j_i (x) j_j (0) \rangle
=\frac{\tau_f}{16\pi^2} 
\frac{P_{ij}}{x^2} +\frac{i\kappa_f}{2\pi} \epsilon_{ijk}\del_k \delta^{(3)}(x)~.
\ee
where $P_{ij}=\delta_{ij}\del^2 -\del_i \del_j$. 
Here we compute $\tau_f$ and $\kappa_f$
associated with the $U(1)$ flavor symmetry 
which assigns charges $+1$ and $-1$ 
to the chiral multiplet $\mathcal{A}_1$ and $\mathcal{A}_2$ respectively
in 3d $\mathcal{N}=2$ language.
Two-point function of the canonically normalized stress tensor 
in 3d CFT on flat space \cite{Osborn:1993cr} takes the form
\begin{eqnarray}
\langle T_{ij}(x) T_{k\ell}(0) \rangle
&=& \frac{c_T}{64} (P_{ik}P_{j\ell} +P_{jk}P_{i\ell} -P_{ij}P_{k\ell}) \frac{1}{16\pi^2 x^2}~ \nonumber\\
&& +\frac{i\kappa_T}{192\pi} \left(
\epsilon_{i km}\del^m P_{j\ell} +\epsilon_{jkm}\del^m P_{i\ell}
+\epsilon_{i\ell m}\del^m P_{jk} +\epsilon_{j\ell m}\del^m P_{ik}
 \right) \delta^{(3)}(x) ,
\end{eqnarray}
where we normalize $c_T$
such that for each free real scalar or Majorana fermion, $c_T =1$.
One may expect that
there is a simple relation between $\tau_f$ and $c_T$ in the ABJ theory
since extended SUSY field theories have non-Abelian $R$-symmetry 
which includes the $U(1)_R$ symmetry and $U(1)$ flavor symmetries.
Indeed it is known \cite{Chester:2014fya} that
$\tau_f$ in the $\mathcal{N}=6$ ABJ theory has the relation
\be
c_T = 4\tau_f~ .
\ee
In app.~\ref{obfproj}
we prove that 
this relation holds also in the $\mathcal{N}=5$ ABJ theory
based on the result of \cite{Chester:2014fya}
and so-called large-$N$ orbifold equivalence \cite{Kovtun:2004bz}.
Therefore, $c_T$ can be obtained once $\tau_f$ is known. 
$\tau_f$ and $\kappa_f$ can be computed 
from the partition function on $S^3$ \cite{Closset:2012vg} deformed by real mass\footnote{
The real mass can be introduced by taking 3d $\mathcal{N}=2$ background vector multiplet associated with the flavor symmetry
to be constant adjoint scalar with flat connection and trivial gaugino.
}:
\be
\tau_f = -8 \left. {\rm Re}\frac{1}{Z(0)}\frac{\del^2 Z(m)}{\del m^2} \right|_{m=0}~,\quad 
\kappa_f=2\pi \left. {\rm Im}\frac{1}{Z(0)}\frac{\del^2 Z(m)}{\del m^2} \right|_{m=0}.
\label{tfkf}
\ee
The mass deformed partition function $Z(m)$
can be exactly computed by SUSY localization \cite{Kapustin:2009kz}
and its explicit form is 
\be
Z(m)
= \frac{1}{|W|}
\int \frac{d^{N_1} \mu}{(2\pi )^{N_1}} \frac{d^{N_2} \nu}{(2\pi )^{N_2}}
e^{\frac{{\rm i}k}{2\pi}\left( \sum_{j=1}^{N_1} \mu_j^2 -\sum_{b=1}^{N_2} \nu_b^2 \right)}
Z_{\rm vec}^O (\mu ) Z_{\rm vec}^{USp} (\nu ) Z_{\rm bi}(\mu ,\nu ,m)~,
\ee
where
\begin{eqnarray}
|W|
&=& 2^{N_2} N_2 ! |W_O |~,\nn\\
Z_{\rm vec}^O (\mu )
&=& \left \{  \begin{matrix}
\prod_{i<j} \Bigl[
2\sinh{\frac{\mu_i -\mu_j}{2}}\cdot 2\sinh{\frac{\mu_i + \mu_j}{2}} \Bigr]^2
 & {\rm for\ even}\ N_1 \cr
\Big[\prod_{j=1}^{N_1} 4\sinh^2{\frac{\mu_j}{2}}\Big]
\prod_{i<j} \Bigl[
2\sinh{\frac{\mu_i -\mu_j}{2}}\cdot 2\sinh{\frac{\mu_i + \mu_j}{2}} \Bigr]^2
 & {\rm for\ odd}\ N_1 \cr
\end{matrix} \right.~,\nn\\
Z_{\rm vec}^{USp} (\nu )
&=&\Big[ \prod_{b=1}^{N_2} 4\sinh^2{\nu_b}\Big]
 \prod_{a<b} \Bigl[
2\sinh{\frac{\nu_a -\nu_b}{2}}\cdot 2\sinh{\frac{\nu_a + \nu_b}{2}} \Bigr]^2~,\nn\\
Z_{\rm bi}(\mu ,\nu ,m)
&=& \left \{  \begin{matrix}
\frac{1}{\prod_{i,b}
2\cosh{\frac{\mu_i -\nu_b +m}{2}}\cdot 2\cosh{\frac{\mu_i + \nu_b +m}{2}}
\times (m\rightarrow -m)
}
& {\rm for\ even}\ N_1 \cr
 \frac{1}
{\prod_{b=1}^{N_2}2\cosh{\frac{\nu_b +m}{2}}
\prod_{i,b}
2\cosh{\frac{\mu_i -\nu_b +m}{2}}\cdot 2\cosh{\frac{\mu_i + \nu_b +m}{2}}
\times (m\rightarrow -m)   }
& {\rm for\ odd}\ N_1 \cr
\end{matrix}\right.~.\nn\\
\label{eq:1loopdet}
\end{eqnarray}
$|W_O|$ is the rank of the Weyl group associated with gauge group $O(N_1)$, which is equal to $2^{N-1}N!$ ($2^{N}N!$) for $O(2N)$ ($O(2N+1)$).

\subsubsection{$O(N)$ internal symmetry}
We first consider the limit $N_1\ll N_2$. In this case, the ${\cal N}=5$ ABJ theory is dual to the bulk ${\cal N}=5$ HS theory with $O(N_1)$ internal symmetry.
We rewrite the partition function as
\be
Z(m)
= \frac{1}{|W_O | } \int \frac{d^{N_1}\mu}{(2\pi )^{N_1} }
e^{\frac{{\rm i}k}{2\pi} \sum_{j=1}^{N_1} \mu_j^2 }
  \prod_{\alpha \in {\rm root}_{O(N_1)}} (\alpha \cdot \mu )
   \left\langle  e^{V(\mu ,\nu )} \right\rangle_{USp(2N_2)_{-k}}~,
\ee
where
\be
V(\mu ,\nu )
= \sum_{\alpha \in {\rm root}_{O(N_1)}}
 \log{\frac{2\sinh{\frac{ \alpha \cdot \mu }{2}} } {\alpha \cdot \mu}  } -\log{Z_{\rm bi}}(\mu ,\nu ,m )~,
\ee
and
$\langle\mathcal{O}\rangle_{USp(2N_2)_{-k}}$ denotes the unnormalized VEV over the $USp(2N_2)$ part
\be
\langle\mathcal{O}\rangle_{USp(2N_2)_{-k}}
=\frac{1}{2^{N_2}N_2 !} \int  \frac{d^{N_2}\nu}{ (2\pi )^{N_2}}
\,\mathcal{O}\, e^{-\frac{1}{2g_s}\sum_a \nu_a^2 }
 \prod_{ a \neq b } \Biggl[
 2\sinh{\frac{\nu_a -\nu_b }{2}} \cdot  2\sinh{\frac{\nu_a +\nu_b }{2}}  \Biggr]
\prod_{b=1}^{N_2} 4\sinh^2{\nu_b}~,
\ee
where
\be
g_s =-\frac{\pi {\rm i}}{k}~.
\ee
This is formally the same as the VEV of $\mathcal{O}$ in the $USp(2N_2 )_{-k}$ CS matrix model
on $S^3$ (without the level shift). When $N_1/k \ll 1$,
the integration over $\mu$ is dominated by the region $\mu\simeq 0$ and
we can approximate $V(\mu ,\nu )$ by small $\mu$ expansion
\bea
V(\mu, \nu )
&=&  -\log{Z_{\rm bi}(\mu =0,\nu ,m)}  +\mathcal{O}(\mu^2 )  \nn\\
&=&
-N_1 \sum_{a=1}^{N_2}\Biggl[ \log{\left( 1+e^{  \nu_a +m}  \right)}
+\log{\left( 1+e^{  \nu_a -m}  \right)} -\nu_a  \Biggr]
+\mathcal{O}(\mu^2 )~.
\eea
Because the integration measure over $\nu$ is an even function of $\nu$,
we find in the limit $N_1/k \ll 1$, the mass deformed partition function is approximately given by
\be
Z(m)
\simeq
Z_{\rm Gauss}^O \left( \frac{\pi {\rm i}}{k},N_1 \right)
  \left\langle
\exp\Bigl[
 -N_1 \sum_{a=1}^{N_2} \log (1+e^{ \nu_a +m})(1+e^{ \nu_a -m}) \Bigr]  \right\rangle_{USp(2N_2)_{-k}}~,
\label{eq:Zm_O}
\ee
where\footnote{
${\rm vol}(O(N)) 
=\frac{2^N \pi^{N(N+1)/4}}{\prod_{n=1}^N \Gamma (n/2)}$
(see e.g.~\cite{Marinov:1980jn}).
}
\begin{equation}
Z_{\rm Gauss}^O (g,N)
\equiv  \frac{1}{|W_O |}
\int \frac{d^{N}x}{(2\pi )^{N} }
 e^{-\frac{1}{2g} \sum_{j=1}^N x_j^2 }
\prod_{\alpha \in {\rm root}_{O(N)}} (\alpha\cdot x)~ 
= \frac{2^{{\rm rank}(O(N))+1}  (2\pi g)^{{\rm dim}(O(N))}}{{\rm vol}(O(N)) } .
\end{equation}
Now we are interested in the planar limit of the $\left\langle  e^{V(\mu ,\nu )} \right\rangle_{USp(2N_2)_{-k}}$ part.
Since the planar limit of the $USp(2N_2)_{-k}$ CS theory
is the same as the one of $O(2N_2 )_{-2k}$ CS theory\footnote{Notice that
$\langle \mathcal{O}\rangle_{USp(2N_2)_{-k},\,{\rm unnormalized}}$
$=$
$Z_{USp(2N_2)_{-k}}$
$\langle\mathcal{O}\rangle_{USp(2N_2)_{-k},\,{\rm normalized}}$ and in the planar limit,
$\langle \mathcal{O}\rangle_{USp(2N_2 )_{-k},\,{\rm normalized,\,planar}}$
$=$
$\langle \mathcal{O}\rangle_{O(2N_2 )_{-2k},\,{\rm normalized,\,planar}}$.},
we can rewrite $Z(m)$ in the higher spin limit as
\be
Z(m)
\simeq
Z_{\rm Gauss}^O \left( \frac{\pi i}{k},N_1 \right) Z_{USp(2N_2)_{-k}{\rm CS}}
  \left\langle
\exp\Bigl[
 -N_1 \sum_{a=1}^{N_2} \log (1+e^{ \nu_a +m})(1+e^{ \nu_a -m}) \Bigr]  \right\rangle_{O(2N_2 )_{-2k},\,{\rm planar}}~,
\ee
where $\langle \cdots \rangle_{O(2N_2 )_{-2k},\,{\rm planar}}$
denotes the normalized VEV in the planar limit of $O(2N_2 )_{-2k}$ CS matrix model.
It can be computed by combining the result of Appendix. \ref{app:OU}
with the technique in \cite{Drukker:2010nc}. Let us introduce
\be
g_O (X;t_2 )
= -\frac{1}{N_2} \left\langle
\sum_{a=1}^{N_2} \log{(1-Xe^{\nu_a})}
\right\rangle_{O(2N_2 )_{-2k},\,{\rm planar}}~,
\label{gO}
\ee
where
\be
t_2 =-\frac{\pi {\rm i} N_2}{k}~.
\ee
Using \eqref{gO}, we find that in the planar limit
\be
\langle e^{V(\mu ,\nu )}\rangle_{USp(2N_2)_{-k},\,{\rm normalized}}
\simeq \exp{\Biggl[
N_1 N_2
\Big(
 g_O (-e^{m};t_2 ) +g_O (-e^{-m};t_2 )
\Big)
\Biggr]}~.
\ee
To compute $g_O (Y;t_2 )$,
we first use the relation between the
single trace VEV in $O(2N)_{-2k}$ CS and $U(N)_{-k}$ CS
in the planar limit,
which is shown in Appendix. \ref{app:OU}.
This relation leads us to
\be
g_O (X;t_2 ) = g_U (X;2t_2)~,
\ee
where
\be
g_U (X;2t_2 )
= -\frac{1}{N_2} \left\langle
\sum_a \log{(1-Xe^{\nu_a})} \right\rangle_{U(N_2)_{-k},\,{\rm planar}}~.
\ee
$g_U (X;t)$ was obtained in \cite{Drukker:2010nc} for arbitrary $X$ as
\bea
g_U (X;t )
&=& \frac{1}{t} \Biggl[ \frac{\pi^2}{6} -\frac{1}{2} \Big( \log{h(X)} \Big)^2
+\log{h(X)} \Big( \log{(1-e^{-t}h(X))} -\log{(1-h(X))} \Big) \nn\\
&& -{\rm Li}_2 (h(X)) +{\rm Li}_2 (e^{-t}h(X)) -{\rm Li}_2 (e^{-t}) \Biggr]~,
\eea
where
\be
h(X) = \frac{1}{2} \left[ 1+X +\sqrt{(1+X)^2 -4e^{t}X} \right]~.
\ee

Using this, we get
\begin{equation}
\left. \frac{\del^2}{\del m^2}\langle e^{V(\mu ,\nu )} \rangle_{USp(2N_2),\,{\rm normalized}} \right|_{m=0}
=-\frac{N_1 M}{\pi t} e^{\frac{\pi {\rm i}t}{2}} \sin{\frac{\pi t}{2}}
e^{ 2N_1 Mg_U (-1;2t_2)}
+\mathcal{O}(1)~,
\label{eq:del2O}
\end{equation}
where $N_2 =|O(N_1 )|+M$, and $t=M/k$.
Notice that $Z(0)$ in the higher spin limit is given by
\be
Z(0)
\simeq   e^{ 2N_1 N_2 g_U (-1;2t_2)}
Z_{\rm Gauss}^O \left( \frac{\pi {\rm i}}{k},N_1 \right)  Z_{USp(2N_2)_{-k}{\rm CS}}~,
\label{z1}
\ee
we finally obtain
\be
\tau_f= -8 {\rm Re}\left( -\frac{N_1 M}{\pi t} e^{\frac{\pi {\rm i}t}{2}} \sin{\frac{\pi t}{2}}   \right)= \frac{4N_1 M\sin{\pi t}}{\pi t}~,\quad
\kappa_f=-\frac{4|O(N_1)|M}t\sin^2\frac{\pi t}2~ .
\label{eq:flavorC_O}
\ee
Then using $c_T =4\tau_f$ immediately leads us to
\be
c_T =  \frac{16N_1 M\sin{\pi t}}{\pi t}~.
\ee
As a consistency check,
let us consider the $t\rightarrow 0$ limit.
Then, since the ${\cal N}=5$ ABJ theory has
$8N_1 N_2$ real scalars and $8N_1 N_2$ Majorana fermions,
$c_T$ should be
$16N_1 N_2 =16N_1 M+\mathcal{O}(1)$,
which is reproduced by our result. 
The result on $\kappa_f$ in \eqref{eq:flavorC_O}
is not apparently invariant under Seiberg duality. 
However, 
we can make $\kappa_f$ invariant under the duality
by shifting $\kappa_f$ by the integer $2|O(N_1)|k$, 
which is the degree of freedom of adding a local CS counterterm 
in the CFT Lagrangian \cite{Closset:2012vp}. 
After the shift, we find
\be
\left. \kappa_f \right|_{\rm shifted}
= \frac{2|O(N_1)|M\cos{\pi t}}{t}~.
\label{eq:shifted_O}
\ee
In app.~\ref{app:beforeG}
we show that 
$\tau_f$ and $\left. \kappa_f \right|_{\rm shifted}$ are
the same as the ones in two-point function of $O(N_1 )$ gauge current in the HS limit.

Utilizing \eqref{z1}, we can compute the free energy \eqref{fvabj} in the limit $N_1\ll N_2$ as
\bea
F^{\rm vec}_{N,M}&=&-\log{\frac{|Z(0)|}{|Z_{\rm CS}^{USp(2M)_{-k}}|}}\Big|_{N_1\ll N_2} \nn\\
&= &-2N_1M {\rm Re}\Big[g_U(-1;2t_2)\Big]
-\log\Big|Z_{\rm Gauss}^O \left(\frac{\pi{\rm  i}}{k},N_1 \right)\Big|
-\log{\frac{|Z_{\rm CS}^{USp(2N_2)_{-k}}|}{|Z_{\rm CS}^{USp(2M)_{-k}}|}}+\mathcal{O}(1)~.
\eea
Using the result of app.~\ref{rcspf} on the third term above
we obtain
\be
F^{\rm vec}_{N,M}
= \frac{4NM}{\pi t} I\left( \frac{\pi t}{2} \right)
+\frac{{\rm dim}[O(N_1)]}{2} \log{M} +\mathcal{O}(1)~,
\label{eq:F_O}
\ee
where $N$ is the rank of the global symmetry group $O(N_1)$ in the higher spin limit and
\[
I(x) = {\rm Im}\Bigl[ {\rm Li}_2 ({\rm i}\tan{x}) \Bigr] -x \log{\tan{x}}~.
\]

\subsubsection{$USp(2N)$ internal symmetry}
We now turn to the other limit  $N_1\gg N_2$. In this case, the ${\cal N}=5$ ABJ theory is dual
to the bulk ${\cal N}=5$ HS theory with $USp(2N_2)$ internal symmetry.
The mass deform partition function can be rewritten as
\be
Z(m)
= \frac{1}{|W_{USp} | } \int \frac{d^{N_2}\nu}{(2\pi )^{N_2} }
e^{-\frac{{\rm i}k}{2\pi} \sum_{a=1}^{N_2} \nu_a^2 }
  \prod_{\alpha \in {\rm root}_{USp(2N_2)} } (\alpha \cdot \nu )
   \left\langle  e^{V(\mu ,\nu )} \right\rangle_{O(N_1 )_{2k}}~,
\ee
where
\be
V(\mu ,\nu )
= \sum_{\alpha \in {\rm root}_{USp(2N_2)} }
 \log{\frac{2\sinh{\frac{ \alpha \cdot \nu }{2}} } {\alpha \cdot \nu}  } -\log{Z_{\rm bi}}(\mu ,\nu ,m )~,
\ee
and
$\langle\mathcal{O}\rangle_{O(N_1 )_{2k}}$
denotes the unnormalized VEV over the $O(N_1 )$ part
\be
\langle\mathcal{O}\rangle_{O(N_1 )_{2k}}
=\frac{1}{|W_O |} \int  \frac{d^{N_1}\mu}{ (2\pi )^{N_1}}
\,\mathcal{O}\, e^{-\frac{1}{2g_s}\sum_j \mu_j^2 }
 \prod_{\alpha \in {\rm root}_{O(N_1)} }  2\sinh{\frac{\alpha\cdot \mu }{2}}~,
\ee
where
\be
g_s =\frac{\pi {\rm i}}{k}~.
\ee
$\langle\mathcal{O}\rangle_{O(N_1 )_{2k}}$ is formally the same as the VEV in the $O(N_1 )_{2k}$ CS matrix model on $S^3$ .
In the higher spin limit $N_2/k\ll1$,
$V(\mu ,\nu )$ can be approximated by small $\nu$ expansion
\bea
V(\mu, \nu )
&=&  -\log{Z_{\rm bi}(\mu ,\nu =0 ,m)}  +\mathcal{O}(\nu^2 )  \nn\\
&=& -2N_2 \sum_j \Biggl[ \log{\left( 1+e^{  \mu_j +m}  \right)}
+\log{\left( 1+e^{  \mu_j -m}  \right)} -\mu_j
+\log{\left( 2\cosh{\frac{m}{2}} \right)} \Biggr]
+\mathcal{O}(\nu^2 )~.\nn\\
\eea
Using the fact that the integration measure over $\mu$ is an even function of $\mu$, we find
\be
Z(m)
\simeq  \frac{ Z_{\rm Gauss}^{USp} \left( -\frac{\pi {\rm i}}{k},N_2 \right) }{\left( 2\cosh{\frac{m}{2}} \right)^{2N_2}}
  \left\langle
\exp\Bigl[
 -2N_2 \sum_j \log (1+e^{ \mu_j +m})(1+e^{ \mu_j -m}) \Bigr]  \right\rangle_{O(N_1 )_{2k}}~,
\label{eq:Zm_Usp}
\ee
where\footnote{
${\rm vol}(USp(2N)) =2^{-2N}{\rm vol}(O(2N+1)) $.
}
\be
Z_{\rm Gauss}^{USp} (g,N)
= \frac{1}{|W_{USp} |}\int \frac{d^{N}x}{(2\pi )^{N} }
 e^{-\frac{1}{2g} \sum_{j=1}^N x_j^2 }
\prod_{\alpha \in {\rm root}_{USp(2N)} } (\alpha \cdot x)= \frac{2^{N+1}  (2\pi g)^{\frac{N(2N+1)}{2}}}{{\rm vol}(USp(2N))}~.
\ee
Now we need to compute the $\left\langle  e^{V(\mu ,\nu )} \right\rangle_{O(N_1 )_{2k}}$ in the planar limit.
Let us rewrite $Z(m)$ in the higher spin limit as
\be
Z(m)
\simeq
\frac{Z_{\rm Gauss}^{USp} \left( -\frac{\pi {\rm i}}{k},N_2 \right) Z_{O(N_1 )_{2k} {\rm CS}} }
{{\left( 2\cosh{\frac{m}{2}} \right)^{2N_2}}}
  \left\langle
\exp\Bigl[
 -2N_2 \sum_j \log (1+e^{ \mu_j +m})(1+e^{ \mu_j -m}) \Bigr]  \right\rangle_{O(N_1)_{2k},\,{\rm normalized}}~.
\ee
Then we find in the planar limit
\be
\left\langle e^{V(\mu ,\nu )}\right\rangle_{O(N_1)_{2k},\,{\rm normalized}}
\simeq \frac{1}{{\left( 2\cosh{\frac{m}{2}} \right)^{2N_2}}}
 \exp{\Biggl[ 2N_2 |O(N_1 )|
\left( g_O (-e^{m};t_1 ) +g_O (-e^{-m};t_1 )  \right)
\Biggr]} ,
\ee
where
\be
t_1 =\frac{\pi {\rm i} |O(N_1 )|}{k}~.
\ee
As in the previous case, we obtain
\begin{equation}
\left. \frac{\del^2}{\del m^2}\langle e^{V(\mu ,\nu )} \rangle_{O(N_1 )\,{\rm normalized}} \right|_{m=0}
= -\frac{2N_2 M  }{\pi t} e^{-\frac{\pi {\rm i}t}{2}} \sin{\frac{\pi t}{2}} e^{ 4N_2 M g_U (-1;2t_1)}
+\mathcal{O}(1)~,
\label{eq:del2USp}
\end{equation}
where we have set $|O(N_1 )| =N_2 +M$ and $t=M/k$.
Since $Z(0)$ in the higher spin limit is given by
\be
Z(0)
\simeq   e^{ 4N_2 M g_U (-1;2t_1)}
Z_{\rm Gauss}^{USp} \left( -\frac{\pi {\rm i}}{k},N_2 \right)  Z_{O(N_1 )_{2k}{\rm CS}}~,
\ee
we finally obtain
\be
\tau_f= -8 {\rm Re}\left( -\frac{2N_2 M}{\pi t} e^{-\frac{\pi {\rm i}t}{2}} \sin{\frac{\pi t}{2}}   \right)
= \frac{4N_2 M\sin{\pi t}}{\pi t}~,\quad
\kappa_f =\frac{4N_2 M}t\sin^2\frac{\pi t}2~ .
\label{eq:flavorC_USp}
\ee
and
\be
c_T
=4\tau_f
=  \frac{32N_2 M\sin{\pi t}}{\pi t}~.
\ee
As a consistency check,
let us consider the $t\rightarrow 0$ limit. In this limit,
$c_T$ should behave as $16N_1 N_2 =32N_2 M+\mathcal{O}(1)$
and this is consistent with our result. 
As in the previous case,
we can make $\kappa_f$ invariant under the duality
by shifting $\kappa_f$ by the integer $-2N_2 k$:
\be
\left. \kappa_f \right|_{\rm shifted}
= -\frac{2N_2 M\cos{\pi t}}{ t}~.
\label{eq:shifted_USp}
\ee
In app.~\ref{app:beforeG}
we find that 
$\tau_f$ and $\left. \kappa_f \right|_{\rm shifted}$ are
given by the same formula as those in two-point function of $USp(2N_2 )$ gauge current in the HS limit.

The free energy in the other higher spin limit $N_1\gg N_2$ is given as
\bea
F^{\rm vec}_{N_2,M}
&=&-\log{\frac{|Z(0)|} {|Z_{\rm CS}^{O(2N)_{2k}}|}}\Big|_{N_1 \gg N_2}\nn\\
&=&-4N_2M {\rm Re}[g_U(-1;2t_1 )]
-\log\Big|Z_{\rm Gauss}^{USp(2N_2)} \left(-\frac{\pi{\rm  i}}{k} \right)\Big|-\log{\frac{|Z_{\rm CS}^{O(N_1)_{2k}}|}{|Z_{\rm CS}^{O(2M)_{2k}}|}}+\mathcal{O}(1)~.
\eea
Using the results of app.~\ref{rcspf},
we obtain
\bea
F^{\rm vec}_{N_2,M}=\frac{4N_2M}{\pi t} I\left( \frac{\pi t}{2} \right)
+\frac{{\rm dim}[USp(2N_2)]}{2} \log{M} +\mathcal{O}(1)~.
\eea

\subsection{Holographic dictionary and prediction of on-shell action}
\label{sec:comparison}
In the previous subsection, 
we have computed $c_T$, $\tau_f$ and $\kappa_f$ associated with the $R$-currents and the free energy \eqref{fvabj} in the two different higher spin limits. 
The results are summarized below
\bea
&& c_T=\frac{32NM\sin\pi t}{\pi t}~,\quad 
\tau_f = \frac{8N M\sin{\pi t}}{\pi t}~,\quad 
\kappa_f= \frac{2N M\cos{\pi t}}{ t}~, \quad M\gg N~, \nonumber\\
&& F^{\rm vec}_{N,M}
= \frac{4NM}{\pi t} I\left(\frac{\pi t}{2} \right)
+\frac{{\rm min}\{{\rm dim}O(N_1 ),{\rm dim}USp(2N_2 )\}}{2}
\log{M} +\mathcal{O}(1)~,\nn\\\label{fvabj2}
\eea
where $N={\rm min}\{|O(N_1 )|,N_2 \}$ and
$M=|N_2 -|O(N_1 )| |$.
We can relate the Newton constant on the bulk 
to the CFT parameters
using the logic in \cite{Honda:2015sxa} and 
$c_T$ computed in the previous subsection.
First let us consider 
usual AdS/CFT correspondence between CFT and Einstein gravity.
If we consider the canonically nomarlized Einstein-Hilbert action,
then the stress tensor two-point function is generated by 
\be
 S[g]
=\left. \frac{1}{16\pi G_N} \int d^4x \sqrt{g}R
 \right|_{\rm quadratic\, term}~.
\ee
In this normalization
the holographic computation shows \cite{Buchel:2009sk}
\be
\left. \frac{G_N}{L_{\rm AdS}^2} \right|_{\rm Einstein\ gravity}
=\frac{32}{\pi c_T}~.
\ee
Now we come back to the Vasiliev theory with internal symmetry
whose fields are matrix valued.
Since the graviton coupling to the CFT stress tensor 
should be singlet under both the bulk $R$-symmetry and internal symmetry,
we have to take the singlet part 
and identify the Newton constant with
\be
\frac{G_N}{L_{\rm AdS}^2}
=\frac{32}{\pi N c_T}
=\frac{t}{M\sin\pi t}~.
\label{idG}
\ee

Next we find the relation between the parity-violating phase $\theta$
and the parameters in the ABJ theory.
The mixed boundary condition \eqref{mbc} for the bulk $USp(4)$ singlet spin-1 gauge field implies that
a bulk CS term should be added to the boundary action
\be
S[A]
=-\frac1{4g_{\rm bulk}^2}\int d^3xdz\sqrt{-g}F_{\mu\nu}F^{\mu\nu}
+\frac{ik_{\rm bulk}}{4\pi}\int d^3x\varepsilon^{ijk}A_i\partial_jA_k~.
\label{Aaction}
\ee
The mixed boundary condition \eqref{mbc} then follows from the variational principle. We find that
\be
\tan2\theta=\frac{2\pi}{g_{\rm bulk}^2 k_{\rm bulk}}~.
\ee
The action \eqref{Aaction} also leads to the holographic two point function for the dual spin-1 current (in the Euclidean signature)
\be
\langle\,J_i(x)J_j(y)\rangle\Big|_{\rm holographic}
=\frac1{2\pi^2g_{\rm bulk}^2}(\delta_{ij}\del^2 -\del_i \del_j ) \frac{1}{x^2}+
\frac{{\rm i} k_{\rm bulk}}{2\pi}\epsilon_{ijk}\del_k \delta^{(3)}(x)~.
\ee
where the parity even term has been read off from \cite{Freedman:1998tz}.
Comparing the holographic result with the CFT result, we obtain
\be
\frac1{g_{\rm bulk}^2 k_{\rm bulk}}=\frac{\tau_f}{8 \kappa_f}~.
\ee
As discussed in app.~\ref{app:beforeG},
the results on $\tau_f$ and $\kappa_f$ 
take the same form as those for the $U(1)$ flavor symmetry 
in the previous subsection up to the integer shift of $\kappa_f$ 
by the local counter term.
Using \eqref{fvabj2}
we arrive at the relation between $\theta$ and $t$
\[
\theta=\frac{\pi t}{2}~.
\]
One can easily show that 
this is true also for the $\mathcal{N}=6$ ABJ theory
using the results in app.~\ref{app:beforeG}.

We can compare the CFT free energy in \eqref{fvabj2} 
with the free energy of the Vasiliev theory. 
First, utilizing the results derived in \cite{Pang:2016ofv}, 
one can check that
the bulk free energy at one-loop is free of UV divergence \cite{Pang:2016ofv}.
The coefficient of the $\log M$ term 
also agrees with the expectation from \cite{Giombi:2013yva},
which states that 
each bulk spin-1 gauge fields obeying the mixed boundary condition 
contribute 
to the one loop free energy of the Vasiliev theory
by $-(1/2)\log{M}$. 
Thus the coefficient of the $\log{M}$ term
should be $(-1/2)$ times the dimension of the weakly gauged symmetry group,
which is $O(N_1)$ for $N_1\ll N_2$ and $USP(2N_2)$ for $N_1\gg N_2$. 
Due to the lack of a bulk HS action, 
it is infeasible to compute the bulk leading free energy and compare it to the CFT one. 
However, one can translate the CFT leading free energy to its bulk counterpart 
by assuming our conjecture.
Using the identifications \eqref{eq:GN} and \eqref{eq:theta}, 
we predict that the leading term 
in the free energy of the ${\cal N}=5$ Vasiliev theory takes the form
\be
F_{\rm HS}^{(0)}=\frac{8L^2_{\rm AdS}I(\theta)}{G_N\pi\sin2\theta}~.
\ee
One should notice that 
$F_{\rm HS}^{(0)}$ 
diverges as $\mathcal{O}(\log{\theta})$ 
in the limit $\theta\rightarrow 0$, which was also observed 
in the ${\cal N}=6$ case \cite{Hirano:2015yha}.
At this moment, due to the lack of a well defined bulk action, 
we are not able to confirm this by a direct evaluation on the bulk 
and postpone the interpretation of this divergence to future work.

\section{Conclusions and discussions}
We have studied
the physical consequences of adding the orientifolds 
to the ${\cal N}=6$ ABJ triality \cite{Chang:2012kt, Giombi:2011kc},
which leads us to the ABJ quadrality.
The ABJ quadrality is
the AdS/CFT correspondence
among the ${\cal N}=5$ ABJ theory 
with the gauge group $O(N_1)_{2k}\times USp(2N_2 )_{-k}$, 
type IIA string in $AdS_4 \times\mathbb{CP}^3 /\mathbb{Z}_2$
and two ${\cal N}=5$ supersymmetric Vasiliev theories in $AdS_4$.
It has turned out that
the $\mathcal{N}=5$ case is more involved 
since there are two formulations of $\mathcal{N}=5$ Vasiliev theory with either $O$ or $USp$ internal symmetry.
Accordingly, we have proposed that 
the two possible vector-like limits of 
the ${\cal N}=5$ $O(N_1 )\times USp(2N_2)$ ABJ theory defined by $N_2 \gg N_1$ and $N_1 \gg N_2 $ 
correspond to the semi-classical ${\cal N}=5$ Vasiliev theories 
with $O(N_1 )$ and $USp(2N_2)$ internal symmetries respectively.
We have also put forward
the precise holographic dictionary 
between the parameters on the both sides
by matching the correlation functions,
where the Newton constant $G_N$ is related to $M$ and $t$ by \eqref{eq:GN}
and the parity violating phase $\theta$ is related to $t$ via \eqref{eq:theta}.

We have provided various evidence for the correspondence
between the $\mathcal{N}=5$ ABJ and Vasiliev theories. 
First, the full spectrum of the ${\cal N}=5$ Vasiliev theory 
has been shown to match 
with that of the higher spin currents in the ${\cal N}=5$ ABJ theory.
Second, we have exhibited the equivalence of the ``orientifold projections" on the HS and CFT sides at the level of the spectrum.  
Third, we have observed 
the SUSY enhancement from ${\cal N}=5$ to $\mathcal{N}=6$ occurs on both sides when the weakly gauged symmetry is $O(2)$.
Finally,
we have proposed that
the free energy of the $\mathcal{N}=5$ Vasiliev theory
should be compared to the combination \eqref{fvabj} on the CFT side,
which has the following properties i) The leading term in the $1/M$-expansion is linear in $M$;
ii) It respects the Seiberg-like duality \eqref{sdua};
iii) The $\mathcal{O}(\log{M})$ term matches the $\mathcal{O}(\log{G_N})$ term in the one-loop free energy of the $\mathcal{N}=5$ Vasiliev theory.
Based on the free energies defined for the vector limits of the ${\cal N}=5$ ABJ theory, we predict the form of the leading free energies of the ${\cal N}=5$ Vasiliev theories in $AdS_4$  upon applying the holographic dictionary.

So far our results on the HS side rely on the linear analysis of Vasiliev equations and HS gauge transformation rules. 
In order to extract three and higher point correlation functions of $3d$ higher spin currents from $4d$ Vasiliev equations,
one must go beyond the linear level and derive the higher order corrections to the linearized equations of motion. 
As observed in \cite{Giombi:2012ms, Boulanger:2015ova}, 
there are subtleties in deriving HS interaction vertices from the Vasiliev equations. 
The standard way of solving the Vasiliev equations order by order in the weak field expansion 
leads to apparent non-localities in certain cubic vertices. 
Especially in the parity violating case,
the bulk computation following the procedure of \cite{Giombi:2012ms}  
cannot reproduce
the three-point correlation functions
in which the three spins do not satisfy the triangle inequality\footnote{
Except for the $\langle 0s_1s_2\rangle$ case 
where although the three spins do not obey the triangle inequality, 
the corresponding HS cubic vertices are local since they are governed by the HS algebra.
Computation of all correlators of type $\langle s_1s_20\rangle$ was recently completed in \cite{Sezgin:2017jgm}.
The $0-s-s$ vertex was also obtained in \cite{Taronna:201606} last June.}. 
It is illustrated in the recent papers [59] that the apparent 
non-locality in the cubic vertices can be circumvented and there 
exists a well defined procedure which gives rise to manifestly local 
quadratic corrections to the free equations of motion for generic $\theta$. 
It was recently shown in \cite{Misuna:2017bjb} that if restricting \cite{Didenko:2017lsn} to bosonic A-model, then the result of \cite{Didenko:2017lsn} agree with the previous result \cite{Sleight:2016dba} obtained by means of reconstructing HS vertices from CFT correlators.
It is interesting to generalize the analysis in \cite{Didenko:2017lsn} 
to the case with extended SUSY and internal symmetry so that
one can compare
the three-point correlators 
computed from the ${\cal N}=5$ Vasiliev theories
with those computed in the ${\cal N}=5$ ABJ theory. 
It is known \cite{Buchbinder:2015wia} that
SUSY Ward identities provide 
a simple relation 
between the three- and two-point correlators  
of the currents within the stress tensor multiplet. 
Therefore, for instance, 
matching $\langle TTT\rangle^2/\langle TT\rangle^3$ on both sides
provides an independent check of the identification between the HS and CFT parameters. 
However, 
one should bear in mind that 
in $AdS_4$, 
the $\Delta=1$ Fefferman-Graham coefficients of the scalars 
and the magnetic components of the spin-1 gauge fields 
can survive at the $AdS$ boundary 
and give finite contributions to the boundary action which may affect three and higher point functions. 
The choice of boundary terms for these fields should be consistent with their boundary conditions. 
As far as we are aware, 
fully HS invariant boundary actions have not been constructed and 
it is illuminating to construct them in future investigation. 
The cubic corrections to the free equations of motion seem to contain 
genuine non-localities \cite{Bekaert:2015tva}. 
However, these non-localities may still be compatible 
with holography and a proper interpretation of them is currently under investigation.

We have shown that 
the $\mathcal{N}=5$ Vasiliev theories with $O$ and $USp$ internal symmetries 
descends from the projections \eqref{eq:orientifoldV}
of the $\mathcal{N}=6$ Vasiliev theory,
which we identify 
with the orientifold projections in the Vasiliev theory.
It is interesting to identify counterparts of orientifold projections
in other ``stringy" HS AdS/CFT correpondences \cite{Gaberdiel:2013vva}. 
One of important open problems is
to link Vasiliev theory to string theory more directly.
Although we expect that
the fundamental string in the string theory
is realized by the ``flux tube solution" in the Vasiliev theory as in \cite{Chang:2012kt},
for the time being, it seems difficult to check this 
because it is not known how to quantize Vasiliev theory.
One of the approaches from string theory is to analyze equations of motion of the dual string field theory in the ``tensionless" limit
and compare to Vasiliev equations.
Finally it is known that 
some supersymmetric quantities in the ABJ theory
are described 
by topological string \cite{Marino:2009jd,Matsumoto:2013nya,Okuyama:2016xke}
(see also \cite{Aganagic:2017tvx} from a slightly different perspective).
This fact may give some insights 
on the relation between string theory and Vasiliev theory.

\subsection*{Acknowledgments}
We are grateful to O.~Aharony, V.~E.~Didenko, K.~Mkrtchyan, E.~D.~Skvortsov, S.~Theisen and R.~Yacoby  
for useful conversations.
We would like to thank O.~Aharony for useful comments on the draft.
Y. P. is supported by the Alexander von Humboldt fellowship.
M.~H.~ would like to thank
Centro de Ciencias de Benasque Pedro Pascual, CERN and Yukawa Institute for Theoretical Physics for hospitalities.

\appendix
\section{{\it Bulk basics}}
\label{conv}
\subsubsection*{\underline{Spinor convention}}
In 4d Minkowski space with isometry $SO(3,1)\simeq SL(2,\mathbb{C})$, we use
\eq{
\spl{
(\sigma^\mu)_\alpha{}^{\dot{\beta}}
=(\mathbf{1},\sigma^i)_\alpha{}^{\dot{\beta}}~,~~
(\bar{\sigma}^\mu)_{\dot{\alpha}}{}^{\beta}
=(-\mathbf{1},\sigma^i)_{\dot{\alpha}}{}^\beta~,~~~
\mu=0,1,2,3~,
}
}
where $\sigma^i$ are the usual Pauli matrices. We also refer to the fourth component of $\sigma^\mu$ as $\sigma^r$. Spinor indices are raised or lowered by $\epsilon={\rm i}\sigma^2$. We also define
\eq{
(\sigma^{\mu\nu}){}_\alpha{}^\beta=\ft{1}{2}(\sigma^\mu\bar{\sigma}^\nu-\sigma^\nu\bar{\sigma}^\mu)_\alpha{}^\beta~,~~
(\bar{\sigma}^{\mu\nu}){}_{\dot{\alpha}}{}^{\dot{\beta}}=\ft{1}{2}(\bar{\sigma}^\mu\sigma^\nu-\bar{\sigma}^\nu\sigma^\mu)_{\dot{\alpha}}{}^{\dot{\beta}}~,
}
with the properties $\sigma^\mu_{\alpha\dot{\beta}}=\bar{\sigma}^\mu_{\dot{\beta}\alpha}$,
$\sigma^{\mu\nu}_{\alpha\beta}=\sigma^{\mu\nu}_{\beta\alpha}$ and
$\bar{\sigma}^{\mu\nu}_{\dot{\alpha}\dot{\beta}}=\bar{\sigma}^{\mu\nu}_{\dot{\beta}\dot{\alpha}}$.

\subsubsection*{\underline{Consistencies of $\tau$- and reality conditions with Vasiliev equations}}
We first show that the reality conditions and the $\tau$-projection conditions are imposed in a consistent way
\eq{
\spl{
&(A^\dag)^\dag=A~,~~(\Phi^\dag)^\dag=\Gamma\bar{\kappa}\star\kappa\star\Phi\star\kappa\star\bar{\kappa}\Gamma=\pi\bar{\pi}\pi_\xi\pi_\eta(\Phi)=\Phi~,\\
&\tau^2(A)=\pi\bar{\pi}\pi_\xi\pi_\eta(A)=A~,~~\tau^2(\Phi)=\pi\bar{\pi}\pi_\xi\pi_\eta(\Phi)=\bar{\pi}^2(\Phi)=\Phi~,
}
}
where we have used the properties of the Kleinians
\bea
\label{kl}
\kappa\star f(y,\bar{y},z,\bar{z})&=&f(z,\bar{y},y,\bar{z})\kappa~,\quad \bar{\kappa}\star f(y,\bar{y},z,\bar{z})=f(y,\bar{z},z,\bar{y})\bar{\kappa}~,\nn\\
f(y,\bar{y},z,\bar{z})\star\kappa&=&f(-z,\bar{y},-y,\bar{z})\kappa~,\quad f(y,\bar{y},z,\bar{z})\star\bar{\kappa}=f(y,-\bar{z},z,-\bar{y})\bar{\kappa}~.
\eea
By field redefinitions that are consistent with field equations, one can put $\hat{V}$ and $\hat{\bar{V}}$ in a simple form
\eq{
\hat{V}=e^{{\rm i}\theta}\Phi\star\kappa\Gamma~,\qquad \hat{\bar{V}}=e^{-{\rm i}\theta}\Phi\star\bar{\kappa}~.
}
We now show that the field equations are invariant under reality and $\tau$-conditions
\al{
&(d_xA+A\star A)^\dag=-(d_xA+A\star A)
=-\frac{{\rm i}}{4}[d\bar{z}^2(\hat{V}^\dag)+dz^2(\hat{\bar{V}}^\dag)]\nn\\
=&-\frac{{\rm i}}{4}[d\bar{z}^2(e^{-{\rm i}\theta}\Gamma\bar{\kappa}\star\kappa\star\Phi\star\kappa\Gamma)+dz^2(e^{{\rm i}\theta}\kappa\star\kappa\star\Phi\star\kappa\Gamma)]
=-\frac{{\rm i}}{4}[d\bar{z}^2(\hat{\bar{V}})+dz^2(\hat{V})]~,\\[0.7em]
&\tau(d_xA+A\star A)=-(d_xA+A\star A)
=\frac{{\rm i}}{4}[-dz^2(\tau(\hat{V}))-d\bar{z}^2(\tau(\hat{\bar{V}}))]\nn\\
=&-\frac{{\rm i}}{4}[dz^2(e^{{\rm i}\theta}\Gamma\kappa\star\bar{\kappa}\star\Phi\star\bar{\kappa})+d\bar{z}^2(e^{-{\rm i}\theta}\bar{\kappa}\star\bar{\kappa}\star\Phi\star\bar{\kappa})]
=-\frac{{\rm i}}{4}[dz^2(\hat{V})+d\bar{z}^2(\hat{\bar{V}})]~,
}
where we used $\pi\bar{\pi}\pi_\xi\pi_\eta A=\Gamma\bar{\kappa}\star\kappa\star A\star\kappa\star\bar{\kappa}\Gamma=A$. Since the Vasiliev's equation of motion for the master 0-form can be derived from the equation of motion of the 1-form by using Bianchi identity, the equation of motion for the 0-form is also be invariant under reality condition and $\tau$-condition.

\section{{\it Relation between $SO(5)$ and $USp(4)$ indices}}
\label{so52usp4}
In addition to the representations of the internal symmetry, 
each HS field carries in the $\mathcal{N}=5$ Vasiliev theory
also the indices of the fundamental representation 
of $SO(5)\simeq USp(4)$ $R$-symmetry group. 
In sec.~\ref{basicHS}, 
we have worked in the $SO(5)$ notation
while in sec.~\ref{6to5bside} we have used the $USp(4)$ notation
for the convenience in the reduction from $\mathcal{N}=6$ to $\mathcal{N}=5$. 
In this appendix
we explain a connection between these two notations. 
The connection is provided by the $SO(5)$ gamma matrices 
obeying the following symmetry properties
\eq{
\label{gms}
\spl{
& C^{T}=-C~,\quad C^{\dagger}C=1~,\quad
(\gamma^iC)^T=-\gamma^iC~,\quad (\gamma^{ij}C)^T=\gamma^{ij}C~,\\
&(\gamma^{ijk}C)^T=\gamma^{ijk}C~,\quad (\gamma^{ijkl}C)^T=-\gamma^{ijkl}C~,\quad
(\gamma^{ijklm}C)^T=-\gamma^{ijklm}C~.
}
}
Here $C$ matrix is the analog of the charge conjugation matrix defined in even dimensions.
Now we are ready to show the equivalence between the $\tau$-condition introduced in sec.~\ref{basicHS} 
and the automorphism projection introduced in sec.~\ref{6to5bside}. 
First of all, by comparing \eqref{tauO} and \eqref{tauUsp} with \eqref{aato}, 
it is straightforward to see that 
the both projection conditions act in the same way on the internal symmetry of the HS fields. 
Moreover, in both cases the spinorial oscillators undergo 
the same transformation $Y\rightarrow {\rm i}Y$, 
which in turn distinguishes the symmetry properties of HS fields with different spins. 
Finally
one can identify the $C$ matrix with the $USp(4)$ invariant matrix $J$, 
and gamma matrices $\gamma^i$ with $\xi^i$. 
The symmetry properties of the two $USp(4)$ indices, e.g. \eqref{pjc} and \eqref{pjcc}, 
as required by the automorphism condition, 
are then translated into the requirements on the $SO(5)$ representations through \eqref{gms}.

\section{{\it Seiberg-like dualities in ABJ theory}}
\label{app:duality}
In this appendix
we provide another argument to support 
the Seiberg-like dualities \eqref{sdua} for the $\mathcal{N}=5$ ABJ theory.
It is known \cite{Kapustin:2010mh} that
the Seiberg-like duality for $S^3$ partition function in the $\mathcal{N}=6$ ABJ theory
can be understood from Giveon-Kutasov duality \cite{Giveon:2008zn},
which is another Seiberg-like duality for $U(N)_k$ SQCD coupled to fundamental hyper multiplets. 
Here we argue that
the dualities \eqref{sdua} for the $\mathcal{N}=5$ ABJ theory
can be also understood as Giveon-Kutasov type dualities  
with the gauge group $O(N)_{2k}$ or $USp(2N)_{k}$. 

\subsubsection*{\underline{$U(N+M)_k \times U(N)_{-k}$ type}}
First we review the argument \cite{Kapustin:2010mh} for the $\mathcal{N}=6$ case.
Let us freeze the path integral over the $U(N)_{-k}$ vector multiplet.
Then the theory becomes
the $U(N+M)_k$ SQCD with $2N$ fundamental hyper multiplets
and the background $U(N)_{-k}$ vector multiplet.
Conversely thinking,
the $\mathcal{N}=6$ ABJ theory can be derived
by gauging $U(N)_{-k}$ in the $U(N+M)_k$ SQCD. 
For $S^3$ partition function,
this gauging procedure is technically equivalent 
to integrating over real mass associated with the $U(N)_{-k}$ symmetry in the localization formula.
For this type of SQCD,
there is a duality called Giveon Kutasov duality \cite{Giveon:2008zn},
which states that the equivalence between the gauge groups
\begin{equation}
U(N_c)_k \ \leftrightarrow\ 
U(N_f -N_c +|k|)_{-k} ,
\end{equation}
where $N_f$ is the number of the fundamental hyper multiples.
Since $N_c =N+M$ and $N_f =2N$ in our SQCD, 
this duality transforms as\footnote{
More precisely, there are induced Chern-Simons terms of flavor symmetry,
which make the other gauge group $U(N)_{-k}$ to $U(N)_k$
}
\begin{equation}
M\rightarrow |k|-M,\quad k\rightarrow -k ,
\end{equation}
which is the same action as the Seiberg-like duality in the $\mathcal{N}=6$ ABJ theory.
Thus if Giveon-Kutasov duality is correct,
then the Seiberg-like duality in the $\mathcal{N}=6$ ABJ theory is also correct.
Fortunately 
there is already a proof of Given-Kutasov duality for the $S^3$ partition function \cite{Willett:2011gp}
and this leads us to the Seiberg-like duality in the $\mathcal{N}=6$ ABJ theory.

\subsubsection*{\underline{$O(N_1 )_{2k} \times USp(2N)_{-k}$ type}}
Let us take ${\rm rank}[O(N_1 )]=N+M$
and freeze the $USp(2N)_{-k}$ vector multiplet similarly.
Then the theory becomes
$O(N_1 )_{2k}$ SQCD with $4N$ fundamental chiral multiplets
and the background $USp(2N)_{-k}$ vector multiplet.
This type of SQCD has the conjectural duality \cite{Kapustin:2011gh}:
\begin{equation}
O(N_c)_{2k} \ \leftrightarrow\ 
O(N_f -N_c +2|k|+2)_{-2k} ,
\end{equation}
where $N_f$ is the number of the fundamental chiral multipltets.
For our SQCD with $N_c =2N+2M$ and $N_f =4N$, 
this duality transforms as
\begin{equation}
M\rightarrow |k|-M+1,\quad k\rightarrow -k ,
\end{equation}
which is the same as the Seiberg-like duality
in the $O(2N+2M)_{2k}\times USp(2N)_{-k}$ ABJ theory.
Next,
for our SQCD with $N_c =2N+2M+1$ and $N_f =4N$, 
the Giveon-Kutasov-like duality acts as
\begin{equation}
M\rightarrow |k|-M,\quad k\rightarrow -k ,
\end{equation}
which is the same action as the Seiberg-like duality
in the $O(2N+2M+1)_{2k}\times USp(2N)_{-k}$ ABJ theory.

\subsubsection*{\underline{$USp(2N+2M)_{k} \times O(N_1)_{-2k}$ type}}
Let us freeze the $O(N_1)_{-2k}$ vector multiplet.
Then the theory becomes
$USp(2N+2M)_{k}$ SQCD with $N_1$ pairs of fundamental chiral multiplet
and the background $O(N_1)_{-2k}$ vector multiplet.
There is a conjectural duality 
for the $USp$-type SQCD \cite{Kapustin:2011gh}:
\begin{equation}
USp(2N_c)_{k} \ \leftrightarrow\ 
USp(2(N_f -N_c -1+|k|))_{-k}
\end{equation}
where $N_f$ is the number of the pairs of the fundamental chiral multipltets.
For our SQCD with $N_c =N+M$ and $N_f =2N$, 
this transforms as
\begin{equation}
M\rightarrow |k|-M-1,\quad k\rightarrow -k ,
\end{equation}
which is the same as the Seiberg-like duality
in the $USp(2N+2M)_{k}\times O(2N)_{-2k}$ ABJ theory.
When $N_1 =2N+1$,
the Giveon-Kutasov-like duality transforms as
\begin{equation}
M\rightarrow |k|-M,\quad k\rightarrow -k ,
\end{equation}
which is the same action as the Seiberg-like duality
in the $USp(2N+2M)_{k}\times O(2N+1)_{-2k}$ ABJ theory.

\section{{\it $O(2N )_{-2k}$ CS v.s. $U(N)_{-k}$ CS theories}}
\label{app:OU}
In this appendix we derive a simple relation 
between eigenvalue densities in the $U(N)$ and $O(2N)$ Chern-Simons matrix models in the planar limit.
The $S^3$ partition function in the $U(N)_{-k}$ CS theory is \cite{Marino:2002fk,Kapustin:2009kz}
\be
Z_{U(N)_{-k}}
=\frac{1}{N!} \int  \frac{d^N \nu}{ (2\pi )^N} e^{-S_U (\nu )}~,
\ee
where\footnote{
The minus sign is just convention for convenience in the main text.
} $g_s^U =-2\pi {\rm i}/k$ and 
\be
S_U (\nu )
=\frac{1}{2g_s^U}\sum_{a=1}^N \nu_a^2
-\sum_{1\leq a<b \leq N} \log{\left( 2\sinh{\frac{\nu_a -\nu_b }{2}} \right)^2}~.
\ee
In the planar limit $N\rightarrow \infty$, $t_U = g_s^U N ={\rm fixed}$,
the matrix integral is dominated by a saddle point determined by
\be
\frac{1}{g_s^U}\nu_a
-2\sum_{b\neq a} \coth{\frac{\nu_a -\nu_b}{2}}=0~.
\ee
Introducing the eigenvalue density
\be
\rho_U (\nu ;t_U ) =\frac{1}{N}\sum_{a=1}^N \delta (\nu -\nu_a ) ,
\ee
the saddle point equation becomes
\be
\frac{1}{t_U}\nu
-2 P\int dx \rho_U (x;t_U ) \coth{\frac{\nu -x}{2}}=0~.
\ee
Under the standard one cut ansatz
$\rho_U (\nu ;t_U )$ 
has been explicitly found (see e.g.~\cite{Marino:2011nm}) and 
satisfies
\be
\rho_U (\nu ;t_U ) = \rho_U (-\nu ;t_U )~.
\ee
This means that $\rho_U (\nu ;t_U )$ also satisfies
\be
\frac{1}{t_U}\nu
-2 P\int dx \rho_U (x;t_U ) \coth{\frac{\nu +x}{2}} = 0~.
\ee
Combining the two saddle point equations,
we also find the equivalent saddle point equation
\be
\frac{2}{t_U}\nu
-2 P\int dx \rho_U (x;t_U ) \coth{\frac{\nu -x}{2}}
-2 P\int dx \rho_U (x;t_U ) \coth{\frac{\nu +x}{2}} = 0~.
\label{eq:saddleU}
\ee

The action for the $O(2N)_{2k}$ CS matrix model is
\be
S_O (\nu )
=\frac{1}{2g^{O}_s}\sum_{a=1}^N \nu_a^2
-\sum_{1\leq a<b\leq N}\Biggl[ 
\log{\left( 2\sinh{\frac{\nu_a -\nu_b }{2}}  \right)^2}
+\log{\left( 2\sinh{\frac{\nu_a +\nu_b }{2}}  \right)^2} \Biggr] ,\quad 
g^{O}_s=-\pi {\rm i}/k~,
\ee
which leads to the saddle point equation
\be
\frac{1}{g^O_s}\nu_a
-2\sum_{b\neq a} \coth{\frac{\nu_a -\nu_b}{2}}
-2\sum_{b\neq a} \coth{\frac{\nu_a +\nu_b}{2}} = 0~.
\ee
Introducing the eigenvalue density $\rho_O (\nu ;t_{O} )$, the saddle point equation takes the form
\be
\frac{1}{t_O}\nu
-2 P\int dx \rho_O (x;t_{ O}) \coth{\frac{\nu -x}{2}}
-2 P\int dx \rho_O (x;t_{ O}) \coth{\frac{\nu +x}{2}} = 0~.
\ee
Comparing this with the \eqref{eq:saddleU}, we find that
the saddle point equation is solved by the following eigenvalue density
\be
\rho_O (\nu ;t_O) =\rho_U (\nu ;t_U=2t_O)~.
\ee
Thus we can use the solution of the $U(N)$ CS matrix model
for the $O(2N)$ CS matrix model.
This can be understood 
as a particular example of the so-called orbifold equivalence for field theories in the planar limit \cite{Kovtun:2004bz}.

\section{{\it Proof of $c_T=4\tau_f$ in the ${\cal N}=5$ ABJ theory}}
\label{obfproj}
%
\begin{table}[t]
\begin{center}
  \begin{tabular}{|c|c|c|c|c|  }
  \hline   & $A_1$ & $A_2$ & $B_1$  & $B_2$    \\\hline
$U(1)_1$   & $+1$   & $+1$   & $-1$ & $-1$      \\\hline
$U(1)_2$   & $+1$   & $-1$   & $+1$ & $-1$      \\\hline
$U(1)_3$   & $+1$   & $-1$   & $-1$ & $+1$      \\\hline
  \end{tabular}
\end{center}
\caption{Assignments of $U(1)^3$ flavor charges in the $\mathcal{N}=6$ ABJ theory in 3d $\mathcal{N}=2$ language.
}
\label{t8}
\end{table}
In this appendix we show the relation $c_T=4\tau_f$ holds in the ${\cal N}=5$ ABJ theory.
We first review the derivation of this relation in the ${\cal N}=6$ ABJ theory \cite{Chester:2014fya}.
It is known \cite{Chester:2014fya} that 
$c_T$ in a 3d ${\cal N}=2$ CFT with a classical SUGRA dual is related to the sphere energy $F_{S^3}=-\log{Z_{S^3}}$ by
\be
\left. c_T \right|_{\rm SUGRA} 
=\left. \frac{64}{\pi^2} F_{S^3} \right|_{\rm SUGRA} .
\label{eq:cTF}
\ee
$\tau_f$ also has a simple relation to $F_{S^3}$ in the ${\cal N}=6$ ABJ theory in the classical SUGRA limit.
Let us consider three $U(1)$ flavor symmetries in the ${\cal N}=6$ ABJ theory
explained in Table~\ref{t8}
and analyze the coefficient $\tau_a$ in the two-point function of $U(1)_a$ flavor current ($a=1,2,3$).
We can compute $\tau_a$ 
in terms of the $S^3$ free energy $F(m)$ deformed by the real mass\footnote{
In \cite{Chester:2014fya}, 
this analysis is done by means of trial $U(1)_R$ charges
but technically this is equivalent to using real masses.
} $m_a$ associated with $U(1)_a$ \cite{Closset:2012vg}:
\be
\tau_a
= 8 \left. {\rm Re}\frac{\del^2 F(m)}{\del m_a^2} \right|_{m_a =0}.
\label{eq:taua}
\ee
One can actually show $\tau_1 =\tau_2 =\tau_3$ \cite{Chester:2014fya}
and we simply denote $\tau_a $ by $\tau_f$ below.
An explicit calculation shows that
$\tau_f$ in the SUGRA limit of the ${\cal N}=6$ ABJ theory is given by \cite{Chester:2014fya}
\be
\left. \tau_f \right|_{\rm SUGRA}
=\left. \frac{16}{\pi^2} F_{S^3}^{\mathcal{N}=6} \right|_{\rm SUGRA} .
\label{eq:tauF}
\ee
Comparing this with \eqref{eq:cTF}, one immediately finds $c_T = 4\tau_f$.

Let us now turn to the ${\cal N}=5$ case.
For this case,
the relation \eqref{eq:cTF} between $c_T$ and $F_{S^3}$ still holds in the SUGRA limit
but we do not know at this moment 
whether \eqref{eq:tauF} is also correct
since there exist no explicit calculations to check \eqref{eq:tauF} in literature.
Here instead of using \eqref{eq:cTF} and \eqref{eq:tauF}  directly,
we adopt the idea of large-$N$ orbifold equivalence or orientifold equivalence \cite{Kovtun:2004bz}
which states that
when theory $B$ is obtained from theory $A$ via a projection by the group $\Gamma$,
then in the planar limit the free energies of these two theory satisfy
\be
\left. F_B \right|_{\rm planar} 
=\left. \frac{F_A}{|\Gamma |} \right|_{\rm planar} ,
\ee
where $|\Gamma |$ is the order of $\Gamma$.
To use the orientifold equivalence,
we regard the ${\cal N}=5$ $O(N_1 )_{2k}\times USp(2N_2)_{-k}$ ABJ theory
as the quotient of the ${\cal N}=6$ $U(N_1 )_{2k}\times U(2N_2)_{-2k}$ theory
by the projection \eqref{CFT65}.
First we consider the relation between $c_T$'s 
in the $\mathcal{N}=5$ and $\mathcal{N}=6$ ABJ theories.
It is known \cite{Closset:2012ru} that
$c_T$ in 3d $\mathcal{N}=2$ superconformal field theory 
is related to the squashed sphere free energy $F_{S^3_b}$ by 
\be
c_T=\frac{32}{\pi^2}{\rm Re}\frac{\partial^2 F_{S^3_b}}{\partial b^2}\Big|_{b=1}~,
\ee
where $b$ is the squashing parameter and $b=1$ corresponds to the round sphere.
Combining this with the orientifold equivalence leads us to\footnote{
One can also  check this explicitly
by using \eqref{eq:cTF} and the result of \cite{Gulotta:2012yd}
on the free energy of the $\mathcal{N}=5$ ABJ theory
in the classical SUGRA limit.
}
\be
\left. c_T^{O(N_1 )_{2k}\times USp(2N_2 )_{-k}}\right|_{\rm planar} 
=\left. \frac{1}{2} c_T^{U(N_1 )_{2k}\times U(2N_2)_{-2k}} \right|_{\rm planar}.  
\label{eq:cT_planar}
\ee

Next let us proceed to $\tau_f$.
Since the orientifold projection breaks the $U(1)_1$ and $U(1)_3$ symmetries,
we consider $\tau_f$ associated with the $U(1)_2$ symmetry,
which assigns charge $+1$ to one chiral multiplet
and charge $-1$ to the other chiral multiplet.
Then using \eqref{eq:taua} and the orientifold equivalence,
we find
\be
\left. \tau_f^{O(N_1 )_{2k}\times USp(2N_2 )_{-k}}\right|_{\rm planar} 
=\left. \frac{1}{2} \tau_f^{U(N_1 )_{2k}\times U(2N_2)_{-2k}} \right|_{\rm planar}.  
\ee
Combining this with \eqref{eq:cT_planar} 
and $c_T=4\tau_f$ for the $\mathcal{N}=6$ case,
we easily see that
$c_T=4\tau_f$ still holds in the ${\cal N}=5$ ABJ theory.

\section{{\it Gauge current correlation functions}}
\label{app:beforeG}
In order to determine the relation between the bulk parity violating phase $\theta$ and parameters in the ABJ theory,
we need the coefficients in the two-point functions of the currents 
associated with the ``smaller" gauge group.
In this appendix, we shall compute these coefficients in the higher spin limits.
We emphasize that
this has not been done even for the $\mathcal{N}=6$ case.

\subsubsection*{\underline{$U(N_1 )$ gauge current in the $U(N_1 )_{k} \times U(N_2 )_{-k}$ theory}}
The $\mathcal{N}=6$ ABJ theory can be viewed as
gauging the $U(N_1 )$ flavor symmetry with the CS gauge field at level $k$
in the $U(N_2 )_{-k}$ SQCD with $2N_1$ fundamental hyper multiplets.
More precisely,
if we parametrize the $2N_1$ hypers by $(Q_j ,Q_j' )$ $(j=1,\cdots ,N_1 )$,
then the gauged flavor symmetry is
$U(N_1 )$ rotation of $Q_j$ and $Q_j'$ simultaneously.
In the higher spin limit,
the $U(N_1 )$ gauge interaction is very weak and
we can approximate the two point function of the $U(N_1 )$ gauge current in the ABJ theory
by the one of the $U(N_1 )$ flavor current in the SQCD, 
which can be computed by localization.

To compute the coefficients in the two point function of the $U(N_1 )$ flavor current,
we need the $S^3$ partition function of the SQCD deformed by the real mass associated with the $U(N_1 )$ flavor symmetry.
We can easily write down
the partition function before gauging 
just by freezing the integrals over the $U(N_1 )$  vector multiplet in the ABJ theory:
\begin{equation}
Z_{U(N_2 )_{-k}\rm SQCD}
= \frac{e^{\frac{ik}{4\pi}\sum_{j=1}^{N_1}\mu_j^2}}{N_2! } 
\int \frac{d^{N_2}\nu}{(2\pi )^{N_2} }
e^{-\frac{ik}{4\pi}\sum_{a=1}^{N_2}\nu_a^2 } 
\frac{  \prod_{a\neq b} 2\sinh{\frac{  \nu_a -\nu_b }{2}} }
 {\prod_{j=1}^{N_1} \prod_{a=1}^{N_2} \left( 2\cosh{\frac{\mu_j -\nu_a }{2}} \right)^2 }  ,
\end{equation}
where $\mu_j$ is the real mass associated with the $U(N_1 )$ flavor symmetry
and the numerator in the first factor is 
the CS term of the background $U(N_1 )$ vector multiplet with the level $k$.
For our purpose, 
it is sufficient to know only one component of $U(N_1 )$ and
hence we take $\mu_j =m$:
\begin{equation}
Z_{U(N_2 )_{-k}\rm SQCD}(m)
= \frac{e^{\frac{iN_1 k}{4\pi}m^2} }{N_2! } \int \frac{d^{N_2}\nu}{(2\pi )^{N_2} }
e^{-\frac{ik}{4\pi}\sum_{a=1}^{N_2}\nu_a^2 } 
\frac{  \prod_{a\neq b} 2\sinh{\frac{  \nu_a -\nu_b }{2}} }
 {\prod_a\left( 2\cosh{\frac{m -\nu_a }{2}} \right)^{2N_1} }  ,
\end{equation}
where $m$ is understood as the real mass 
associated with the diagonal part of $U(N_1 )$ symmetry.
We can calculate the coefficients $\tau_f$ and $\kappa_f$
by \cite{Closset:2012vg}
\begin{eqnarray}
\tau_f 
&=& -8 \left. {\rm Re}
\frac{1}{Z_{U(N_2 )_{-k}\rm SQCD}(0)}
\frac{\del^2 Z_{U(N_2 )_{-k}\rm SQCD}(m)}{\del m^2} \right|_{m=0} ,\nonumber\\
\kappa_f
&=& 2\pi {\rm Im} \left. \frac{1}{Z_{U(N_2 )_{-k}\rm SQCD}(0)}
\frac{\del^2 Z_{U(N_2 )_{-k}\rm SQCD}(m)}{\del m^2} \right|_{m=0} .
\end{eqnarray}

Similar to the procedure adopted in sec.~\ref{sec:correlation},
we can rewrite 
the mass deformed partition function in the planar limit as
\begin{equation}
Z_{U(N_2 )_{-k}\rm SQCD}(m)
=e^{\frac{iN_1 k}{4\pi}m^2} Z_{U(N_2 )_{-k}{\rm CS}}
\exp{\Biggl[ 
 2N_1 N_2 g_U (-e^{-m};t_2 ) \Biggr]} .
\end{equation}
Now recall that
the mass deformed partition function of ABJ theory 
in the HS limit is \cite{Honda:2015sxa}
\begin{equation}
Z_{\mathcal{N}=6 \rm ABJ}(m)
= Z_{\rm Gauss}^U \left( \frac{2\pi i}{k} ,N_1 \right)
 Z_{U(N_2 )_{-k}{\rm CS}}
\exp{\Biggl[ 
 N_1 N_2 \left( g_U (-e^{m};t_2 ) +g_U (-e^{-m};t_2 ) \right)\Biggr]} .
\end{equation}
Comparing the above two equations,
we find
\begin{equation}
\left. \frac{1}{Z_{U(N_2 )_{-k}\rm SQCD}(0)}
\frac{\del^2 Z_{U(N_2 )_{-k}\rm SQCD}(m)}{\del m^2} \right|_{m=0} 
=\frac{ikN_1 }{2\pi}
+\left.\frac{1}{Z_{\mathcal{N}=6 \rm ABJ}(0)}\frac{\del^2 Z_{\mathcal{N}=6 \rm ABJ}(m)}{\del m^2} \right|_{m=0} .
\end{equation}
Hence, using the result of \cite{Honda:2015sxa},
we obtain 
\begin{equation}
\tau_f  =  \frac{4N M\sin{\pi t}}{\pi t} ,\quad
\kappa_f = \frac{N M \cos{\pi t}}{t} .
\end{equation}
$\tau_f $ is the same as the one of $U(1)$ flavor symmetries 
obtained in \cite{Honda:2015sxa}
while $\kappa_f $ is diffent only by the integer $kN_1$.

\subsubsection*{\underline{$O(N_1 )$ gauge current in the $O(N_1 )_{2k}\times USp(2N_2 )_{-k}$ theory} }
We can compute the gauge current two point function as in the $\mathcal{N}=6$ case.
We regard the ABJ theory as
gauging the $O(N_1 )$ flavor symmetry with CS level $2k$
of $USp(2N_2 )_{-k}$ SQCD with $N_1$ fundamental hyper multiplets.
Then the partition function before gauging is 
\begin{equation}
Z_{USp(2N_2 )_{-k}\rm SQCD}
= \frac{e^{\frac{ik}{2\pi}\sum_{j=1}^{|O(N_1 )|} \mu_j^2 } }{2^{N_2} N_2 !} 
\int  \frac{d^{N_2} \nu}{(2\pi )^{N_2}} 
e^{-\frac{ik}{2\pi}\sum_{b=1}^{N_2} \nu_b^2 }   Z_{\rm vec}^{USp} (\nu ) Z_{\rm bi}(\mu ,\nu ,0)~,
\end{equation}
where the functions in the integrand are defined in \eqref{eq:1loopdet}
and $\mu_j$ plays a role as the real mass associated with the $O(N_1 )$ flavor symmetry
and the first exponential factor is the background $O(N_1 )$ CS term with the level $2k$.
When $\mu_j =m$,
we can rewrite the mass deformed partition function
in the planar limit as
\begin{eqnarray}
&& Z_{USp(2N_2 )_{-k}\rm SQCD}(m) \nonumber\\
&& =e^{\frac{ik|O(N_1 )|}{2\pi} m^2 } Z_{USp(2N_2 )_{-k}{\rm CS}} \times
  \left\{ \begin{matrix}
e^{ 2|O(N_1 )| N_2 \left( g_O (-e^{m} )  +g_O (-e^{-m} ) \right) }
& {\rm for\ even}\ N_1, \cr
e^{ 2|O(N_1 )| N_2 \left( g_O (-e^{m} )  +g_O (-e^{-m} ) \right)
 +2N_2 g_O (-1;t_2 ) }
& {\rm for\ odd}\ N_1.\cr
\end{matrix}\right.
\end{eqnarray}
Recalling \eqref{eq:Zm_O} and \eqref{eq:del2O},
we find
\begin{equation}
\left. \frac{1}{Z_{USp(2N_2 )_{-k}\rm SQCD}(0)}
\frac{\del^2 Z_{USp(2N_2 )_{-k}\rm SQCD}(m)}{\del m^2} \right|_{m=0} 
=\frac{ik|O(N_1 )|}{\pi}
+\left.\frac{1}{Z_{\mathcal{N}=5 \rm ABJ}(0)}\frac{\del^2 Z_{\mathcal{N}=5 \rm ABJ}(m)}{\del m^2} \right|_{m=0,N_1 \rightarrow 2|O(N_1 )|}~.
\end{equation}
Thus we obtain
\begin{equation}
\tau_f =  \frac{8|O(N_1)| M\sin{(\pi t)}}{\pi t} ,\quad
\kappa_f = \frac{2|O(N_1)| M \cos{\pi t}}{t} ~.
\end{equation}
$\tau_f $ is the same as the one of the $U(1)$ flavor symmetry \eqref{eq:flavorC_O}
while $\kappa_f$ is the same as the shifted one \eqref{eq:shifted_O}.

\subsubsection*{\underline{$USp(2N_2 )$ gauge current in the  $O(N_1 )_{2k}\times USp(2N_2 )_{-k}$ theory}}
In this case, the theory before gauging $USp(2N_2 )$ is
$O(N_1 )_{2k}$ SQCD with $N_2$ fundamental hyper multiplets
and the back ground CS term of $USp(2N_2 )_{-k}$,
whose partition function is 
\begin{equation}
Z_{O(N_1 )_{2k}\rm SQCD}
= \frac{e^{-\frac{ik}{2\pi}\sum_{a=1}^{N_2}\nu_a^2}}{|W_O|} 
\int  \frac{d^{N_1}\mu}{(2\pi )^{N_1}} 
e^{\frac{ik}{2\pi}\sum_j \mu_j^2 }  
 Z_{\rm vec}^{O} (\nu ) Z_{\rm bi}(\mu ,\nu ,0)~,
\end{equation}
where $\nu_a$ is now understood as the real mass associated with $USp(2N_2 )$
and the exponential prefactor is the background $USp(2N_2 )$ CS term of the level $-k$.
If we take $\nu_a =m$,
the mass deformed partition function in the planar limit becomes
\begin{equation}
Z_{O(N_1 )_{2k}\rm SQCD}(m)
=e^{-\frac{ikN_2 }{2\pi}m^2} Z_{O(N_1 )_{2k}{\rm CS}} 
e^{ 2|O(N_1 )| N_2 \left( g_O (-e^{m} )  +g_O (-e^{-m} ) \right) }
\times
  \left\{ \begin{matrix}
1 & {\rm for\ even}\ N_1~,\cr
\frac{1}{\left( 2\cosh{\frac{m }{2}} \right)^{2N_2}}
& {\rm for\ odd}\ N_1~. \cr
\end{matrix}\right.
\end{equation}
Recalling \eqref{eq:Zm_Usp} and \eqref{eq:del2USp},
we find
\begin{equation}
\left. \frac{1}{Z_{O(N_1 )_{2k}\rm SQCD}(0)}
\frac{\del^2 Z_{O(N_1 )_{2k}\rm SQCD}(m)}{\del m^2} \right|_{m=0} 
=\frac{ikN_2}{\pi}
+\left. \frac{1}{Z_{\mathcal{N}=5 \rm ABJ}(0)}\frac{\del^2 Z_{\mathcal{N}=5 \rm ABJ}(m)}{\del m^2} \right|_{m=0}~,
\end{equation}
which immediately gives rise to 
\begin{equation}
\tau_f (t)
=  \frac{8N_2 M\sin{(\pi t)}}{\pi t} , \quad
\kappa_f 
= -\frac{2N_2 M\cos{\pi t}}{t} ~ .
\end{equation}
$\tau_f $ is the same as the one of the $U(1)$ flavor symmetry \eqref{eq:flavorC_USp}
while $\kappa_f$ is the same as the shifted one \eqref{eq:shifted_USp}.

\section{{\it The ratio of the pure CS partition functions}}
\label{rcspf}
In sec.~\ref{sec:correlation}, we have utilized the ratio of 
the $S^3$ partition functions of the pure CS theory  in the large $M$-expansion.  
In this appendix, we present more details of computing the ratio.
The pure CS partition function with gauge group $G$ is
\begin{equation}
Z_{\rm CS}^{\rm G} (g)
= \frac{1}{|W| }
\int \frac{d^{|G|}x}{(2\pi )^{|G|} }
 e^{-\frac{1}{2g}{\rm tr} x^2 }
     \prod_{\alpha\neq 0 } 2\sinh{\frac{\alpha\cdot x}{2}} 
= ({\rm det}C)^{\frac{1}{2}}
\frac{i^{-(\sum_{\alpha >0} 1 )-\frac{|G|}{2}}}{k^{\frac{|G|}{2}}}
e^{\frac{2\pi i}{k}\rho^2} \prod_{\alpha >0}2\sin{\frac{\pi\alpha\cdot\rho}{k}}~,
\end{equation}
where $C$ is Cartan matrix and $\rho$ is Weyl vector $\rho = \frac{1}{2}\sum_{\alpha >0}\alpha$.

\subsubsection*{\underline{$U(N)$ type}}
When $G=U(N)$, we have
\be
\left| Z_{\rm CS}^{U(N)} (g) \right|
=\frac{1}{k^{N/2}} 
\prod_{1\leq \ell <m\leq N}2\sin{\frac{\pi (m-\ell )}{k}}
=\frac{1}{k^{N/2}}
\prod_{\ell =1}^N \left( 2\sin{\frac{\pi\ell}{k}} \right)^{N-\ell}~.
\ee
Now we would like to expand
$\log{ |Z_{\rm CS}^{U(N+M)_k}| /|Z_{\rm CS}^{U(M)_k}|  }$
up to $\mathcal{O}(\log{M})$.
This can be done by using the technique in \cite{Gopakumar:1998ii}.
We first rewrite the pure CS free energy as
\be
\log{|Z_{\rm CS}^{U(M)_k}|  }
= -\frac{M}{2}\log{k} +\sum_{j=1}^{M-1} (M-j) \log{\left( 2\sin{\frac{\pi j}{k}} \right)}~.
\ee
To expand this, we use the formula
\be
\sin{(\pi z)} =\pi z \prod_{m=1}^\infty \left( 1-\frac{z^2}{m^2} \right)\,, \quad
\sum_{j=1}^{N-1} (N-j)\log{j} =\log{G_2 (N+1)}~,
\ee
where $G_2 (z)$ is the Barnes G-function.
Then the free energy becomes
\begin{equation}
\log{|Z_{\rm CS}^{U(M)_k}|  }
= -\frac{M}{2}\log{k} +\frac{M}{2}(M-1) \log{\frac{2\pi }{k}} +\log{G_2 (M+1)}
        +\sum_{j=1}^{M-1} (M-j) \sum_{m=1}^\infty \log{\left( 1-\frac{j^2}{m^2 k^2} \right)}~.
\end{equation}
The last term is often referred to as perturbative piece:
\be
\log{Z^P (M)_k}
= \sum_{j=1}^{M-1} (M-j) \sum_{m=1}^\infty \log{\left( 1-\frac{j^2}{m^2 k^2} \right)}
=\sum_{m=1}^\infty \frac{\zeta (2m)}{m} \frac{1}{k^{2m}}
 \sum_{j=1}^{M-1} (M-j) j^{2m}~,
\ee
which has the expansion \cite{Gopakumar:1998ii}
\be
\log{Z^P (M)_k}
= \sum_{g=0}^\infty \sum_{h=2}^\infty F_{g,h}^P \left( \frac{2\pi}{k} \right)^{2g-2+h} M^h~.
\ee
Here we need only the $g=0$ part,
whose coefficients are
\be
F_{0,\,h\leq 3}^P = 0~,\quad
F_{0,\,h\geq 4}^P = -\frac{|B_{h-2}|}{(h-2)h!}~ .
\ee
Using the above formula, we find
\begin{equation}
\log{ \frac{|Z_{\rm CS}^{U(N+M)_k}| }{|Z_{\rm CS}^{U(M)_k}| } }
= -NM +NM\log{(2\pi t)}
 +\frac{N}{M}\sum_{h=4}^\infty h F_{0,h}^P
\left( \frac{2\pi}{k} \right)^{-2+h} M^h +\mathcal{O}(1)~,
\end{equation}
where we have used
\be
\log{G_2 (N+1)}
=  \frac{N^2}{2} \log{N} -\frac{1}{12}\log{N} -\frac{3}{4}N^2
+\mathcal{O}(1)~.
\ee
Performing the sum over $h$ explicitly,
we finally obtain
\begin{equation}
\log{ \frac{|Z_{\rm CS}^{U(N+M)_k}| }{|Z_{\rm CS}^{U(M)_k}| } }
= NM\left(  \log{t}
        +\frac{\zeta ^{(1,0)}(-1,1-t) -\zeta ^{(1,0)}(-1,1+t)}{t} \right)
 +\mathcal{O}(1)~.
\end{equation}

\subsubsection*{\underline{ $O(2N)$ type}}
For the $O(2N)$ gauge group, we have
\bea
\left| Z_{\rm CS}^{O(2N)_{2k}}  \right|
&=& \frac{2}{(2k)^{N/2}}
\prod_{1\leq \ell <m \leq N}2\sin{\frac{\pi (m-\ell )}{2k}} \cdot
2\sin{\frac{\pi (m+\ell )}{2k}} \nn\\
&=& \frac{2^{3/4}}{k^{1/2}}
\frac{|Z_{\rm CS}^{U(N)_{2k}}|}{| Z_{\rm CS}^{U(N+1)_{2k}}|}
\Biggl[
\frac{| Z_{\rm CS}^{U(N)_k} |\cdot | Z_{\rm CS}^{U(2N+1)_{2k}}|}
{| Z_{\rm CS}^{U(N+1)_k}|}
 \Biggr]^{1/2}~.
\eea
Using the result for the $U(N)$ case, we find
\bea
\log{ \frac{|Z_{\rm CS}^{O(2N+2M)_{2k}}| }{|Z_{\rm CS}^{O(2M)_{2k}}| } }
&=&\frac{1}{2} \log{ \frac{|Z_{\rm CS}^{U(2N+2M+1)_{2k}}| }
{|Z_{\rm CS}^{U(2M+1)_{2k}}| } } +\mathcal{O}(1)\nn\\
&=& 2NM\left(  \log{t}
        +\frac{\zeta ^{(1,0)}(-1,1-t) -\zeta ^{(1,0)}(-1,1+t)}{t} \right)
 +\mathcal{O}(1)~.
\eea
%

\subsubsection*{\underline{$O(2N+1)$ type}}
For the $O(2N+1)$ gauge group, the CS partition function is
\bea
\left| Z_{\rm CS}^{O(2N+1)_{2k}}  \right|
&=&\frac{\sqrt{2}}{(2k)^{N/2}}
\prod_{1\leq \ell <m \leq N}2\sin{\frac{\pi (m-\ell )}{2k}} \cdot 2\sin{\frac{\pi (m+\ell )}{2k}}
\prod_{\ell =1}^N 2\sin{\frac{\pi \ell}{2k}} \nn\\
&=& k^{1/2}
\left| Z_{\rm CS}^{O(2N)_{2k}}  \right|
\frac{| Z_{\rm CS}^{U(N+1)_{2k}}|} {| Z_{\rm CS}^{U(N)_{2k}}|}~.
\eea
Then the result for the $U(N)$ case leads us to
\bea
\log{ \frac{|Z_{\rm CS}^{O(2N+2M+1)_{2k}}| }{|Z_{\rm CS}^{O(2M+1)_{2k}}| } }
&=& \log{ \frac{|Z_{\rm CS}^{O(2N+2M)_{2k}}| }{|Z_{\rm CS}^{O(2M)_{2k}}| } }
  +\mathcal{O}(1)\nn\\
&=& 2NM\left(  \log{t}
        +\frac{\zeta ^{(1,0)}(-1,1-t) -\zeta ^{(1,0)}(-1,1+t)}{t} \right)
 +\mathcal{O}(1)~.
\eea
%

\subsubsection*{\underline{$USp(2N)$ type}}
The CS partition function for this case is given by
\bea
\left| Z_{\rm CS}^{USp(2N)_k}  \right|
&=& \frac{\sqrt{2}}{(2k)^{N/2}}
\prod_{1\leq \ell <m \leq N}2\sin{\frac{\pi (m-\ell )}{2k}} \cdot 2\sin{\frac{\pi (m+\ell )}{2k}}
\prod_{\ell =1}^N 2\sin{\frac{\pi \ell}{k}} \nn\\
&=& \frac{k^{1/2}}{\sqrt{2}}
\left| Z_{\rm CS}^{O(2N)_{2k}}  \right|
\frac{| Z_{\rm CS}^{U(N+1)_{k}}|} {| Z_{\rm CS}^{U(N)_{k}}|}~.
\eea
Thus using the above results, we obtain
\bea
\log{ \frac{|Z_{\rm CS}^{USp(2N+2M)_{k}}| }{|Z_{\rm CS}^{USp(2M)_{k}}| } }
&=& \log{ \frac{|Z_{\rm CS}^{O(2N+2M)_{2k}}| }{|Z_{\rm CS}^{O(2M)_{2k}}| } }
  +\mathcal{O}(1)\nn\\
&=& 2NM\left(  \log{ t}
        +\frac{\zeta ^{(1,0)}(-1,1-t) -\zeta ^{(1,0)}(-1,1+t)}{t} \right)
 +\mathcal{O}(1)~.
\eea
%

{}

\end{document}